\documentclass[11pt,a4paper]{article}

\pdfoutput=1
\usepackage{jheppub}

\hypersetup{pdfencoding=unicode, bookmarksopen=true, bookmarksnumbered}
\usepackage[nomath]{lmodern}
\usepackage{microtype}
\usepackage{amsthm,amsbsy,amsfonts,mathrsfs,enumerate,float,wrapfig,amsmath}
\usepackage{multirow}
\usepackage{subfigure}
\usepackage{longtable}
\usepackage{adjustbox} 
\usepackage{physics}
\usepackage{enumitem}
\usepackage{framed}
\usepackage{booktabs}
\usepackage {dashrule}
\usepackage{tikz}
\usetikzlibrary{arrows.meta, positioning}
\newcommand{\be}{\begin{equation}}
\newcommand{\ee}{\end{equation}}

\usepackage{amsmath}
\usepackage{xcolor}

\definecolor{light_blue}{rgb}{0.15, 0.35, 0.95}
\definecolor{kit_green}{rgb}{
    0, 
    0.58823, 
    0.50980  
}

\title{6d Supergravity Blocks}

\author[a,b,c]{Yuta Hamada}
\author[d]{Seongmin Jeon}
\author[d,e]{Hee-Cheol Kim}

\affiliation[a]{Theory Center, IPNS, High Energy Accelerator Research Organization (KEK), 1-1 Oho, Tsukuba, Ibaraki 305-0801, Japan}
\affiliation[b]{Graduate Institute for Advanced Studies, SOKENDAI, 1-1 Oho, Tsukuba, Ibaraki 305-0801, Japan}
\affiliation[c]{RIKEN Center for Interdisciplinary Theoretical and Mathematical Sciences (iTHEMS), RIKEN, Wako 351-0198, Japan}
\affiliation[d]{Department of Physics, POSTECH, Pohang 37673, Korea}
\affiliation[e]{Asia Pacific Center for Theoretical Physics, Postech, Pohang 37673, Korea}

\abstract{We propose a systematic framework for constructing six-dimensional supergravity theories with eight supercharges that respect all known consistency constraints, including anomaly cancellation and the non-perturbative {\it H}-string constraints recently discovered by Kim, Vafa, and Xu.  The basic objects in this framework are {\it supergravity blocks}, which are minimal collections of tensor multiplets consisting of a single little string theory sharing the {\it H}-string charge together with additional tensors whose string charges intersect it positively. A characteristic feature of each supergravity block is that its Gram matrix has exactly one positive eigenvalue, and therefore it necessarily contains gravitational BPS strings that cannot become tensionless anywhere in tensor moduli space. Any consistent 6d $(1,0)$ supergravity theory can then be obtained by gluing compatible blocks and subsequently enhancing the gauge algebras and matter content. As a first step toward establishing this framework concretely, we provide a complete classification of the {\it non-Higgsable supergravity blocks}, (or {\it non-Higgsable gravity blocks} for short) namely those built from tensor multiplets that support only non-Higgsable gauge algebras.}

\preprint{KEK-TH-2849,~RIKEN-iTHEMS-Report-26}
\begin{document}

\maketitle

\section{Introduction}

Six-dimensional supergravity theories with eight supercharges provide a particularly constrained framework for studying quantum gravity. Their low-energy effective descriptions are highly constrained by local consistency requirements, including anomaly cancelation, supersymmetry, and unitarity, while simultaneously exhibiting intrinsically non-perturbative phenomena such as tensionless strings and strongly coupled fixed points. These features make 6d ${\cal N}=(1,0)$ supergravity an especially sharp setting for investigating the Swampland program \cite{vafa2005string}, whose aim is to delineate the space of effective field theories that admit a consistent ultraviolet completion in quantum gravity.

A central lesson arising from recent work is that, although the perturbative consistency conditions in six dimensions are remarkably restrictive, they are not sufficient to fully characterize the set of consistent quantum gravity theories. A crucial additional ingredient is provided by non-perturbative effects, most notably the emergence of tensionless BPS strings at the boundaries of the tensor moduli space where the effective field theory description breaks down. Such boundaries are not arbitrary singularities of the effective theory, but are instead associated with well-defined UV descriptions in terms of 6d superconformal field theories (SCFTs), little string theories (LSTs), or critical string theories. From the Swampland perspective, the existence and consistency of these tensionless strings furnish indispensable diagnostics of consistent UV completions, particularly in the context of the Emergent String Conjecture proposed in \cite{Lee:2019wij}.

Recent advances in \cite{Kim:2024eoa} have promoted this insight to a concrete organizing principle based on the conjecture that every boundary of the tensor moduli space of a consistent 6d ${\cal N}=(1,0)$ supergravity must be associated with a tensionless BPS string. Together with the classification of 6d SCFTs and LSTs in \cite{Heckman:2013pva,Heckman:2015bfa,Bhardwaj:2015xxa,Bhardwaj:2015oru,Bhardwaj:2019hhd}, this principle imposes strong constraints on the tensor charge lattice and its intersection structure. In particular, the tensor moduli space (more precisely its dual BPS cone) is generated by a set of primitive BPS string charges, whose mutual intersections encode the global consistency of the tensor moduli space. This viewpoint has already led to striking consequences, such as the existence of a distinguished BPS string, the {\it H}-string associated with a critical heterotic string and LSTs, and stringent bounds on the rank of admissible gauge algebras arising from worldsheet unitarity constraints. An especially significant recent result in this context is the proof of the finiteness of the number of massless fields in consistent 6d ${\cal N}=(1,0)$ supergravity theories \cite{Kim:2024eoa}, which was mainly motivated by the finiteness conjecture proposed in \cite{douglasprivate,vafa2005string,acharya2006finite,Hamada:2021yxy}.
See also other attempts to constrain the six-dimensional supergravity landscape \cite{Avramis:2005hc,Kumar:2009ae,Kumar:2009ac,Kumar:2010ru,Bedroya:2021fbu,Kim:2019vuc,Lee:2019skh,Tarazi:2021duw,Cvetic:2021vsw,Lee:2022swr,Hayashi:2023hqa,Hamada:2023zol,Kim:2024tdh,Hamada:2025vga,Lockhart:2025lea,Baykara:2025gcc,Lockhart:2026xml,Lockhart:2026gvb}. 
Complementary recent work has also pursued explicit uniform bounds on the spectrum using methods motivated by birational geometry \cite{Birkar:2025rcg,Birkar:2025gvs}.

The purpose of this work is to exploit these new developments in order to formulate a systematic framework of constructing admissible six-dimensional ${\cal N}=(1,0)$ supergravity theories. Our construction makes extensive use of the refined tensor intersection structures developed in \cite{Kim:2024eoa}. A key ingredient in this framework is the presence of the {\it H}-string, whose charge is shared by a special class of LSTs, together with the fact that the tensor intersection form of a general 6d supergravity theory can be realized geometrically in terms of the topology of K\"ahler surfaces such as $\mathbb{P}^2$, Hirzebruch surfaces $\mathbb{F}_n$, and their blowups, which in geometric constructions can serve as bases of elliptic Calabi-Yau (CY) 3-folds \cite{Vafa:1996xn,Morrison:1996na,Morrison:1996pp,Bershadsky:1996nh,Morrison:2012np,Morrison:2012js,Taylor:2015isa}. In turn, the BPS generators, which are primitive charges for BPS strings, are organized into two classes: those contained in the LSTs associated with the {\it H}-string, and those that positively intersect the {\it H}-string, which we call {\it external generators}. A tensor base of any supergravity theory is specified by the intersections of these BPS generators. For a given tensor base, we can construct a supergravity theory, more precisely the massless sector of a supergravity, by introducing gauge algebras supported on tensor charges and charged hypermultiplets.

We propose that the natural unit in this construction is the notion of a {\it supergravity block}. A supergravity block is a minimal subsystem consisting of a single {\it H}-string LST sector together with a set of external generators attached to it. The essential feature of this subsystem is that the Gram matrix of its tensor charges contains exactly one positive eigenvalue. This guarantees that the BPS strings supported on the subsystem cannot all become simultaneously tensionless anywhere in tensor moduli space. Consequently, the subsystem contains tensor charges that are intrinsically gravitational and can give rise to BPS black strings when embedded in a supergravity. In this sense, supergravity blocks are the minimal subsystems in this framework that are intrinsically gravitational, rather than belonging to a local SCFT or LST sector. They therefore provide the natural building blocks for assembling general supergravity theories.

Once the complete set of supergravity blocks is known, the construction of a general theory can be organized into five operations:  (i) choose a set of external generators; (ii) collect from the classification table certain supergravity blocks compatible with that choice;  (iii) glue these blocks together along the chosen external generators to form a candidate tensor base; (iv) verify that this tensor base admits a blowdown to a Hirzebruch surface; and (v) decorate the tensors of positive self-intersection with gauge algebras and matter content in such a way that both gauge and gravitational anomalies are canceled. A key point is that this entire procedure is formulated purely in terms of the data in the effective field theory and, especially, does not assume the existence of a geometric realization such as F-theory. The resulting framework therefore applies equally well to geometric constructions and to intrinsically non-geometric 6d supergravity theories.

In this paper, as a first concrete implementation of this framework, we focus on a particularly tractable subclass, namely \emph{non-Higgsable gravity blocks}.  These are the supergravity blocks built entirely from non-Higgsable clusters (NHCs) familiar from the classification of 6d SCFTs and LSTs. NHCs are basic tensor building blocks supporting the smallest possible gauge algebras, and they are completely classified \cite{Morrison:2012np}. We give a complete classification of these non-Higgsable blocks and explain how they can be used as inputs for constructing more general supergravity theories. Restricting to this subclass allows us to separate the topology of the tensor base from the subsequent choice of gauge algebras and charged matter. Once the non-Higgsable gravity blocks are classified, more general supergravity theories can be constructed by gluing them into tensor bases and then enhancing the gauge algebras and matter content on those bases, including Abelian sectors subject to their anomaly constraints \cite{Park:2011wv,Taylor2018}. In this sense, the present work offers a systematic organization of the space of consistent 6d $(1,0)$ supergravity theories in terms of tensor-base building blocks and provides a natural starting point for a full classification of six-dimensional supergravity theories, which we leave for future investigation.

The structure of the paper is as follows. In Section~\ref{sec:supergravity-review} we review the basic ingredients of 6d ${\cal N}=(1,0)$ supergravity and the role of BPS strings in the tensor moduli space. Section~\ref{sec:construction} introduces the refined structure of the tensor moduli space and supergravity blocks. In Section~\ref{sec:classification} we present our main classification results. Section~\ref{sec:conclusion} concludes with discussions and future directions.

\section{6d \texorpdfstring{$\mathcal{N}=(1,0)$}{N=(1,0)} supergravity}\label{sec:supergravity-review}
In this subsection we briefly review several basic aspects of six-dimensional $\mathcal{N}=(1,0)$ supergravity theories. First, the massless field content consists of a gravity multiplet,  $T$ tensor multiplets, $V$ vector multiplets, and $H$ hypermultiplets. The gravity multiplet comprises the graviton $g_{\mu\nu}$ and a self-dual two-form field $B^+_{\mu\nu}$. Each tensor multiplet contains an anti-self-dual tensor field $B^-_{\mu\nu}$ together with a real scalar field. The vector multiplets gauge a symmetry of the general form $G=\prod_iG_i$, and the hypermultiplets transform in representations of this gauge symmetry. In what follows we restrict to theories without Abelian gauge factors, as these are the primary cases relevant for the analysis presented below.

The scalar fields in the tensor multiplets, collectively denoted by $J \in \mathbb{R}^{1,T}$, parametrize the tensor moduli space. This moduli space is the coset space $SO(1,T)/SO(T)$ subject to the conditions $J^0 > 0$ and $J \cdot J \equiv \Omega_{\alpha\beta} J^\alpha J^\beta = 1$. Here the matrix $\Omega_{\alpha\beta}$ is the tensor intersection form with Lorentzian signature $(+,(-)^T)$ and it provides the canonical pairing of charges in the tensor (string) charge lattice $\Gamma$.

Strong constraints on consistent 6d supergravity theories arise from the requirement of gravitational, gauge, and mixed anomaly cancellation. These local anomalies must be canceled via the Green-Schwarz-Sagnotti mechanism \cite{Green:1984sg,Green:1984bx,Sagnotti:1992qw,Erler:1993zy}, which requires the one-loop anomaly polynomial $I_8$ to factorize as
\begin{align}
    I_8  = \frac{1}{2}\Omega_{\alpha\beta}X^\alpha_4 X^\beta_4 \ , \quad 
    X_4^\alpha = -\frac{1}{2}b_0^\alpha {\rm tr}R^2 + \frac{1}{2}\sum_i\frac{b_i^\alpha}{\lambda_i} {\rm tr} F_i^2 \ .
\end{align}
Here $R$ and $F_i$ denote the curvature and gauge field strengths, respectively. The coefficients $b_0$ and $b_i^\alpha$, referred to as anomaly vectors, lie in the string charge lattice $\Gamma$. The constants $\lambda_i$ are normalization constants for the non-Abelian gauge groups $G_i$ listed in Table \ref{tb:Normalization}, and the trace `${\rm tr}$' without a subscript is taken in the fundamental or defining representation.

The requirement of this factorization imposes strong relations among the anomaly vectors and the massless matter spectrum. For non-Abelian gauge symmetries, the anomaly cancellation conditions take the form
\begin{align}\label{eq:GS-conds-non-abelian}
    &H - V = 273-29 T \ , \quad b_0\cdot b_0 = 9-T \ , \nonumber \\
    & B_{\bf adj}^i = \sum_{\bf r}n^i_{\bf r}B^i_{\bf r} \ , \quad b_0\cdot b_i = \frac{\lambda_i}{6}\left(\sum_{\bf r}n_{\bf r}^iA_{\bf r}^i-A^i_{\bf adj}\right) \ , \nonumber \\
    &b_i\cdot b_i = \frac{\lambda_i^2}{3}\left(\sum_{\bf r}n^i_{\bf r}C^i_{\bf r}-C^i_{\bf adj}\right) \ , \quad b_i\cdot b_j = 2\lambda_i\lambda_j\sum_{\bf r,s}n_{\bf r,s}^{ij}A^i_{\bf r}A^j_{\bf s} \quad i\neq j \ ,
\end{align}
where $n^i_{\bf r}$ denotes the number of hypermultiplets in the representation ${\bf r}$ of the gauge group $G_i$, and $n_{\bf r,s}^{ij}$ denotes the number of hypermultiplets transforming in the bifundamental representation $({\bf r},{\bf s})$ under $G_i\times G_j$. The group-theoretic coefficients $A^i_{\bf r}, B^i_{\bf r},$ and $ C^i_{\bf r}$ are defined by ${\rm tr}_{\bf r}F^2=A_{\bf r}{\rm tr}F^2 , {\rm tr}_{\bf r}F^4 = B_{\bf r}{\rm tr}F^4 + C_{\bf r}({\rm tr}F^2)^2 $. Inner products are taken with respect to $\Omega_{\alpha\beta}$ via
$v\cdot w = \Omega_{\alpha\beta} v^\alpha w^\beta$.

\begin{table}[t]
\centering
\begin{tabular}{|c|c|c|c|c|c|c|c|c|}
    \hline
    $G_i$ & $SU(N)$ & $SO(N)$ & $Sp(N)$ & $G_2$ & $F_4$ & $E_6$ & $E_7$ & $E_8$ \\
    \hline 
    $\lambda_i$ & $1$ & $2$ & $1$ & 2 & 6 & 6 & 12 & 60 \\ 
    \hline
\end{tabular}
\caption{Normalization constants for non-Abelian gauge groups}\label{tb:Normalization}
\end{table}

The string charge lattice $\Gamma$ is a Lorentzian unimodular lattice of signature $(1,T)$ \cite{Seiberg:2011dr}.  The quantization of anomaly coefficients in this lattice is further constrained in \cite{Monnier2018}. The anomaly cancellation conditions then impose strong discrete constraints on the anomaly vectors. The vectors $b_I=(b_0,b_i)$ span a sublattice $\mathfrak{L}\subset \Gamma$, which we refer to as the anomaly lattice. The associated Gram matrix is a symmetric matrix defined by $\mathcal{G}_{IJ}\equiv b_I\cdot b_J$. Although the group-theoretic coefficients in the factorization conditions are rational, the Green–Schwarz conditions \eqref{eq:GS-conds-non-abelian} ensure that $\mathfrak{L}$ is in fact integral and that the matrix $\mathcal{G}$ has integer entries. Moreover, the Gram matrix $\mathcal{G}_{IJ}$ must have signature compatible with that of the ambient Lorentzian lattice $\Gamma$. In particular, the signature satisfies \cite{Hamada:2023zol,Hamada:2024oap}
\begin{align}
    n_+^\mathfrak{L} \le 1 \ , \quad n_-^\mathfrak{L} \le T \ ,
    \label{eq:eigenvalue_constraint}
\end{align}
where $n_\pm^\mathfrak{L}$ denote the numbers of positive and negative eigenvalues of $\mathcal{G}_{IJ}$, respectively. These conditions encode highly nontrivial relations between the matter spectrum and the gauge content of the theory to the topological structure of the tensor charge lattice.

\subsection{Tensor moduli space and BPS strings}

The tensor moduli space, parametrized by the scalar fields $J$, is bounded by singularities at which the effective field theory description breaks down. Following the main assumption of \cite{Kim:2024eoa}, we will assume that every boundary of the tensor moduli space, whether at finite or infinite distance, corresponds to a point where a BPS string becomes tensionless. These BPS strings serve as charged sources for the two-form gauge fields $B_{\mu\nu}^\pm$ and preserve one-half of the supersymmetry. Unitarity requires their tensions, $T_Q\sim J\cdot Q$ for a charge $Q$, to be non-negative throughout the tensor moduli space. Consequently, the tensor moduli space $\mathcal{M}$ is defined by the condition
\begin{align}\label{eq:positive-tension}
    J\cdot Q_i\ge 0\quad {\rm with} \quad J\cdot J=1 \ , 
\end{align}
where $Q_i$ denotes the 2-form tensor charge of the $i$-th BPS string. The assumption of \cite{Kim:2024eoa} therefore implies that every boundary point $J_{\partial\mathcal{M}}$ must host a BPS string of charge $Q$ that becomes tensionless with a condition $J_{\partial\mathcal{M}}\cdot Q=0$.

In addition, the scalar fields $J$ must satisfy
\begin{align}
    J\cdot b_0\ge 0 \ , \quad J\cdot b_i\ge 0 \ ,
\end{align}
in the tensor moduli space.
The latter condition ensures the positivity of the gauge kinetic terms and, equivalently, the non-negativity of the tensions of instantonic BPS strings associated with the gauge algebra $G_i$, whose string charges are proportional to $b_i$. The former condition was justified in \cite{Kim:2024eoa} using the strong form of the cobordism conjecture \cite{McNamara:2019rup} applied to a supersymmetric compactification on $K3$, which implies the existence of a BPS string with charge $Q=24b_0$ and thus $J\cdot b_0\ge0$. This requirement is also consistent with the arguments of \cite{Cheung:2016wjt,Hamada:2018dde}, which demand the coefficient of the Gauss–Bonnet term $J\cdot b_0$ to be non-negative. More details are provided in Section 2.4 of \cite{Kim:2024eoa}.

These conditions also constrain the structure of BPS string charges. Together with the assumption that tensionless BPS strings emerge at the boundaries of the tensor moduli space, the condition \eqref{eq:positive-tension} implies that the charges of BPS strings form a cone called the BPS cone. This cone is dual to the tensor cone. Here, the tensor cone is defined by the condition \eqref{eq:positive-tension} with $J\cdot J>0$, while the tensor moduli space corresponds to its slice at $J\cdot J=1$. 

The BPS cone is spanned by generators $\mathcal{C}_i$ representing charges of primitive BPS strings whose tensor charge cannot be decomposed into non-negative combinations of other BPS charges \cite{Kim:2024eoa}. Each generator $\mathcal{C}_i$ saturates the condition \eqref{eq:positive-tension} either at a finite distance or at an asymptotic infinity in the tensor moduli space, thereby giving rise to a tensionless BPS string at that location. This immediately implies $\mathcal{C}_i^2 \le 0$.  The structure of the BPS cone is thus determined by these primitive BPS strings and the intersection properties of their tensor charges.

In a 6d $(1,0)$ supergravity, the only BPS strings that can become tensionless within tensor moduli space are strings associated with local SCFTs, LSTs, or critical strings, whose classifications are systematically developed in \cite{Heckman:2015bfa,Bhardwaj:2015xxa,Bhardwaj:2015oru,Bhardwaj:2018jgp,Bhardwaj:2019hhd}. Adopting the assumption of \cite{Kim:2024eoa}, we assume that these classifications provide a complete list of tensionless BPS strings that can appear in 6d supergravity. This also means that we have a complete set of BPS generators $\mathcal{C}_i$, along with their detailed properties.  We will rely heavily on this set of generators and their properties when constructing tensor bases for the supergravity theories.

\begin{table}[t]
\centering
\begin{tabular}{|c|l|c|c|c|}
    \hline
    $\mathfrak{g}_i$ & $H_i$ & $b_i^2$ & $b_0\cdot b_i$ & Notes \\
    \hline 
    $\mathfrak{g}$ & {\bf Adj} & 0 & 0 & \\ 
    \hline
    $\mathfrak{su}_N$ & $(2N)\times {\bf N}$ & $-2$ & 0 &\\
    $\mathfrak{su}_N$ & $(N-8)\times {\bf N}\oplus{\bf \frac{N(N+1)}{2}}$ & $-1$ & $-1$ & $N\ge8$\\
    $\mathfrak{su}_N$ & $(N+8)\times {\bf N}\oplus{\bf \frac{N(N-1)}{2}}$ & $-1$ & 1 &\\
    $\mathfrak{su}_N$ & $16\times {\bf N}\oplus2\times{\bf \frac{N(N-1)}{2}}$ & $0$ & 2 &\\
    $\mathfrak{su}_N$ & ${\bf \frac{N(N-1)}{2}}\oplus{\bf \frac{N(N+1)}{2}}$ & $0$ & 0 &\\
    \hline
    $\mathfrak{su}_6$ & $15\times {\bf 6}\oplus\frac{1}{2} {\bf 20}$ & $-1$ & 1 &\\
    $\mathfrak{su}_6$ & $17\times {\bf 6}\oplus{\bf 15}\oplus\frac{1}{2} {\bf 20}$ & $0$ & $2$ & \\
    $\mathfrak{su}_6$ & $18\times {\bf 6}\oplus{\bf 20}$ & 0 & 2 & \\
    $\mathfrak{su}_6$ & ${\bf 6}\oplus\frac{1}{2} {\bf 20}\oplus{\bf 21}$ & $0$ & 0 &\\
    \hline
    $\mathfrak{so}_N$ & $(N-8)\times {\bf N}$ & $-4$ & $-2$ & $N\ge 8$ \\
    $\mathfrak{so}_N$ & $(N-7)\times {\bf N}\oplus(2^{\lfloor \frac{10-N}{2}\rfloor})\times{\bf 2^{\lfloor\frac{N-1}{2}\rfloor}}$ & $-3$ & $-1$ & $12\ge N\ge 7$ \\
    $\mathfrak{so}_N$ & $(N-6)\times {\bf N}\oplus(2\times 2^{\lfloor \frac{10-N}{2}\rfloor})\times{\bf 2^{\lfloor\frac{N-1}{2}\rfloor}}$ & $-2$ & $0$ & $13\ge N\ge 6$ \\
    $\mathfrak{so}_N$ & $(N-5)\times {\bf N}\oplus(3\times 2^{\lfloor \frac{10-N}{2}\rfloor})\times{\bf 2^{\lfloor\frac{N-1}{2}\rfloor}}$ & $-1$ & $1$ & $12\ge N\ge 5$ \\
    $\mathfrak{so}_N$ & $(N-4)\times {\bf N}\oplus(4\times 2^{\lfloor \frac{10-N}{2}\rfloor})\times{\bf 2^{\lfloor\frac{N-1}{2}\rfloor}}$ & $0$ & $2$ & $14\ge N\ge 4$ \\
    \hline
    $\mathfrak{sp}_N$ & $(2N+8)\times {\bf 2N}$ & $-1$ & 1 &\\
    $\mathfrak{sp}_N$ & $16\times {\bf 2N}\oplus{\bf (N-1)(2N+1)}$ & $0$ & 2 &\\
    $\mathfrak{sp}_3$ & $\frac{35}{2} {\bf 6}\oplus \frac{1}{2} {\bf 14}' $ & $0$ & 2 &\\
    \hline
    $\mathfrak{e}_8$ &  & $-12$ & 10 &\\
    $\mathfrak{e}_7$ & $\frac{k}{2}\times {\bf 56}$ & $k-8$ & $k-6$ & $k\le 8$\\
    $\mathfrak{e}_6$ & $k\times {\bf 27}$ & $k-6$ & $k-4$ & $k\le 6$\\
    $\mathfrak{f}_4$ & $k\times {\bf 26}$ & $k-5$ & $k-3$ & $k\le 5$\\
    $\mathfrak{g}_2$ & $(3k+1)\times {\bf 7}$ & $k-3$ & $k-1$ & $k\le 3$\\
    \hline
\end{tabular}
\caption{Gauge algebras $\mathfrak{g}_i$ and types of charged hypermultiplets $H_i$ supported on $b_i$ with $b^2\le0$.}\label{tb:generators}
\end{table}

Notably, all anomaly vectors $b_i$ with $b_i^2<0$, as summarized in Table \ref{tb:generators}, are the generators. Each one corresponds to the unit instanton string for the associated gauge algebras $\mathfrak{g}_i$. In addition to these, there exist four more generators characterized by
\begin{align}\label{eq:more-generators}
     &1) \ \mathcal{C}^2=-1, \, b_0\cdot\mathcal{C}=1 \, , \quad  2)\ \mathcal{C}^2=-2, \, b_0\cdot\mathcal{C}=0 \, , \nonumber \\
     &3)\  \mathcal{C}^2=0, \, b_0\cdot\mathcal{C}=2 \, , \quad 4)\ \mathcal{C}^2=0, \, b_0\cdot\mathcal{C}=0 \ ,
\end{align}
which do not carry any gauge algebra.
These correspond respectively to the E-string, M-string, and the critical heterotic and Type II strings. The first two strings can become tensionless at finite-distance points in the tensor moduli space, while the latter two admit tensionless limits only at infinite distance. Moreover, the third charge serves as a generator only when $T=1$, and the fourth only when $T=9$ with $b_0=0$ \cite{Kim:2024eoa}. 

The intersection properties of the generators determine the structure of the BPS cone and, in turn, the geometry of the tensor moduli space. The ways in which two or more generators can intersect have been systematically studied in the SCFT and LST classifications of \cite{Heckman:2015bfa,Bhardwaj:2015xxa,Bhardwaj:2015oru,Bhardwaj:2018jgp,Bhardwaj:2019hhd}. These intersection rules are dictated by the types of charged hypermultiplets $H_i$ and by the gauge anomaly cancellation conditions \eqref{eq:GS-conds-non-abelian} when the intersecting generators support nontrivial gauge algebras, or by their associated flavor symmetries otherwise. For example, the E-string possesses an $E_8$ flavor symmetry, so it may intersect another generator only if the gauge algebra supported on that generator embeds into $\mathfrak{e}_8$ symmetry. The full set of intersection rules, except those for the critical strings, is summarized in \cite{Heckman:2015bfa,Bhardwaj:2015xxa,Bhardwaj:2015oru,Bhardwaj:2018jgp,Bhardwaj:2019hhd}.

As we discuss in more detail below, the critical heterotic string corresponding to case 3) in \eqref{eq:more-generators} becomes a generator only when $T=1$. For $T>1$, its charge $f$ is shared by little strings in certain LSTs and can therefore be written as a positive linear combination of the LST generators. The appearance of this charge in the BPS spectrum imposes strong constraints on the intersection structure of generators $\mathcal{C}_i$ with $f\cdot\mathcal{C}_i>0$

This string is a supergravity string studied in \cite{Kim:2019vuc}, with left- and right-moving central charges $c_L=20$ and $c_R=6$. Gauge algebras $\mathfrak{g}_i$ supported on generators $\mathcal{C}_i$ that intersect $f$ positively induce current algebras at level $k_i = f \cdot \mathcal{C}_i$ on the two-dimensional worldsheet CFT. A current algebra associated with a gauge algebra $\mathfrak{g}$ at level $k$ contributes to the left-moving central charge as \cite{DiFrancesco:1997nk}
\begin{align}\label{eq:current-algebra}
    c_{\mathfrak{g}}= \frac{k\,{\rm dim}(\mathfrak{g})}{k+h^\vee}\ ,
\end{align}
where $\mathrm{dim}(\mathfrak{g})$ and $h^\vee$ denote the dimension and the dual Coxeter number of $\mathfrak{g}$.
As shown in \cite{Kim:2019vuc,Lee:2019skh,Kim:2024eoa}, this implies the bound
\begin{align}\label{eq:f-bound}
    \sum_ic_{\mathfrak{g}_i} \le c_L=20 \ ,
\end{align}
for the gauge algebras $\mathfrak{g}_i$ on $\mathcal{C}_i$ with $f\cdot \mathcal{C}_i>0$.
In particular, this constrains the total rank of these gauge algebras $\mathfrak{g}_i$ to be at most 20. This central charge bound was further improved to 16 in \cite{Baykara:2025gcc} when $T>1$ or when the tensor multiplet with charge $f$ itself supports a gauge algebra. Thus, the generators intersecting strings in the LSTs whose little strings share the charge $f$ are strongly restricted by this constraint.

\subsection{Refined structure of tensor moduli space}\label{sec:refined_structure_of_tensor_moduli_space}

More refined structure of the tensor multiplets and their intersections was analyzed in  \cite{Kim:2024eoa}, and we briefly summarize the relevant results here.

\begin{itemize}
\item M-strings \cite{Haghighat:2013gba} may be removed from the set of generators. The point in tensor moduli space at which an M-string becomes tensionless implements an Weyl reflection that is a $\mathbb{Z}_2$ symmetry acting on the M-string charge as $Q\rightarrow -Q$. This symmetry allows the tensor moduli space to be extended across this locus, which gives rise to an extended tensor moduli space. The associated BPS cone contains no such M-string charge. Consequently, all M-string charges, unless the $\mathfrak{su}_2$ flavor symmetry of the corresponding string is gauged, can be removed from the BPS cone dual to the extended tensor moduli space. In contrast, the instantonic strings for nontrivial gauge algebras with $b_i^2=-2$ and $b_0\cdot b_i=0$, as well as the M-strings whose $\mathfrak{su}_2$ flavor symmetry is gauged (e.g., in the $223$ NHC), cannot be removed through this procedure and must remain in the list of generators.

\item Any two distinct generators must intersect non-negatively:
\begin{align}
    \mathcal{C}_i\cdot \mathcal{C}_j \ge 0 \quad {\rm for} \quad i\neq j \ .
\end{align}
This condition follows either from the mixed gauge anomaly cancellation when both generators support gauge algebras, or from the intersection properties of the E-string and critical strings otherwise.

\item Every 6d ${\cal N}=(1,0)$ supergravity with $T\ge1$ must contain an {\it H}-string, except in the special cases with $T=9$ and $b_0=0$. An {\it H}-string is a distinguished BPS string whose charge $f$ satisfies
\begin{align}
    f^2=0 \ , \quad b_0\cdot f=2 \ ,
    \label{eq:H-string-charge}
\end{align}
which agrees with the intersection property of the third generator, the critical heterotic string, in \eqref{eq:more-generators}.
The argument proceeds first by defining a projection, known as a ``blowdown'', of the tensor moduli space parametrized by $J$ to a subspace with $J'\subset J$ that is orthogonal to a generator with $\mathcal{C}^2=-1$ and $b_0\cdot \mathcal{C}=1$. Concretely, the blowdown projection is defined by imposing the condition $J'\cdot \mathcal{C}=0$ on the subspace parametrized by $J'$. Under this projection, every BPS charge, including $b_0$ and $b_i$, transforms as
\begin{align}
    Q \quad \rightarrow \quad Q'=Q+(\mathcal{C}\cdot Q)\mathcal{C} \ ,
\end{align}
which guarantees that $Q'\cdot \mathcal{C}=0$.
After performing a blowdown, the resulting BPS cone may again contain another generator with $\mathcal{C}'^2=-1$ and $b_0'\cdot \mathcal{C}'=1$. We can repeat the blowdown procedure on this $\mathcal{C}'$.
Iterating such blowdowns yields either a reduced BPS cone in which no further generators can be blown down, or a two-dimensional cone. In the latter case, a simple analysis of the two-dimensional BPS cone shows that it must include a charge $f$. In the former case, the only remaining generators that intersect positively with $b_0$ are necessarily of charge $f$. Together with the positivity condition $J\cdot b_0\ge 0$ throughout the tensor moduli space, this proves the existence of an {\it H}-string.

\item There is a containment relation: any generator $\mathcal{C}_i$ orthogonal to an {\it H}-string charge $f$, i.e., $\mathcal{C}_i\cdot f=0$, must belong to an LST whose little string carries the charge $f$. This follows from the fact that the ratio of string tensions $(J\cdot \mathcal{C}_i)/(J\cdot f)$ remains finite throughout the tensor moduli space, together with the fact that the moduli space is locally polyhedral near the infinite-distance point at which the {\it H}-string becomes tensionless.  In particular, the existence of such orthogonal generators implies that the charge $f$ can be expressed as $f=\sum_i n_i \mathcal{C}_i$ with $\mathcal{C}_i\cdot f=0$ and integers $n_i>0$.

\item Lastly, the tensor moduli space is contained within the BPS cone, and the point at which the {\it H}-string becomes tensionless corresponds to an asymptotic infinity $J\sim f$. The local polyhedral nature of the moduli space near this asymptotic limit implies that it can only intersect a single boundary face of the BPS cone. As a result, the charge $f$ can be a generator only in the case $T=1$, since for $T>1$ all generators must reside at intersections of boundary faces of the BPS cone. Moreover, for $T=1$ the BPS cones coincide with the Mori cones of Hirzebruch surfaces $\mathbb{F}_n$, which arise as K\"ahler bases of elliptic CY threefolds used in F-theory compactifications.

\end{itemize}

Taken together, these results imply the following:
\begin{framed}  
    \centering
    {\it Intersection structure of tensor multiplets in any 6d supergravity theory matches that of K\"ahler surfaces: $\mathbb{P}^2$ for $T=0$, $\mathbb{F}_n$ for $T=1$, or their blowups for $T>1$.}
\end{framed}
Here, the intersection structure is encoded by the intersection matrix $\Omega_{\alpha\beta}$ and the anomaly vectors $b_0, b_i$. The only exceptions to this statement occur when $b_0=0$, which is possible only for $T=9$. Conversely, any intersection pairing in a supergravity can always be reduced, via a sequence of blowdowns, to that of $\mathbb{P}^2$ or one of the Hirzebruch surfaces $\mathbb{F}_n$, specified by
\begin{align}\label{eq:P2-Fn}
    \Omega = \left(\begin{array}{cc}0& 1 \\ 1 & -n\end{array}\right)\,, \ b_0=(2\!-\!n,2) \quad {\rm for} \ \ \mathbb{F}_n \ , \qquad \Omega = 1 \,, \ b_0=3 \quad {\rm for} \ \ \mathbb{P}^2 \ ,
\end{align} 
again except $b_0=0$ cases.

Using this refined structure of the tensor moduli space together with the known classifications of 6d SCFTs and LSTs, the finiteness of the massless spectrum in 6d ${\cal N}=(1,0)$ supergravity was proven in \cite{Kim:2024eoa}. This analysis further yields the bounds
\begin{align}\label{eq:bounds}
    T \le 193 \ , \quad r(V) \le 480 \ ,
\end{align}
where $r(V)$ denotes the rank of gauge algebras of the 6d theory.

We will make extensive use of these properties of the tensor moduli space in the constructions and classifications of 6d supergravity theories presented in the following sections.

\section{Construction of 6d supergravities}\label{sec:construction}

The refined structure of the tensor moduli space provides a systematic framework for constructing 6d ${\cal N}=(1,0)$ supergravity theories. In this section, we present our construction method for theories with $T\ge1$. The construction and classification of $T=0$ theories can be found in \cite{Kumar:2010am,Hamada:2024oap}, and the exceptional case with $b_0 = 0$ and $T = 9$ will be addressed separately at the end of this section. 

We begin by introducing a general construction scheme for 6d supergravity theories. We then focus on a distinguished class of building blocks, which we call {\it non-Higgsable gravity blocks}, and explain how they can be used as the basic ingredients in a systematic construction of tensor bases for general 6d supergravity theories.

\subsection{General construction}\label{sec:general_construction}

The construction of a general 6d supergravity theory can be organized into the following three steps:
\begin{itemize}
    \item Step 1: Construct LSTs associated with an {\it H}-string charge $f$.
    \item Step 2: Add {\it external} BPS generators and their associated matter fields.
    \item Step 3: Place gauge algebras and charged hypermultiplets on tensor multiplets with positive self-intersection.
\end{itemize}
Here, the {\it external} BPS charges (equivalently, tensor multiplets) are those that positively intersect the {\it H}-string charge $f$. This construction is guided by the structure of the tensor moduli space discussed in the previous section, in particular the existence of an {\it H}-string, and the fact that the intersection pairing of tensor multiplets can be blown down to that of a Hirzebruch surface $\mathbb{F}_n$. 
Below, we elaborate on the precise rules and constraints that we will apply at each stage.

In Step 1, we first note that LSTs associated with an {\it H}-string charge $f$ necessarily have a {\it P}-type endpoint, which is one of the possible endpoints classified in \cite{Bhardwaj:2015oru}.\footnote{In this type, the LST blows down to a single generator $b_\text{end}$ with properties $b_\text{end}^2=0$ and $b_\text{end}\cdot b_0=2$. The generator $b_\text{end}$ is identified as the charge class $f$.}
This follows directly from the defining properties of the charge $f$, namely $f^2 = 0$ and $b_0 \cdot f = 2$.  In this case, the {\it H}-string charge can be expressed as $f=\sum_i n_i\mathcal{C}_i$ with $n_i\ge0$, where each $\mathcal{C}_i$ is a BPS generator satisfying $f\cdot \mathcal{C}_i=0$ in one of these LSTs. 

A key feature of LSTs with a {\it P}-type endpoint is that they admit a geometric realization without frozen singularities. From the geometric perspective, all such LSTs can be blown down to a rational fiber curve $f$ embedded in a Hirzebruch surface. 
By contrast, LSTs involving frozen singularities, classified in \cite{Bhardwaj:2015oru,Bhardwaj:2019hhd}, always possess non–{\it P}-type endpoints. Consequently, the LSTs associated with the charge $f$ are highly constrained, and their complete classification is provided in \cite{Bhardwaj:2015oru}.  In particular, circular intersections among tensor multiplets are excluded for LSTs with {\it P}-type endpoints. We emphasize that this does not imply that supergravity theories cannot contain non–{\it P}-type LSTs. Such LSTs can arise from suitable combinations of tensor multiplets involved in the LSTs of the charge $f$ together with external tensor multiplets and matter fields. We summarize the essential features of LSTs in the Appendix \ref{app:SCFTsLSTs}.

When LSTs with a {\it P}-type endpoint are embedded into supergravity by sharing the {\it H}-string charge $f$, they are further constrained by the bounds on the number of massless multiplets given in \eqref{eq:bounds}. The tensor bound $T \le 193$ restricts the number of tensor multiplets in such an LST to $T_{\rm LST} \le 192$, since at least one additional tensor multiplet is required to complete the LST into a consistent supergravity theory. Moreover, the rank bound $r(V)\le 480$ severely restricts the allowed gauge algebras and matter content, excluding arbitrarily long tensor chains or arbitrarily large gauge groups. These bounds thus imply that the set of LSTs associated with an {\it H}-string that can be consistently embedded in supergravity is finite, and their complete classification is achievable. In Section~\ref{sec:LST}, we present the full list of such {\it P}-type LSTs composed solely of NHCs that can share the charge $f$. 

The next step, Step 2, is to complete a given collection of LSTs associated with an {\it H}-string into a supergravity. This step determines the full structure of the tensor moduli space and the associated BPS cone. We begin by selecting candidate LSTs that can share the {\it H}-string charge class $f$, and then couple them to an appropriate set of external BPS generators. In doing so, the following three conditions must be satisfied:
\begin{enumerate}
    \item Each external generator must intersect all LSTs sharing $f$ with the same intersection number.
    \item Intersections between external generators and LST generators must obey the rank bound in \eqref{eq:f-bound}, together with the intersection rules among BPS generators arising in SCFT/LST classifications.
    \item The resulting intersection form must have signature $(1,T)$ and admit a sequence of blowdowns to the intersection pairing of a Hirzebruch surface given in \eqref{eq:P2-Fn}.
\end{enumerate}

The first condition ensures a consistent intersection structure between an {\it H}-string and external generators. As explained earlier, the little strings in LSTs sharing an {\it H}-string carry a common charge $f$, which must be decomposed as $f=\sum_in_i^a\mathcal{C}^a_i$ with $n^a_i>0$ for BPS generators $\mathcal{C}^a_i$ in the $a$-th LST. An external generator $\mathcal{C}_{\rm ext}$ must satisfy the same intersection relation with each LST, namely $\mathcal{C}_{\rm ext}\cdot f = \mathcal{C}_{\rm ext} \cdot \sum_in_i^{a}\mathcal{C}_i^a$ for all $a$.

The second condition imposes strong restrictions on the allowed external generators supporting non-trivial gauge algebras. The rank bounds of $20$ for $T=1$ and $16$ for $T>1$ on gauge algebras supported on generators $\mathcal{C}_i$ with $f\cdot \mathcal{C}_i>0$, discussed near \eqref{eq:f-bound}, apply directly to external generators and hence bound the number of such external generators. For $T=1$, the intersection form is uniquely fixed to be \eqref{eq:P2-Fn}, which allows only a single external generator intersecting $f$ once. We therefore focus on the cases with $T>1$. 

The intersection between external and LST generators, systematically studied in \cite{Heckman:2015bfa,Bhardwaj:2015xxa,Bhardwaj:2015oru,Bhardwaj:2018jgp,Bhardwaj:2019hhd}, is determined by gauge and gravitational anomaly cancellation and by the requirement that the gauge algebra on a given generator must embed into the flavor symmetry of its neighboring generators. These constraints lead to the following gluing rules:
\begin{align}
    &1 - [12] \ , \quad 1 - [8]  \ , \quad 1 - [7]  \ , \quad 1 - [6]  \ , \quad 1 - [5] \ , \quad 1 \overset{n\le2}{-} [4] \ , \quad 2 \overset{n\le2}{-} [4] \nonumber \\
    & 1 \overset{n}{-} [3]   \ , \quad 2 - [3] \ , \quad 1 \overset{n}{-} [2] \ , \quad 2 \overset{n}{-} [2]  \ , \quad 1 \overset{n}{-} [1] \ , \quad 1 - [\hat{1}] \ , \quad 2 - [\hat{1}] \ ,
\end{align}
where $p \overset{n}{-} [q]$ denotes an LST generator of self-intersection $-p$ and an external generator of self-intersection $-q$, or vice versa, with mutual intersection number $n$. The condition $n\le2$ indicates that the mutual intersection number can be either 1 or 2, while the absence of $n$ implies a single intersection. For the gluing $1\overset{n}-[3]$, the number $n$ is in principle unbounded if the `$-3$' tensor supports an $\mathfrak{su}_3$ algebra. For the gluings $1-[2], 2-[2]$ and $1 - [1]$, the intersection number $n$ is unbounded when either the `$-1$' tensor carries no gauge algebra and the other tensor supports a small gauge algebra $\mathfrak{g}=\emptyset, \mathfrak{su}_2, \mathfrak{su}_3$, or when the `$-2$' tensor has no gauge algebra and the other tensor supports $\mathfrak{g}=\emptyset, \mathfrak{su}_2$, where $\emptyset$ denotes no gauge algebra. The symbol $\hat{1}$ denotes the generator with $\mathcal{C}^2=-1$ and $b_0\cdot \mathcal{C}=-1$. This generator can only be realized as an external generator, since all LSTs containing it are known to have non-{\it P}-type endpoints \cite{Bhardwaj:2019hhd,Kim:2024eoa}.

The third condition ensures that the tensor moduli space and the BPS cone are consistent with the refined intersection structure explained in the previous section. It also places strong constraints on the intersections of BPS generators. In particular, after quotienting by the equivalence relations explained in Section~\ref{sec:equivalence_relations}, it bounds all mutual intersection numbers that were previously unbounded.\footnote{While these equivalence relations imply that the mutual intersection numbers are bounded, they do not determine the precise bounds. We leave the calculation of these bounds for future work.} Together with the bound $T\le 193$, this leaves only finitely many admissible tensor intersection structures.

Step 1 and Step 2 fully fix the tensor intersection structure, with respect to a reference {\it H}-string charge $f$, and matter fields associated to the BPS generators. In particular, all BPS generators $\mathcal{C}_i$ and their mutual intersections $\mathcal{C}_i \cdot \mathcal{C}_j$ for $i\neq j$ are determined at this stage. The last step, Step 3, is to assign gauge algebras and matter fields to tensor charges with non-negative self-intersection that are neither BPS generators nor equal to $f$. Tensor charges $Q_\alpha$ of this type can be expressed as non-negative sums of BPS generators as $Q_\alpha=\sum_i n_{\alpha i}\mathcal{C}_i$ with $n_{\alpha i}\ge0$. Also, since $Q_\alpha^2\ge0$ and $Q_\alpha\neq f$, it follows that this charge must intersect the charge $f$ positively, $Q_\alpha\cdot f > 0$, and likewise intersect any other such charge $Q_\beta$ positively, $Q_\alpha\cdot Q_\beta>0$.

The gauge algebras $\mathfrak{g}_\alpha$ supported on these tensors $Q_\alpha$, together with the associated charged hypermultiplets, are constrained by the anomaly cancellation conditions \eqref{eq:GS-conds-non-abelian}. In particular, Abelian gauge symmetries can be supported only on these charges $Q_\alpha$, so that $U(1)\subset \prod_\alpha \mathfrak{g}_\alpha$, while neither BPS generators nor the {\it H}-string charge $f$ can host Abelian gauge algebras. Furthermore, the condition $Q_\alpha\cdot f > 0$ implies that these gauge algebras must embed into the common flavor symmetry of the LSTs associated with $f$, and their ranks are bounded by 20 for $T=1$ and 16 for $T>1$, as explained around \eqref{eq:f-bound}.

After this step, additional constraints may arise from the unitarity requirements associated with supergravity strings \cite{Kim:2019vuc}, as well as from discrete anomaly cancellation \cite{Dai:1994kq,Basile:2023zng,Dierigl:2025rfn,Hamada:2025vga}. Taking these constraints into account, one eventually obtains a supergravity theory consistent with the refined structure of the tensor moduli space and all other known constraints.

\paragraph{Counting tensors and dummy LSTs}

In Step 2, we determine the tensor intersection structure by identifying the external generators that can be consistently attached to a chosen collection of LSTs sharing an {\it H}-string. In practice, however, we restrict this analysis to external generators that support gauge algebras. The reason is that, once the number of tensor multiplets becomes large, there can be infinitely many generators without gauge algebra, namely E-strings and M-strings, and these can intersect one another in arbitrarily many ways. This makes a direct classification of all external generators and their intersections intractable. While M-strings can be eliminated from the charge lattice by Weyl reflections, one still needs a systematic way to infer the intersections of the E-strings from the finite set of generators carrying gauge algebras and their intersections.  We propose such a method here that, once the external generators supporting gauge algebras are fixed, determines the total number of tensor multiplets and reconstructs the configuration of the external E-string sectors.

We begin by constructing the Gram matrix $\mathcal{G}_{IJ}$ for the anomaly vectors $b_0$ and $b_i$ and computing the number of positive eigenvalues, $n^{\mathfrak{L}}_+$. In a consistent supergravity theory this number must satisfy $n^{\mathfrak{L}}_+ \le 1$, which typically cannot be satisfied for small $T$. Consequently, the requirement $n^{\mathfrak{L}}_+ \le 1$ imposes a lower bound $T_{\rm min}$ on the number of tensor multiplets for a given configuration of generators, while the gravitational anomaly constraint determines the upper bound $T_{\rm max}$.

As discussed above, each primitive tensor multiplet in an LST corresponds to a generator. Combined with the fact that any tensor base can be blown down to that of a Hirzebruch surface, this implies that the total number of tensor multiplets in supergravity is
\begin{align}\label{eq:number-of-tensors}
    T = \sum_a T^{H}_a+1 \ ,
\end{align}
where the index $a$ labels LSTs sharing the {\it H}-string charge $f$, and $T^H_a$ denotes the number of dynamical tensor multiplets in the $a$-th LST. This $T$ matches the number of negative eigenvalues of the intersection matrix $\Omega$. We note that the tensor multiplet associated with the little string tension becomes non-dynamical once gravity is decoupled and is excluded from $T^H_a$. Instead, the tensor multiplet for the little string tension in supergravity is accounted for by the last $+1$ in the formula.

There is a special class of LSTs with little string charge $f$ that contains only tensor multiplets, without any gauge algebra or charged matter. We will refer to these as \emph{dummy LSTs}. These theories consist solely of $-1$ and $-2$ generators.\footnote{In particular, we refer to $1\underbrace{2\cdots2}_{n}1$ as an \emph{$A$-type dummy LST} and to $2\overset{2}{2}\underbrace{2\cdots2}_n1$ as a \emph{$D$-type dummy LST}.}
Adding a dummy LST increases the number of tensor multiplets without modifying the gauge algebra or charged matter content. Hence, models with $T_{\rm min} \le T \le T_{\rm max}$ can be obtained by adding an appropriate number of dummy LSTs. For instance, a model with $T_{\rm max}$ is constructed from the model with $T_{\rm min}$ by adding $T_{\rm max}-T_{\rm min}$ dummy LSTs of type `$11$' (representing two E-strings intersecting at a point with the little string charge $f$). Hence, once the types of LSTs and the vectors $b_i$ on the external generators are specified, the addition of dummy LSTs fixes the total number of tensor multiplets as well as their intersection structure. If the resulting intersection form $\Omega$ satisfies unimodularity and the condition~\eqref{eq:eigenvalue_constraint}, the corresponding tensors and their intersections define a consistent tensor base for a supergravity.

The external $-1$ generators that do not support gauge algebras are then determined automatically. If no other external generators are present, one needs a single $-1$ generator intersecting the {\it H}-string charge $f$ in order to recover a Hirzebruch base after repeated blowdowns.\footnote{We do not need to consider the cases where a $-1$ external generator intersects the {\it H}-string charge $f$ at more than one point. Such an external cannot be a generator after blowdown, so it is enough to consider intersection number one with the {\it H}-string.} When external generators supporting gauge algebras are present, on the other hand, all charges $Q$ satisfying $Q^2=-1$ and $Q\cdot b_0=1$ that intersect non-negatively with all other generators and positively with $f$ correspond to external $-1$ generators. Since these generators are uniquely fixed by the rest of the configuration, we will not specify them separately below.

\subsection{Equivalence relations on tensor bases}\label{sec:equivalence_relations}

There may exist several supergravity theories that share the same gauge algebras, the same charged matter representations, and even the same total number of tensor multiplets, yet differ in their tensor intersection forms.  However, such theories are not always physically distinct. In particular, different tensor bases may correspond to the same low-energy theory once one identifies them under suitable equivalence relations acting on the tensor charge lattice. Here we introduce two such equivalence relations: the first comes from Weyl reflections associated with $-2$ tensors (or M-strings), and the second from an infinite duality group.

Let us first discuss the equivalence relation generated by Weyl reflection. As explained above, when a $-2$ tensor multiplet does not support any gauge algebra, it corresponds to an M-string, and such a tensor can be removed from the BPS cone without affecting the low-energy physics, provided that its flavor $\mathfrak{su}_2$ symmetry is not gauged. This removal is realized as a simple Weyl reflection on the tensor charge lattice at the locus where the corresponding string tension vanishes. By contrast, if the $-2$ tensor supports a gauge algebra, the same operation is no longer allowed, since the associated gauge instanton string also becomes tensionless at that locus. Similarly, a Weyl reflection is not permitted when the $\mathfrak{su}_2$ flavor symmetry of an M-string is gauged, as occurs in $223$ NHCs. It follows that whenever two tensor bases are related by a Weyl reflection of a $-2$ tensor without gauge algebra, they should be regarded as describing the same low-energy supergravity theory.

Let us illustrate this equivalence relation with a few concrete examples. A first class of examples arises from transitions among dummy LST configurations. Two representative sequences are
\begin{align}\label{eq:dummy-transition}
    &1\underbrace{222\cdots 222}_{n}1 \quad \rightarrow \quad 1\underbrace{22\cdots 22}_{n-1}1+11 \quad \rightarrow \quad 1\underbrace{22\cdots 22}_{n-2}1 +(11)^2\quad \rightarrow\ \cdots \ \rightarrow \quad (11)^{n+1} \ , \nonumber\\
    & 2\underbrace{\overset{2}{2}2\cdots 22}_{n} 1 \quad \rightarrow \quad 2\underbrace{\overset{2}{2}2\cdots 22}_{n-1} 1 +11 \quad \rightarrow \quad \cdots \quad \rightarrow \quad 212 +(11)^n \quad \rightarrow \quad (11)^{n+2}
\end{align}
where all resulting $122\cdots 221$ and $2\overset{2}{2}2\cdots 21$ LSTs have the same little string of charge $f$. These transitions correspond physically to removing a $-2$ tensor multiplet via a Weyl reflection, while maintaining massless matter content, and geometrically to smoothly rearranging blown-up curves in the K\"ahler base that is generically realized by complex structure deformations associated to neutral hypermultiplets. These configurations when embedded in supergravity describe an identical low-energy physics. Thus, we treat theories that differ only by dummy LST arrangements connected by such transitions as the same gravity theory.

Weyl reflections of $-2$ tensors can also relate tensor configurations of LSTs that look quite different at first sight. For example, the following LST configurations are all equivalent under Weyl reflections:
\begin{align}
    {1 \overset{1}{\underset{1}{5}} 1 2},\ \ \
    {2 1 \overset{1}{5} 1 2},\ \ \
    {2 2 1 \overset{1}{5} 1},\ \ \
    {2 2 1 5 1 2},\ \ \
    {2 2 2 1 5 1},\ \ \
    {2 2 2 2 1 5} \ ,
\end{align}
where each $-2$ tensor is an M-string.
These configurations are in fact all equivalent to an LST consisting of a single $-5$ tensor intersecting five mutually non-intersecting $-1$ tensors. Likewise, external $-2$ tensor multiplets without gauge algebra can be removed in the same manner. The simplest example is
\begin{align}
    2\, 0 \quad \rightarrow \quad  0\, 0 \ ,
\end{align}
which represents a transition between tensor bases in $T=1$ supergravity theories. Geometrically, this is the standard complex structure deformation from the Hirzebruch surface $\mathbb{F}_2$ to $\mathbb{F}_0$.

The second type of equivalence relation arises from duality groups with infinitely many elements. When the number of tensor multiplets is sufficiently large, the tensor charge lattice may admit such an infinite duality group. This phenomenon is familiar from geometry: Hirzebruch bases with enough blow-ups are known to have infinite duality groups. For instance, the del Pezzo surface ${\rm dP}_9$, and more generally bases with even more blow-ups, exhibit this behavior. 

In the supergravity setting, these dualities can generate infinite families of tensor-basis presentations with the same massless matter content and the same relevant {\it H}-string structure. Thus two supergravity bases related by such a duality may still look different if one inspects only finitely many BPS generators and their intersections. In particular, the mutual intersection number between an external $-2$ or $-3$ tensor and a $-1$ generator in an LST may become arbitrarily large in such presentations. In the cases with an infinite duality group, this unbounded intersection number does not distinguish inequivalent low-energy theories; rather, it is a parameter along an infinite duality orbit. After quotienting by the duality group, the allowed intersection data of inequivalent tensor bases are therefore finite.

\begin{figure}
 \center
 \includegraphics[width=0.8\textwidth]{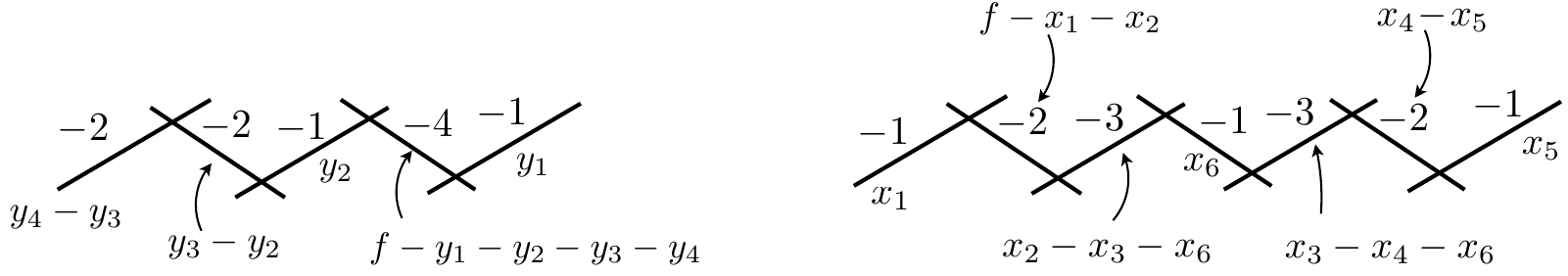} 
\caption{Tensor intersection structure of two LSTs in $\mathbb{F}_0$, with a fiber curve $f$ and 10 blownup points $x_{1,\cdots,6}, y_{1,2,3,4}$. The left figure corresponds to ${\rm LST}_1$ and the right figure to ${\rm LST}_2$. }
 \label{fig:lst-geometry}
 \end{figure}

For example, one may consider the following $T=11$ supergravity:
\begin{align}
    {\rm LST}_1 \ : \ 2\,2\,1^{r}\, \overset{\mathfrak{so}_8}{4}\,1^r \ , \qquad {\rm LST}_2 \ : \ 1^{2r}\,\overset{\mathfrak{su}_2}{2}\,\overset{\mathfrak{g}_2}{3}\,1\,\overset{\mathfrak{so}_7}{3^1} \, \overset{\mathfrak{su}_2}{2} \, 1^{2r-1} \ , \qquad C^{\rm ext} \ : \ \overset{\mathfrak{su}_2}{2} \ ,
\end{align}
where the superscript on a tensor indicates its intersection number with the external $-2$ tensor. This tensor base can be obtained by blowing up ten points on a Hirzebruch surface $\mathbb{F}_0$. The geometric realization of the two LST sectors is shown in Fig. \ref{fig:lst-geometry}. In this example, the external $-2$ tensor supporting an $\mathfrak{su}_2$ gauge algebra can be represented by the infinite family of classes
\begin{align}
    C^{\rm ext} = 4rh+(3r-1)f-2r(x_1+x_2+x_3)-(2r-1)(x_4+x_5)-r\sum_{i=1}^4y_i \ ,
\end{align}
for $r\ge1$, where $h$ and $f$ are two distinct fiber classes of $\mathbb{F}_0$, and $x_i, y_j$ are the blown-up points. These classes are not meant to be simultaneous generators in a single tensor cone. Instead, they are related by an infinite-order Weyl transformation, as in the standard Cremona/Weyl action on rational surfaces \cite{Dolgachev:1983,Sakai:2001}. Indeed, if we define $\delta=C^{\rm ext}_{r+1}-C^{\rm ext}_{r}$ and $a=x_5-x_1$, then both $a$ and $a+\delta$ are $-2$ roots orthogonal to the canonical class, equivalently to $b_0$, and the Weyl element $\tau=s_{a+\delta}s_a$ maps $C^{\rm ext}_{r}$ to $C^{\rm ext}_{r+1}$. Here $s_\alpha(C)=C+(C\cdot \alpha)\alpha$ denotes the Weyl reflection with respect to a $-2$ tensor charge $\alpha$. This transformation preserves the canonical class $b_0$, the massless matter content, and the relevant {\it H}-string structure, while changing the presentation of the intersections between the external tensor and the $-1$ generators in the LST sectors. Thus all these configurations satisfy the gluing rules described above and lie in the same infinite duality orbit. This does not imply the existence of infinitely many distinct consistent tensor bases with the same massless matter content. Rather, after quotienting by the duality group, only finitely many inequivalent intersection data remain, including the finite intersection numbers between the external $-2$ and $-1$ generators.

\subsection{Supergravity blocks}\label{sec:blocks}

\begin{figure}
 \center
 \includegraphics[width=0.8\textwidth]{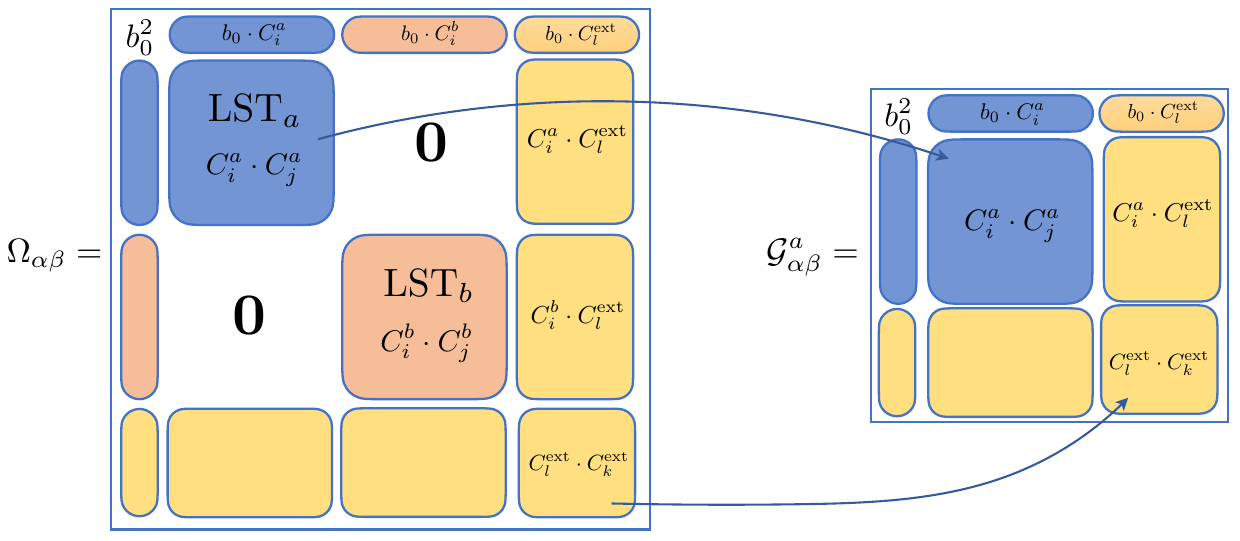} 
\caption{General block structure of the tensor intersection matrix $\Omega_{\alpha\beta}$ determined by the BPS generators and $b_0$. Here two LSTs share the little string charge $f$ associated with the {\it H}-string. The generators $C_i^a$ and $C_j^b$ (satisfying $C_i^{a,b}\cdot f=0$) in LST${}_a$ and LST${}_b$, together with the external generators $C_l^{\rm ext}$ (where $C_l^{\rm ext}\cdot f>0$), form a basis of BPS generators for the supergravity theory. The matrix $\mathcal{G}_{\alpha\beta}^a$ is the Gram matrix of the supergravity block formed by $C_i^a$ and $C_l^{\rm ext}$, and it has a single positive eigenvalue.}
 \label{fig:Intersection-matrix}
 \end{figure}

The existence of the {\it H}-string allows the tensor intersection matrix to be organized into a natural block structure. Once an {\it H}-string charge $f$ is fixed, the BPS generators naturally split into two classes. The first class consists of generators belonging to LSTs that carry the charge $f$. These generators form separate groups labeled by the corresponding LST sectors, and generators from different groups do not intersect. The second class consists of external BPS generators, which intersect the {\it H}-string sector by meeting at least one generator in each LST group. Because the LSTs carrying the {\it H}-string charge are classified, and because the allowed external generators are also all known, Step 2 of the supergravity construction reduces to selecting a consistent set of external generators and attaching them to the chosen LST groups. If the resulting tensor intersection form can be blown down to a Hirzebruch surface, then the full tensor structure of the 6d supergravity is determined.  The intersection matrix therefore takes the general block form shown in Fig.~\ref{fig:Intersection-matrix}. The final step is then to enhance the gauge algebras on tensor multiplets within this fixed tensor structure, which in turn determines the vector multiplets and charged hypermultiplets.

Our aim in this paper is to analyze systematically all possible external generators and the ways they can intersect the LST sectors associated with an {\it H}-string. This is the main content of Step 2 in the construction of the tensor structure. To make this problem manageable, we introduce a special class of tensor subsystems that we call {\it supergravity blocks}.

Recall that a tensor base of a 6d supergravity theory is specified by a set of LSTs sharing an {\it H}-string charge, together with a set of external generators and their intersection pattern. A supergravity block in this setting is constructed by combining all external generators with the generators of a single LST from that collection. Thus a supergravity block is, in general, a proper subset of the full tensor base. For instance, in Fig.~\ref{fig:Intersection-matrix}, the Gram matrix $\mathcal{G}_{\alpha\beta}^a$ on the right-hand side is a supergravity block, while the full tensor structure of the theory is described by the larger intersection matrix $\Omega_{\alpha\beta}$ containing $\mathcal{G}_{\alpha\beta}^a$ as a submatrix.

From this point of view, a complete tensor base can be constructed by consistently arranging supergravity blocks, or equivalently their Gram matrices $\mathcal{G}_{\alpha\beta}^{A}$ with $A=a,b,\cdots$, inside the full tensor intersection form $\Omega_{\alpha\beta}$. Note that all supergravity blocks within a given supergravity share the same set of external generators $C_l^{\rm ext}$, as well as the same $b_0$. In this sense, supergravity blocks are the natural building blocks of supergravity tensor bases. Their most important common feature is that each Gram matrix $\mathcal{G}^a_{\alpha\beta}$ has exactly one positive eigenvalue. As a result, every supergravity block contains BPS strings that can never all become simultaneously tensionless anywhere in tensor moduli space, including at asymptotic infinity. Equivalently, each block contains intrinsically gravitational tensor charges, whose multiples can give rise to BPS black strings. This is what distinguishes supergravity blocks from arbitrary subsets of tensor multiplets. 

In most cases, a single supergravity block can itself be completed into a consistent tensor base of a supergravity with the help of dummy LSTs, although more general tensor bases may also be assembled from several supergravity blocks. We will present explicit examples of this below and in the next section.

These observations suggest that classifying supergravity blocks is the natural first step toward a systematic construction of 6d supergravity theories beyond the mere classification of primitive BPS generators.  Once the supergravity blocks are known, the construction of a full theory proceeds in a systematic way. One first chooses a set of external generators for the desired tensor base. One then collects from the classification table wanted supergravity blocks containing those generators. By putting these blocks together, one obtains a candidate tensor intersection matrix $\Omega_{\alpha\beta}$, as illustrated in Fig.~\ref{fig:Intersection-matrix}. Because the intersection pattern between external generators and the LST generators inside each block is already fixed, and because LST generators belonging to different blocks do not intersect, this step merely reduces to arranging the blocks consistently inside the full matrix rather than solving again for the detailed intersections. If this matrix can be blown down to the tensor base of a Hirzebruch surface, it defines a consistent tensor structure for a supergravity theory. Thus, once the supergravity blocks have been classified, the construction of the tensor base becomes simple and straightforward. Finally, one chooses gauge algebras and charged hypermultiplets on this tensor base so that all gauge and gravitational anomalies are canceled. This will determine the full massless spectrum of the 6d supergravity theory. In this way, the block formalism provides a systematic approach to the construction and the classification of 6d supergravity theories, including non-geometric ones. We postpone the full classification of general theories to future work.

Let us present some concrete examples from F-theory. A simple class of examples are 6d supergravity theories obtained from F-theory compactified on elliptically fibered CY threefolds over a Hirzebruch base $\mathbb{F}_n$ with one blow-up. These theories have $T=2$, and the minimal gauge algebra is supported on the rational curve of self-intersection $-n$, for example $\mathfrak{su}_3$ for $n=3$ and $\mathfrak{e}_6$ for $n=6$. The K\"ahler base of the threefold contains a rational fiber curve $f$, which can degenerate into two intersecting $-1$ curves. These two $-1$ curves form the LST component of the configuration, with little string charge $f$. The remaining $-n$ curve is external to this LST component. Thus the BPS generators consist of the $-n$ curve together with the two $-1$ curves. Including $b_0$, their intersection pairing is
\begin{align}
    \Omega_{\alpha\beta} = \left(\begin{array}{cc}b_0^2 & b_0\cdot \mathcal{C}_j \\ b_0\cdot \mathcal{C}_i & \mathcal{C}_i\cdot \mathcal{C}_j\end{array}\right) = \left(\begin{array}{cccc}7 & 2-n & 1 & 1 \\ 2-n & -n & 1 & 0 \\ 1 & 1 &-1 & 1 \\ 1 & 0 & 1 & -1 \end{array}\right) \ .
\end{align}
This provides a minimal example of a supergravity block. There is only one LST sharing the little string charge $f$, so no further gluing is required. The full intersection matrix $\Omega$ is therefore itself the matrix for a supergravity block with ${\rm sig}(\Omega)=(+,(-)^2,0)$. This case clearly illustrates that an individual supergravity block can realize a full supergravity theory.

More non-trivial examples can be constructed from elliptic threefolds whose base is $B_4=T^4/\mathbb{Z}_m\times \mathbb{Z}_n$, as studied in \cite{Hayashi:2018iqb}. Let us focus on the case $m=n=2$.  In this geometry, the base contains a natural rational fiber class $f$, which is shared by four distinct LST sectors. Each of these LSTs has the same tensor configuration, schematically denoted by
\begin{align}
    {\rm LST}_{a=1,2,3,4} \ : \ 1\, \overunderset{1}{1}4\, 1 \ .
\end{align}
In addition, the base contains four external $-4$ generators. Each external $-4$ curve intersects one of the four $-1$ generators inside a given LST sector at a single point. Thus the theory naturally decomposes into four supergravity blocks. Each block contains one LST sector, together with the four external $-4$ curves attached to it, and its intersection matrix is 
\begin{align}
    \mathcal{G}_{\alpha\beta}^{a=1,2,3,4} = \left(\begin{array}{cccccccccc}-8 & -2 & 1 & 1 & 1 & 1 & -2 & -2 & -2 & -2 \\ 
    -2 & -4 & 1 & 1 & 1 & 1 & 0 & 0 & 0 & 0 \\
    1 & 1 & -1 & 0 & 0 & 0 & 1 & 0 & 0 & 0 \\
    1 & 1 & 0 & -1 & 0 & 0 & 0 & 1 & 0 & 0 \\
    1 & 1 & 0 & 0 & -1 & 0 & 0 & 0 & 1 & 0 \\
    1 & 1 & 0 & 0 & 0 & -1 & 0 & 0 & 0 & 1 \\
    -2 & 0 & 1 & 0 & 0 & 0 & -4 & 0 & 0 & 0 \\
    -2 & 0 & 0 & 1 & 0 & 0 & 0 & -4 & 0 & 0 \\
    -2 & 0 & 0 & 0 & 1 & 0 & 0 & 0 & -4 & 0 \\
    -2 & 0 & 0 & 0 & 0 & 1 & 0 & 0 & 0 & -4  \end{array} \right) \ ,
\end{align}
which has exactly one positive eigenvalue, as required for a supergravity block. Here, the first entry corresponds to $b_0$, and the second through sixth entries correspond to the curves belonging to one LST sector, while the last four entries correspond to the external $-4$ generators. The full intersection matrix $\Omega$ is obtained by gluing four such blocks along the common gravitational vector $b_0$ and along the same four external $-4$ generators. The resulting supergravity theory has $T=17$ with eight $\mathfrak{so}_8$ gauge algebra supported on eight $-4$ curves.

\subsection{Non-Higgsable gravity blocks}

In this paper, we focus on a special class of 6d supergravity blocks built entirely from NHCs appearing in 6d SCFTs and LSTs. NHCs are special configurations of at most three tensor multiplets, each supporting only a minimal gauge algebra that cannot be Higgsed to a simpler one \cite{Morrison:2012np}. We refer to the corresponding class of blocks as {\it non-Higgsable gravity blocks}. The supergravity blocks discussed at the end of the previous subsection provide examples of this class. Restricting to this subclass allows us to bypass the additional complications associated with the choice of gauge algebras and charged matter on tensor charges, and to focus directly on the construction of tensor bases. 

Since all NHCs appearing in SCFTs and LSTs are completely classified, the classification of non-Higgsable gravity blocks can be carried out in a controlled and systematic way. More general gravity blocks can then be obtained from these non-Higgsable ones by enhancing the gauge algebras. We first describe the construction of non-Higgsable gravity blocks and then present their full classification in the next section.

The NHCs that can be realized in F-theory models are given by the following list \cite{Morrison:2012np}:
\begin{align}
    1 \,, \quad 2 \,,\quad \overset{\mathfrak{su}_3}{3} \,,\quad \overset{\mathfrak{so}_8}{4} \,,\quad \overset{\mathfrak{f}_4}{5} \,,\quad \overset{\mathfrak{e}_6}{6} \,,\quad \overset{\mathfrak{e}_7'}{7} \,,\quad \overset{\mathfrak{e}_7}{8} \,,\quad \overset{\mathfrak{e}_8}{12} \,,\quad \overset{\mathfrak{su}_2}2 \ \overset{\mathfrak{g}_2}{3} \,,\quad \overset{\mathfrak{su}_2}2 \ \overset{\mathfrak{so}_7}{3} \ \overset{\mathfrak{su}_2}{2} \,,\quad 2 \ \overset{\mathfrak{sp}_1}2 \ \overset{\mathfrak{g}_2}{3} \ .
\end{align}
Here, each symbol $\overset{\mathfrak{g}}{n}$ denotes a tensor multiplet with self-intersection $-n$ hosting a gauge algebra $\mathfrak{g}$, and $\mathfrak{e}_7'$ represents an $\mathfrak{e}_7$ algebra with a fundamental half-hypermultiplet. At each intersection of two tensor multiplets, there is a bifundamental half-hypermultiplet for the associated pair of gauge algebras.

There are three additional NHCs that do not admit realizations in usual geometric models:
\begin{align}
    \overset{\mathfrak{su}_8}{\hat{1}} \,, \quad \overset{\mathfrak{su}_8}{2} \ \overset{\mathfrak{su}_{16}}{\hat{1}} \,, \quad \overset{\mathfrak{su}_8}{2} \!\!=\!\!\! \overset{\mathfrak{so}_{16}}{4} \ ,
\end{align}
where the notation $p\!=\!q$ indicates that the mutual intersection number is equal to 2. The $\mathfrak{su}_N$ gauge algebra on $\hat{1}$ couples to a symmetric hypermultiplet, and each pair of adjacent gauge algebras supports a bifundamental (full-)hypermultiplet. These clusters can nevertheless be constructed in F-theory by allowing frozen singularities, as described in \cite{Bhardwaj:2019hhd}. We further note that LSTs with a {\it P}-type endpoint cannot contain any of these additional clusters, and therefore they do not appear in LSTs associated with an {\it H}-string. Accordingly, the $\hat{1}$ tensor and, in the last cluster, either the $2$ and $4$ tensors (or both) must appear only as external generators intersecting the charge $f$ positively.

A non-Higgsable gravity block is then a collection of these NHCs. For a fixed {\it H}-string charge $f$, such a block consists of the BPS generators of an LST for the little string charge $f$ together with all external generators having positive intersection with $f$. These generators must be glued together in a way consistent with the structure of the NHCs. Therefore, a non-Higgsable gravity block is specified by the types of BPS generators it contains and by their mutual intersection numbers. A crucial requirement is that the Gram matrix of these generators has exactly one positive eigenvalue as discussed in the previous section.

We now classify all allowed intersections between external generators and tensor multiplets in an LST associated with an {\it H}-string. In the case of non-Higgsable gravity blocks, every such external generator is itself a component of an NHC. The basic constraints on the generators appearing in a block are the following:
\begin{itemize}
    \item External generators do not  intersect one another, except for $-1$ generators and the cases $2 \, 3$, $2 \!=\! 4$.

    \item Aside from $-1$ generators and the configurations $2\,3$, $2\!=\!4$, external generators can intersect only $-1$ generators (or $f$ when $T=1$). In such cases, the intersection numbers are constrained by $\sum_i c_{\mathfrak{g}_i} \le 8$ (or $\sum_i c_{\mathfrak{g}_i} \le 20$ when $T=1$).
\end{itemize}
In the second constraint, $c_{\mathfrak{g}_i}$ represents the central charge contribution, defined in \eqref{eq:current-algebra}, from the current algebra of the gauge algebra $\mathfrak{g}_i$ intersecting the $-1$ generator. The bound
\begin{align}\label{eq:E-string-bound}
    \sum_i c_{\mathfrak{g}_i} \le 8 \ ,
\end{align}
reflects the fact that the E-string theory associated with a $-1$ generator has total left-moving central charge $8$.

Under these conditions, the external generators with self-intersection $\mathcal{C}^2\le -5$, which supports exceptional gauge algebras $\mathfrak{e}_6, \mathfrak{e}_7, \mathfrak{e}_8$ or $\mathfrak{f}_4$, can be attached to $-1$ generators located only at an end in the LST bases or those appearing in an isolated link. The allowed tensor intersections take the form
\begin{align}
    &\textcolor{red}{\overset{\mathfrak{e}_{678},\mathfrak{f}_4}{\mathcal{C}}} \, 1223\cdots , && \textcolor{red}{\overset{\mathfrak{e}_{67},\mathfrak{f}_4}{\mathcal{C}}} \, 123\cdots ,  && \textcolor{red}{\overset{\mathfrak{e}_{6},\mathfrak{f}_4}{\mathcal{C}}} \, 131\cdots , &&  \textcolor{red}{\overset{\mathfrak{f}_4}{\mathcal{C}}} \, 132\cdots , &&   \cdots 2\overset{\textcolor{red}{\overset{\mathfrak{f}_4}{\mathcal{C}}}}13\cdots , \nonumber \\
    & \textcolor{red}{\overset{\mathfrak{e}_{678},\mathfrak{f}_4}{\mathcal{C}}}1222\cdots 2221 , && \textcolor{red}{\overset{\mathfrak{e}_{678},\mathfrak{f}_4}{\mathcal{C}}}1222\cdots 2\overset{2}{2}2 , && \textcolor{red}{\overset{\mathfrak{e}_{678},\mathfrak{f}_4}{\mathcal{C}}}11\ , &&  2\overset{\textcolor{red}{\overset{\mathfrak{e}_{678},\mathfrak{f}_4}{\mathcal{C}}}}12 \ ,
\end{align}
where the red $\mathcal{C}$ denotes an external generator, and the gauge algebra shown above it indicates the allowed gauge algebra. The ellipsis $\cdots$ stands for the remainder of the LST tensor chain. In every case, the tensor multiplets in the LST base support minimal gauge algebras only. In the second line, the latter two configurations arise as special cases of the first two. Few more special configurations also follow from the second line, subject to further restrictions on the allowed external gauge algebras:
\begin{align}
    \textcolor{red}{\overset{\mathfrak{e}_{67},\mathfrak{f}_4}{\mathcal{C}}}\, 1 \overset{\mathfrak{su}_2}{2} 1 \ , && \textcolor{red}{\overset{\mathfrak{e}_{67},\mathfrak{f}_4}{\mathcal{C}}}\, 1 \overset{\mathfrak{su}_2}{2} 21 \ , &&\textcolor{red}{\overset{\mathfrak{e}_{67},\mathfrak{f}_4}{\mathcal{C}}}\, 1 \overunderset{\mathfrak{su}_2}{2}{2} 2 \ , && \overset{\mathfrak{su}_2}2\overset{\textcolor{red}{\overset{\mathfrak{e}_{67},\mathfrak{f}_4}{\mathcal{C}}}}12 \ , && \overset{\mathfrak{su}_2}2\,\overset{\textcolor{red}{\overset{\mathfrak{f}_4}{\mathcal{C}}}}1\,\overset{\mathfrak{su}_2}2 \ .
\end{align}
In these cases, every $-2$ generator carrying an $\mathfrak{su}_2$ gauge algebra must additionally intersect a $-3$ external generator.

An external $-4$ generator can intersect either $-1$ generators or a $-2$ generator in the LST base. If it intersects a $-2$ generator, the supported gauge algebra must be $\mathfrak{so}_{16}$, and the mutual intersection number is required to be two. In all other cases, the external generator supports an $\mathfrak{so}_8$ gauge algebra. This external generator can intersect arbitrarily many $-1$ generators, but it can intersect at most one $-2$ generator. The allowed intersection configurations are
\begin{align}
    &\textcolor{red}{\mathfrak{so}_{8,16}}\,12\cdots , && \textcolor{red}{\mathfrak{so}_{8,16}}\, \overset{2}1 2 \, , && \textcolor{red}{\mathfrak{so}_8}\,13\cdots , && \textcolor{red}{\mathfrak{so}_8}\,14\cdots , && \textcolor{red}{\mathfrak{so}_8\!\!=\!}12\cdots , && \textcolor{red}{\mathfrak{so}_{8}\!\!=}\, \overset{2}1 2 \, ,  \nonumber \\
    & \cdots 2\overset{\textcolor{red}{\mathfrak{so}_8}}13\cdots , && \cdots 2\overset{\textcolor{red}{\mathfrak{so}_8}}14\cdots , && \textcolor{red}{\mathfrak{so}_{16}\!\!=\!\!}\, \overunderset{\mathfrak{su}_8}{1}2 1 \, , && \textcolor{red}{\mathfrak{so}_{16}\!\!=\!\!}\,\overset{\mathfrak{su}_8}{2}1\, 2 \ ,
    \label{eq:-4_external}
\end{align}
where `$=$' again denotes an intersection number of two, and the red algebras are the gauge algebras supported on the external generator. In the first two cases, if the external generator carries $\mathfrak{so}_{16}$, it must additionally intersect another $-2$ generator. Moreover, while the tensor multiplets adjacent to the $-1$ tensor support only minimal gauge algebras unless specified, the $-2$ tensors in the first two cases may carry an $\mathfrak{su}_2$ gauge algebra when the external gauge algebra is $\mathfrak{so}_8$. In the second line, the first case can occur only in an isolated link, and the second case is allowed only in a side link.

Similarly, an external generator $\hat{1}$ may intersect either $-1$ generators or a $-2$ generator in the LST base, with mutual intersection number one. If it intersects a $-1$ generator, the supported gauge algebra must be $\mathfrak{su}_8$, whereas an intersection with a $-2$ generator requires the gauge algebra $\mathfrak{su}_{16}$. While this external generator may intersect arbitrarily many $-1$ generators, it can intersect at most one $-2$ generator. The allowed intersection configurations are
\begin{align}
    &\textcolor{red}{\mathfrak{su}_{8}}\,12\cdots , && \textcolor{red}{\mathfrak{su}_{8}}\, \overset{2}1 2 \, , && \textcolor{red}{\mathfrak{su}_{16}}\, \overunderset{\mathfrak{su}_8}{1}2 1 \, , && \textcolor{red}{\mathfrak{su}_{16}}\,\overset{\mathfrak{su}_8}{2}1\, 2 \ .
\label{eq:hat1_external}
\end{align}
However, the latter two configurations, which involve an $\mathfrak{su}_{16}$ gauge algebra on the external generator, cannot arise in any supergravity theory. 
For instance, the Gram matrix $\mathcal{G}_{IJ}$ corresponding to the third configuration is
\begin{align}
    \mathcal{G}_{IJ} = \left(\begin{array}{ccccc}
    9-T & -1 & 0 & 1 & 1 \\
    -1 & -1 & 1 & 0 & 0 \\
    0 & 1 & -2 & 1 & 1 \\
    1 & 0 & 1 & -1 & 0 \\
    1 & 0 & 1 & 0 & -1
    \end{array}\right) \ ,
\end{align}
leading to $\det \mathcal{G} = T-11$, and $\mathrm{tr}\,\mathcal{G} = 4-T$. This indicates that $T>4$ is necessary to ensure the correct signature~\eqref{eq:eigenvalue_constraint}.
For $T>4$, the presence of either configuration would imply an additional LST sharing the same {\it H}-string charge, and require the external generator to intersect a $-1$ generator within this second LST.\footnote{The external $\hat{1}$ generator cannot intersect a $-2$ generator in the second LST supporting $\mathfrak{su}_8$ gauge algebra since there is not enough flavor symmetry to gauge on the $\hat{1}$ generator.} This is impossible, since a $-1$ generator cannot intersect the $\hat{1}$ generator supporting the $\mathfrak{su}_{16}$ gauge algebra. So the last two configurations cannot be realized in supergravity.

A $-3$ external generator can attach either to $-1$ generators or $-2$ generators in the LST. In the latter case, the mutual intersection number is fixed to be one. When the external generator supports $\mathfrak{g}_2$ or $\mathfrak{so}_7$, it is required to intersect a $-2$ generator carrying an $\mathfrak{su}_2$ gauge algebra, after which it may also intersect additional $-1$ generators. The allowed intersections are as follows:
\begin{align}
    &\overset{\textcolor{red}{\overset{\mathfrak{su}_{3},\mathfrak{g}_2,\mathfrak{so}_7}{|\,n}}}1 \!\!\!\!2\cdots  , \quad  \overset{\textcolor{red}{\overset{\mathfrak{su}_{3},\mathfrak{g}_2,\mathfrak{so}_7}{|\,n}}}1 \!\!\!\!3\cdots \ , \quad \overset{\textcolor{red}{\overset{\mathfrak{su}_{3},\mathfrak{g}_2,\mathfrak{so}_7}{|\,n}}}1 \!\!\!\!4\cdots , \quad \cdots 2\!\!\!\!\overset{\textcolor{red}{\overset{\mathfrak{su}_{3},\mathfrak{g}_2,\mathfrak{so}_7}{|\,n}}}1 \!\!\!\!3\cdots , \qquad\!\!\!\! \cdots 2\!\!\!\!\overset{\textcolor{red}{\overset{\mathfrak{su}_{3},\mathfrak{g}_2,\mathfrak{so}_7}{|\,n}}}1\!\!\!\!4\cdots  , \quad 2\!\!\!\!\overset{\textcolor{red}{\overset{\mathfrak{su}_{3},\mathfrak{g}_2,\mathfrak{so}_7}{|\,n}}}1 \!\!\!\!2 \ , \nonumber \\
    & \overset{\textcolor{red}{\mathfrak{su}_{3},\mathfrak{g}_2}} 1 5 \cdots , \quad \overset{\textcolor{red}{\mathfrak{su}_{3}}}1 6 \cdots , \qquad \!\!\!\! \cdots 2\overset{\textcolor{red}{\mathfrak{su}_3,\mathfrak{g}_2}}15\cdots , \qquad\!\!\!\! \cdots 2\overset{\textcolor{red}{\mathfrak{su}_3}}16\cdots , \qquad \cdots 3\!\overset{\textcolor{red}{\mathfrak{su}_3,\mathfrak{g}_2}}1\!3\cdots  , \nonumber \\
    & \overset{\textcolor{red}{\mathfrak{g}_2,\mathfrak{so}_7}}21\cdots  , \qquad
    \overset{\textcolor{red}{\mathfrak{g}_2}}221\cdots  , \qquad
    2\overset{\textcolor{red}{\mathfrak{g}_2}}21\cdots  , \qquad
    1\!\overset{\textcolor{red}{\mathfrak{g}_2,\mathfrak{so}_7}}2\!1 ,
    \qquad \textcolor{red}{\mathfrak{g}_2}\, \overset{1}{2}21 \ ,
\end{align}
where `$|n$' denotes a mutual intersection number $n$ constrained by the condition \eqref{eq:E-string-bound} for the gauge algebras intersecting the common $-1$ generator.

An external $-2$ generator must support either an $\mathfrak{su}_2$ or an $\mathfrak{su}_8$ gauge algebra, or be part of a $223$ NHC. Otherwise, it can be removed from the set of generators, as discussed in the previous section. When the supported algebra is $\mathfrak{su}_2$, the external generator necessarily intersects a $-3$ tensor multiplet in an LST, and it can additionally intersect other $-1$ generators. If instead it carries an $\mathfrak{su}_8$ algebra, it must first intersect a $-4$ tensor multiplet with intersection number two, and it may intersect other $-1$ generators. The allowed intersections for such cases are
\begin{align}
    & \textcolor{red}{\mathfrak{su}_{2}}\,31\cdots , &&
    \cdots1\overset{\textcolor{red}{\mathfrak{su}_2}}31\cdots ,
    && \textcolor{red}{\mathfrak{sp}_{1}}\,321\cdots ,
    && \cdots1\overset{\textcolor{red}{\mathfrak{sp}_{1}}}{3}21\cdots ,
    && \textcolor{red}{\mathfrak{su}_{8}\!=}\overset{\mathfrak{so}_{16}}412231 ,
   \nonumber \\
    & \textcolor{red}{\mathfrak{su}_{8}\!=}\overset{\mathfrak{so}_{16}}41222\ , && 1\overset{\textcolor{red}{\overset{\mathfrak{su}_8}{||}}}4122\ , && 21\overset{\textcolor{red}{\overset{\mathfrak{su}_8}{||}}}412 \ .
\end{align}
Once one of these configurations is present, the external $-2$ generator can further intersect other $-1$ generators in the LSTs, provided that the bound \eqref{eq:E-string-bound} is satisfied for each attached $-1$ generator. We also note that the three LSTs listed in the second line are in fact equivalent under the transition described around \eqref{eq:dummy-transition}, which rearranges the E- and M-string tensor multiplets.

Following these intersection rules together with the classification of LSTs, we can systematically construct and classify all non-Higgsable gravity blocks.

To perform the classification, it is helpful to understand the allowed intersections between external generators and LSTs.  For small $T$, the patterns are further constrained compared to the general cases discussed in Section~\ref{sec:blocks}.  For instance, a $-4$ external generator can intersect a $-1$ LST generator at two points, giving a Gram matrix of the form
\begin{align}
    \mathcal{G}_{IJ}=
    \left(\begin{array}{ccccc}
    9-T & -2 & 1 & 0 & \cdots \\
    -2 & -4 & 2 & 0 & \cdots \\
    1 & 2 & -1 & 1 & \cdots \\
    0 & 0 & 1 & -2 & \cdots \\
    \cdots & & & &
    \end{array}\right) .
\end{align}
This is the local pattern appearing in the third base of Eq.~\eqref{eq:-4_external}.  Imposing the signature constraint~\eqref{eq:eigenvalue_constraint} on this Gram matrix rules out such a configuration for $T\leq 9$.  Applying the same test to the other small-$T$ exceptional attachments gives
\begin{align}
    &\textcolor{red}{\mathfrak{so}_{8}\!\!=}\, \overset{2}1 2 : T\geq 10,
    &&\textcolor{red}{\mathfrak{so}_{16}\!\!=\!\!}\, \overunderset{\mathfrak{su}_8}{1}2 1 : T\geq 11,
    &&\textcolor{red}{\mathfrak{so}_{16}\!\!=\!\!}\, \overset{\mathfrak{su}_8}{2}1 2 : T\geq 11,
    \nonumber\\
    &\textcolor{red}{\mathfrak{su}_{8}}\,12\cdots : T\geq 10,
    &&\textcolor{red}{\mathfrak{su}_{8}}\,\overset{2}1 2 : T\geq 10,
    \nonumber\\
    &\overset{\textcolor{red}{\overset{\mathfrak{su}_{3},\mathfrak{g}_2,\mathfrak{so}_7}{|\,n\geq2}}}{1\phantom{X}}\!\!\!\!2\cdots : T\geq 10,
    &&\overset{\textcolor{red}{\overset{\mathfrak{su}_{3},\mathfrak{g}_2,\mathfrak{so}_7}{|\,n\geq2}}}{1\phantom{X}}\!\!\!\!3\cdots : T\geq 10,
    &&\overset{\textcolor{red}{\overset{\mathfrak{su}_{3},\mathfrak{g}_2,\mathfrak{so}_7}{|\,n\geq2}}}{1\phantom{X}}\!\!\!\!4\cdots : T\geq 11,
    \nonumber\\
    &2\!\!\!\overset{\textcolor{red}{\overset{\mathfrak{su}_{3},\mathfrak{g}_2,\mathfrak{so}_7}{n\geq2\,|}}}{\phantom{X}1}3\cdots : T\geq 10,
    &&\cdots 2\!\!\!\!\overset{\textcolor{red}{\overset{\mathfrak{su}_{3},\mathfrak{g}_2,\mathfrak{so}_7}{n\geq2\,|}}}{\phantom{X}1}4\cdots : T\geq 11,
    &&2\!\!\!\!\overset{\textcolor{red}{\overset{\mathfrak{su}_{3},\mathfrak{g}_2,\mathfrak{so}_7}{n\geq2\,|}}}{\phantom{X}1}2 : T\geq 10 .
    \label{eq:external_intersection_patterns}
\end{align}
These restrictions explain why several otherwise locally allowed attachments first appear only at $T=10$ or later.

\subsection{\texorpdfstring{$T=9$ and $b_0=0$ case}{T=9 and b0=0 case}}\label{sec:b0_zero}

Let us finally discuss the exceptional case with $T=9$ and $b_0=0$. This case is not covered by the construction above, because the argument for the existence of an {\it H}-string does not apply. In particular, there is no distinguished tensor charge $f$ obeying $b_0\cdot f=2$, and hence there is no  decomposition of the tensor base into LST sectors sharing an {\it H}-string and external generators attached to them. Instead, the tensor base should be described directly in terms of the BPS cone and its BPS generators.

Since $b_0$ is the zero vector in this case, the tensor charge lattice is even. Together with unimodularity and signature $(1,9)$, this identifies the tensor charge lattice with
\begin{align}
    \Gamma^{1,9} \simeq U\oplus E_8(-1) \ .
\end{align}
Here, $U$ denotes the rank-two hyperbolic lattice with intersection matrix $\Omega$ of $\mathbb{F}_0$ in \eqref{eq:P2-Fn}, and $E_8(-1)$ denotes the negative-definite $E_8$ root lattice. This conclusion uses only the charge-lattice constraints and does not assume a geometric realization. A useful geometric model is provided by an Enriques surface, whose numerical divisor lattice is the same lattice $U\oplus E_8(-1)$ \cite{CossecDolgachev1989}. In that realization the canonical class is torsion and hence numerically trivial, matching $b_0=0$.

The possible primitive BPS generators are therefore simple. The generator list in \eqref{eq:more-generators} and the condition $b_0=0$ leave only two types of non-positive primitive charges that can appear as generators:
\begin{align}
   1)\ \mathcal{C}^2=-2 \ , \ b_0\cdot\mathcal{C}=0 \qquad {\rm or} \qquad
    2) \ \mathcal{F}^2=0 \ , \ b_0\cdot\mathcal{F}=0 \ .
\end{align}
The first type is a $-2$ tensor, while the second type is the fourth generator in \eqref{eq:more-generators}, namely the Type II critical string charge, which can become a genuine generator precisely in the present $T=9$, $b_0=0$ case. When a geometric realization exists, these two cases are represented respectively by rational $-2$ curves and genus-one curves.

The tensor cone is then classified as follows. Choose a connected component of the positive cone
\begin{align}
    \mathcal{C}^+ = \{ J\in \Gamma^{1,9}\otimes \mathbb{R} \mid J^2>0\} \ .
\end{align}
Every $-2$ generator defines a Weyl wall $J\cdot \mathcal{C}\ge0$. The tensor cone is a chamber cut out by these walls,
\begin{align}
    \mathcal{T}
    =
    \{J\in \mathcal{C}^+ \mid J\cdot \mathcal{C}\ge0
    \ {\rm for\ all}\ -2\ {\rm generators}\}\ .
\end{align}
Its dual BPS cone is generated by the corresponding $-2$ tensors together with primitive isotropic boundary rays $\mathcal{F}^2=0$. In an Enriques surface, these isotropic rays are genus-one fiber classes. If such a fiber is reducible, its components are $-2$ curves arranged in an affine ADE configuration, and the total fiber class is their positive null combination. If there are no $-2$ generators, the tensor cone is simply the full positive cone, and the boundary rays are Type II string charge classes.

Therefore the $T=9$, $b_0=0$ tensor bases are classified, without assuming geometry, by chambers of the positive cone of $U\oplus E_8(-1)$, or equivalently by the possible collections of Weyl walls of $-2$ generators together with the primitive null rays on their boundaries. Different chamber presentations related by automorphisms of the lattice describe the same tensor cone up to the duality group.

\section{Classification}\label{sec:classification}

In this section, we classify {\it 6d non-Higgsable gravity blocks} introduced in the previous section. We begin by presenting the complete list of LSTs associated with an {\it H}-string that can be embedded in a non-Higgsable block. We then enumerate all non-Higgsable gravity blocks that are anomaly-free and compatible with the refined structure of the tensor moduli space described in Section~\ref{sec:refined_structure_of_tensor_moduli_space}. To illustrate the classification, we also explicitly list all allowed supergravity bases up to $T=10$ and discuss several examples involving frozen singularities. The complete datasets of LST bases and non-Higgsable gravity blocks obtained in our analysis are provided in a separate file linked below.

\subsection{Little string theories}\label{sec:LST}
The classification of non-Higgsable gravity blocks begins with identifying the LST sectors that can share a common {\it H}-string charge $f$. In this subsection, we focus on LST sectors built exclusively from NHCs, as these are the sectors relevant to the non-Higgsable gravity blocks. All such LSTs have a {\it P}-type endpoint, meaning that they can be blown down to the rational fiber class of a Hirzebruch surface. Consequently, they admit geometric realizations in F-theory without frozen singularities \cite{Heckman:2015bfa,Bhardwaj:2015oru}.

There are two types of such LST bases: short bases and long bases. The long bases are those that can be arbitrarily extended by adding conformal matter links before coupling to gravity, while the short bases cannot be extended in this way. When embedded into supergravity, the tensor charge associated with the LST scale, which becomes non-dynamical in the gravity decoupling limit, in these LST bases will be identified with the {\it H}-string charge $f$. The gravitational anomaly cancellation further constrains the number of tensor multiplets in these theories. Notably, the number of tensor multiplets $T^H$, which is defined as the number of tensor fields excluding the one associated with the LST scale, must obey the tensor bound $T^H<193$ in a supergravity theory. As a result, the number of internal links in long bases becomes finite, and only finitely many LST bases can be consistently embedded into supergravity.

We further note that, beyond the bound $T^H<193$, the LST bases that can actually appear in a supergravity base are subject to a stronger restriction. In a consistent supergravity theory the gravitational anomaly fixes $H-V=273-29T$, so that each non-Higgsable gravity block must satisfy $29\,(T^H+1)+(H_{\rm ch}-V)\le 273$, where $H_{\rm ch}$ denotes the number of charged hypermultiplets, once its external generators are attached while the LST sector by itself may have $29\,(T^H+1)+(H_{\rm ch}-V)>273$.
The external generators carrying $\mathfrak{e}_8$, $\mathfrak{e}_7$, $\mathfrak{e}_7'$, $\mathfrak{e}_6$, and $\mathfrak{f}_4$ always intersect a $-1$ curve, so for a given LST the maximal number of such generators that can be attached is calculable. Attaching these $\mathfrak{e}_8,\mathfrak{e}_7,\mathfrak{e}_7',\mathfrak{e}_6,\mathfrak{f}_4$ generators maximally, consistently with the E-string and {\it H}-string central-charge bounds, \eqref{eq:E-string-bound} and \eqref{eq:f-bound} respectively, we conclude that an admissible LST base must obey $29\,(T^H+1)+(H_{\rm ch}-V)\le 273+28\times4=385$, where $H_{\rm ch}$ and $V$ include the contributions from the external generators and $28$ corresponds to the dimension of $\mathfrak{so}_8$ gauge algebra.
We therefore restrict our attention to the LST bases satisfying this bound.\footnote{For the $A$- and $D$-type dummy LSTs in particular, it is immediately clear that they can appear as part of a supergravity base only up to $T^H=25$ and $T^H=16$, respectively, so we include them up to these values.}

\begin{figure}[t]
\centering
\includegraphics[width=\textwidth]{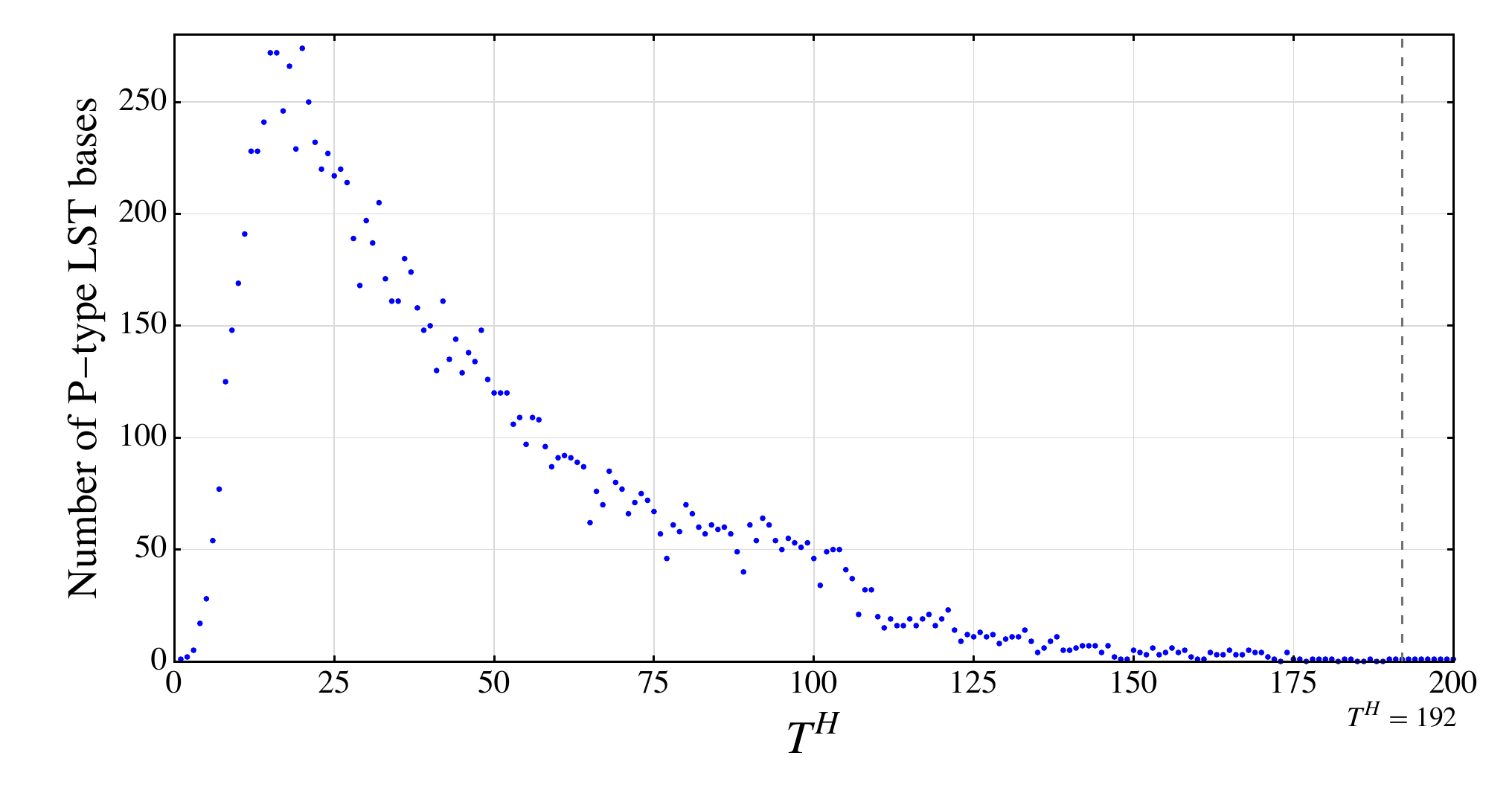}
\caption{The number of {\it P}-type LST bases as a function of the number of tensor multiplets $T^H<193$.}
\label{fig:lst_base_count}
\end{figure}

In Appendix~\ref{app:long_base}, we present a complete classification of long bases for LSTs associated with an {\it H}-string. The full list of short LST bases with up to ten curves is provided in Appendix~\ref{app:LST_base}. The data for the complete classification of LST bases for an {\it H}-string in non-Higgsable blocks are available in the accompanying data repository~\cite{Github:LSTbases}. Altogether, we find that there are only 12,762 such LST bases, whose distribution is shown in Fig.~\ref{fig:lst_base_count}. We emphasize that the bound above is only a coarse check: even among the LST bases retained here, some may not give rise to a consistent supergravity theory.

For reference, we list the {\it H}-string LST bases with a small number of tensor multiplets. In the lists below, $\mathfrak{so}$ denotes $\mathfrak{so}_8$. The complete list for  $T^H \le 6$ is as follows:
\begin{flalign*}
    T^H\!=\!1 \ : \ &11, & \nonumber \\
    T^H\!=\!2 \ : \ &{121}, \ {2 1 2}, \nonumber \\
    T^H\!=\!3 \ :\ &{1221}, \ {2 \overset{2}{2} 1}, \ {1 \overset{1}{3} 1}, \
    {2 1 3 1},\ {2 2 1 3},\nonumber
    \\
    T^H\!=\!4 \ :\ &{12221}, \ {2 \overset{2}{2}2 1}, \ {2 \overset{1}{\underset{1}{3}} 2},
    {1 2 \overset{1}{3} 1},\
    {2 \overset{1}{2} 3 1},\
    {2 \overset{2}{3} 1 2},\
    {3 1 \overset{1}{3} 2},\
    {1 2 3 1 2},\
    {1 3 1 3 1},\
    {2 3 1 2 3}\
    \nonumber\\
    & {3 1 2 3 1},\
    {3 1 3 1 3},\
    {1 \overset{1}{\underset{1}{\mathfrak{so}}} 1},\
    {\mathbf{I}^{\oplus 2} \overset{1}{\mathfrak{so}} 1},\
    {\mathbf{I}^{\oplus 2} \mathfrak{so} \mathbf{I}^{\oplus 2}},\
    {\mathbf{I}^{\oplus 3} \mathfrak{so} 1},\
    {\mathbf{I}^{\oplus 4} \mathfrak{so}},\
    \\
    T^H\!=\!5 \ :\ & {122221}, \ {2 \overset{2}{2}22 1},\ \mathbf{I}^{\otimes 5}5, \
    {1 2 2 \overset{1}{3} 1},\
    {1 2 \overset{1}{3} 2 1},\
    {2 \overset{2}{3} 1 3 1},\
    {3 1 \overset{2}{3} 2 1},\
    {\overset{4,2}{\otimes} 2},\
    {1 3 1 3 2 1},\
    {3 1 2 3 2 1}\nonumber \\
    &
    {1 \overset{1}{\underset{1}{5}} 1 2},\
    {2 1 \overset{1}{5} 1 2},
    {2 2 1 \overset{1}{5} 1},\
    {2 2 1 5 1 2},\
    {2 2 2 1 5 1},\
    {2 2 2 2 1 5},\
    {\overset{2,2}{\otimes} \overset{1}{\mathfrak{so}} 1},\
    {2 2 \overset{1}{3} 1 \mathfrak{so}},\
    {2 \overset{1}{3} 1 \mathfrak{so} 1},\
    {2 \overset{1}{3} 2 1 \mathfrak{so}}
    \nonumber\\
    &
    {3 1 \overset{1}{\mathfrak{so}} 1 3},\ {\overset{2,2}{\otimes} \mathfrak{so} \mathbf{I}^{\oplus 2}},\
    {\overset{2,3}{\otimes} \mathfrak{so} 1},\
    {\overset{2,4}{\otimes} \mathfrak{so}},\
    {2 3 1 3 1 \mathfrak{so}},\
    {3 \overset{2,2}{\otimes} \mathfrak{so} 1},\
    {3 1 3 2 1 \mathfrak{so}},\
    {3 2 1 \mathfrak{so} 1 3},\
\end{flalign*}
\begin{flalign*}
T^H\!=\!6 \ :\ &{1222221}, \ {2 \overset{2}{2}222 1},\  {2 \overset{2}{3} 1 3 2 1},\
{1 2 3 1 3 2 1},\
{1 3 1 3 2 2 1},\
{1 \overset{1}{5} 1 \overset{1}{3} 2},\
{1 \overset{1}{\underset{1}{5}} \overset{2,2}{\otimes}},\
{3 1 \overset{1}{\underset{1}{5}} 1 3},\
{3 1 \overset{\mathbf{I}^{\oplus 2}}{5} 1 3},&\nonumber \\
&
{3 1 \overset{3}{\overset{1}{5}} 1 2},\
{1 5 1 2 \overset{1}{3} 2},\
{1 5 1 \overset{1}{3} 2 2},\
{1 \overset{1}{5} \overset{3,2}{\otimes}},\
{1 \overset{1}{5} 1 3 1 3},\
{2 1 5 1 \overset{1}{3} 2},\
{2 1 \overset{1}{5} \overset{2,2}{\otimes}},\
{3 1 \overset{1}{5} 1 2 3},\
{1 5 1 2 2 3 1},\nonumber \\
&
{1 5 1 2 3 1 3},\
{2 1 5 1 2 3 1},\
{2 1 5 1 3 1 3},\
{2 2 1 5 1 3 1},\ 
{2 2 3 1 3 1 5},\
{2 3 1 3 1 5 1},\
{2 3 1 3 2 1 5},\
{3 2 1 5 1 2 3}, &
\nonumber\\
&
{3 2 2 1 5 1 3},\
{5 1 2 2 3 1 3},\
{2 \overset{2}{3} 1 \overset{1}{\mathfrak{so}} 1},\
{\overset{3,2}{\otimes} \overset{1}{\mathfrak{so}} 1},\
{1 2 \overset{2}{3} 1 \mathfrak{so} 1},\
{2 \overset{2}{3} 1 \mathfrak{so} \mathbf{I}^{\oplus 2}},\
{\overset{3,2}{\otimes} \mathfrak{so} \mathbf{I}^{\oplus 2}},\
{\overset{3,3}{\otimes} \mathfrak{so} 1},\
{\overset{2,2}{\otimes} \mathfrak{so} \overset{2,2}{\otimes}},
\nonumber\\
&
{\mathbf{I}^{\oplus 2} \overset{1}{\underset{1}{\mathfrak{e}_6}} \mathbf{I}^{\oplus 2}},\
{\mathbf{I}^{\oplus 2} \overset{\mathbf{I}^{\oplus 2}}{\mathfrak{e}_6} \mathbf{I}^{\oplus 2}},\
{\mathbf{I}^{\oplus 3} \overset{1}{\underset{1}{\mathfrak{e}_6}} 1},\
{\mathbf{I}^{\oplus 3} \overset{1}{\mathfrak{e}_6} \mathbf{I}^{\oplus 2}},\
{\mathbf{I}^{\oplus 4} \overset{1}{\mathfrak{e}_6} 1},\
{\mathbf{I}^{\oplus 3} \mathfrak{e}_6 \mathbf{I}^{\oplus 3}},\
{\mathbf{I}^{\oplus 4} \mathfrak{e}_6 \mathbf{I}^{\oplus 2}},\
{\mathbf{I}^{\oplus 5} \mathfrak{e}_6 1},
\nonumber\\
&
{\mathbf{I}^{\oplus 6} \mathfrak{e}_6},\
{1 \overset{1}{\mathfrak{so}} 1 \overset{1}{\mathfrak{so}} 1},\
{1 \overset{1}{\mathfrak{so}} 1 \mathfrak{so} \mathbf{I}^{\oplus 2}},\
{2 3 1 \overset{1}{\mathfrak{so}} 1 \mathfrak{so}},\
{3 1 \overset{1}{\mathfrak{so}} 1 \mathfrak{so} 1},\
{1 \mathfrak{so} \overset{2,2}{\otimes} \mathfrak{so} 1},\
{1 \mathfrak{so} 1 \mathfrak{so} 1 2 3},
\nonumber\\
&
{\mathbf{I}^{\oplus 2} \mathfrak{so} 1 \mathfrak{so} \mathbf{I}^{\oplus 2}},\
{2 3 2 1 \mathfrak{so} 1 \mathfrak{so}},\
{\mathfrak{so} \overset{3,2}{\otimes} \mathfrak{so} 1},\
{\mathfrak{so} \overset{3,3}{\otimes} \mathfrak{so}},\
{\mathfrak{so} \overset{2,2}{\otimes} \mathfrak{so} 1 3}\ .
\end{flalign*}
For the notation used in this list, we refer the reader to Appendix~\ref{app:SCFTsLSTs} and \ref{app:long_base}. 

These bases are the LST-sector inputs for the non-Higgsable gravity blocks. Their tensor multiplets support only the minimal gauge algebras, and their mutual intersection numbers are either one or zero. To construct a supergravity tensor base, one chooses one or more LST sectors from this list, identifies their little string charges with the same {\it H}-string charge $f$, and imposes the bound on the number of tensors using the convention in \eqref{eq:number-of-tensors}. The next step, which we carry out in the following subsections, is to attach the allowed external generators to these {\it H}-string sectors to form candidate non-Higgsable gravity blocks. In the following subsections, we construct non-Higgsable gravity blocks by attaching the allowed external generators to these {\it H}-string sectors and then imposing the Gram-matrix signature and blowdown constraints discussed in Section~\ref{sec:blocks}.

\subsection{Non-Higgsable gravity blocks}\label{sec:non_higgsable_gravity_blocks_numerics}

\begin{figure}[t]
\centering
\resizebox{\textwidth}{!}{%
\begin{tikzpicture}[
  node distance=1.6cm and 2.2cm,
  box/.style={
    draw,
    rounded corners,
    align=center,
    minimum width=4.2cm,
    minimum height=1.4cm,
    inner sep=8pt
  },
  arrow/.style={-{Latex[length=2.5mm]}, very thick}
]

\node[box] (lst) {\textbf{$P$-type LST bases}\\Section~\ref{sec:LST}};

\node[box, right=of lst] (attach) {\textbf{Attachment of external generators}\\Tables~\ref{tab:multi}--\ref{tab:cluster_chain}};

\node[box, below=of attach] (check) {\textbf{Consistency checks}\\
Anomaly cancellation~\eqref{eq:GS-conds-non-abelian}\\
E-string bound~\eqref{eq:E-string-bound}, {\it H}-string bound~\eqref{eq:f-bound}\\
Gram signature~\eqref{eq:eigenvalue_constraint}, blowdown~\eqref{eq:P2-Fn}};

\node[box, left=of check] (catalog) {\textbf{Non-Higgsable}\\\textbf{gravity blocks}~\cite{Zenodo}\\Figures~\ref{fig:gravity_block_count_byT}--\ref{fig:Number_distinct_gauge_patterns}, Tables~\ref{tab:simple-lie-algebra-externals}--\ref{tab:distribution-by-number-of-externals}};

\draw[arrow] (lst) -- (attach);
\draw[arrow] (attach) -- (check);
\draw[arrow] (check) -- node[above]{pass} (catalog);

\end{tikzpicture}%
}
\caption{Schematic flow of the numerical construction of non-Higgsable gravity blocks used in Section~\ref{sec:non_higgsable_gravity_blocks_numerics}. Starting from the $P$-type LST bases of Section~\ref{sec:LST}, the code attaches the allowed external generators listed in Tables~\ref{tab:multi}--\ref{tab:cluster_chain} and retains the configurations passing the consistency checks.}
\label{fig:nhgb-flowchart}
\end{figure}
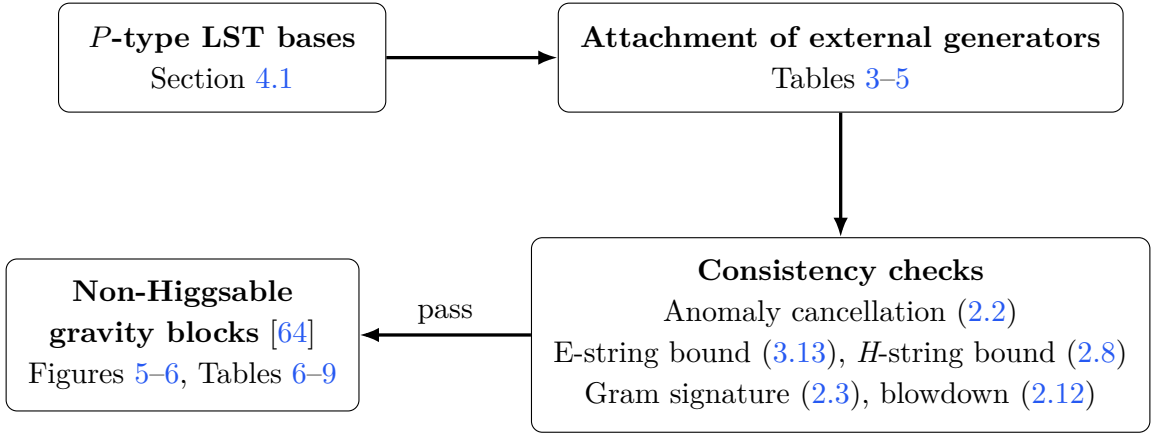

Based on the construction strategy outlined so far, we now explain the numerical implementation of the classification of non-Higgsable gravity blocks and summarize the resulting data.  The computation was performed with the C++ pipeline in the directory \texttt{SUGRA/sugra\_pipeline\_share} of the accompanying code repository~\cite{Github:SUGRAbases}.  The input is the catalog of {\it P}-type LST bases discussed in Section~\ref{sec:LST}. 
The scan then attaches allowed external generators to these LST sectors and retains the configurations that pass the supergravity consistency tests described in Section~\ref{sec:blocks}.
The overall flow of this construction is summarized in Figure~\ref{fig:nhgb-flowchart}.

The computation was submitted to the KEK Central Computing System (KEKCC).  
The first sweep used the serial version of the code and was used to obtain the data up to around $T=90$.  The same pipeline was then run with OpenMP
for the more expensive high-$T$ continuation.  
The OpenMP run parallelizes the same enumeration steps and does not change the catalog definition.
As the $T=1$ Hirzebruch-surface cases are elementary, they are treated analytically in the next subsection and are thus not included in the numerical totals in this subsection.

Regarding the runtime, the completed serial jobs in the first sweep took $6{,}764.01$ CPU hours in total.  
For larger $T$, where OpenMP parallelization was used to reduce wall time, completed OpenMP jobs took $6{,}441.64$ CPU core-hours in total while their elapsed time was $496.85$ hours. This implies an average effective speed-up of about $13$.
We ran these jobs parallelized over $\mathcal{O}(100)$ cores in KEKCC, and eventually the total wall time is a few weeks including the queue waiting time.

\begin{table}[t]
\centering
\renewcommand{\arraystretch}{1.3}
\begin{tabular}{@{}lccc@{}}
\toprule
\multirow{2}{*}{\textbf{External generator $C^{\rm ext}$}} & \multicolumn{2}{c}{\textbf{multi-target}} & \textbf{single-target} \\
\cmidrule(lr){2-3}\cmidrule(l){4-4}
 & \textbf{\# of $(-1)$ targets} & \textbf{\#(int.)\ on each} & \textbf{\#(int.)\ on the $(-1)$} \\
\midrule
$\mathfrak{su}_2$ on a $-2$    & $2$ to all & $1$--$5$ & $1$--$2$ \\
$\mathfrak{su}_3$ on a $-3$    & $2$ to all & $1$--$5$ & $1$--$2$ \\
$\mathfrak{so}_8$ on a $-4$    & $2$ to $9$ & $1$--$2$ & $1$--$2$ \\
$\mathfrak{f}_4$ on a $-5$     & $2$ to $5$ & $1$ & $1$ \\
$\mathfrak{e}_6$ on a $-6$     & $2$ to $4$ & $1$ & $1$ \\
$\mathfrak{e}_7$ on a $-7/{-}8$& $2$ to $3$ & $1$ & $1$ \\
$\mathfrak{e}_8$ on a $-12$    & $2$ to $2$ & $1$ & $1$ \\
$\mathfrak{su}_8$ on a $-1$    & $2$ to $6$ & $1$ & $1$ \\
\midrule
$\mathfrak{su}_8$ on a $-2$    & none       & none & $2$ on the $-4$ \\
$\mathfrak{su}_{16}$ on a $-1$   & none       & none & $2$ on the $-2$\\
$\mathfrak{so}_{16}$ on a $-4$   & none       & none & $2$ on the $-2$\\
\midrule
single $(-1)$                  & \multicolumn{3}{c}{$1$ target, $C^{\rm ext}\cdot f=1$ for any generator} \\
\bottomrule
\end{tabular}
\caption{\textbf{Single- and multi-target attachment.} A single external generator is attached to
several base $(-1)$ generators with an independent intersection number on
each. The light $\mathfrak{su}_2,\mathfrak{su}_3$ run up to \#(int.)$=5$ on every
target; a heavier gauge's per-target \#(int.) and target number are limited by the rank bound. The single-target column lists the \#(int.) on a single $(-1)$, and the last row summarizes the case where the external generator is single $(-1)$. Here $f$ denotes the {\it H}-string charge~\eqref{eq:H-string-charge}, and the condition $C^{\rm ext}\cdot f=1$ for a single $(-1)$ was discussed at the end of Section~\ref{sec:general_construction}.}
\label{tab:multi}
\end{table}

\begin{table}[t]
\centering
\renewcommand{\arraystretch}{1.3}
\begin{tabular}{@{}lccc@{}}
\toprule
\multirow{2}{*}{\textbf{External generator}} & \textbf{core: attached to} & \multicolumn{2}{c}{\textbf{plus extra $(-1)$ legs}} \\
\cmidrule(l){3-4}
 & \textbf{(base generator, \#(int.))} & \textbf{\# legs} & \textbf{\#(int.)\ on each} \\
\midrule
$\mathfrak{su}_2$ on a $-2$    & one $-3$ \ (\#(int.) $1$)        & $1$ to $4$ & $1$--$5$ \\
$\mathfrak{g}_2$ on a $-3$     & one $-2$ \ (\#(int.) $1$)        & $1$ to $4$ & $1$--$5$ \\
$\mathfrak{so}_7$ on a $-3$    & two $-2$ \ (\#(int.) $1$ each)   & $1$ to $4$ & $1$--$5$ \\
gauge-less $-2$                & one $-2$ of a $23$ (\#(int.) $1$)& $0$ & $0$ \\
\bottomrule
\end{tabular}
\caption{\textbf{Mixed attachment.} An external generator is attached to one ``core'' base
generator and, simultaneously, to several base $(-1)$ legs, each at its own
intersection number. The leg count is hard-capped at $4$ for practical cutoff, and
the per-leg \#(int.) runs $1$--$5$. In the scan the gauge-less ``2'' carries no extra
$(-1)$ legs (both entries set to $0$), though the code~\cite{Github:SUGRAbases} also provides an option to
turn these on (up to $2$ legs at \#(int.) $1$--$3$).}
\label{tab:mixed}
\end{table}

\begin{table}[t]
\centering
\renewcommand{\arraystretch}{1.3}
\begin{tabular}{@{}llcc@{}}
\toprule
\textbf{Cluster} & \textbf{Gauge} & \textbf{each cluster generator glued to} & \textbf{\#(int.)} \\
\midrule
$23$  & $\mathfrak{su}_2\!\oplus\!\mathfrak{g}_2$                            & $1$ to $3$ base $(-1)$'s, or one $-2$ & $1$--$2$, or $1$ \\
$232$ & $\mathfrak{su}_2\!\oplus\!\mathfrak{so}_7\!\oplus\!\mathfrak{su}_2$  & $1$ to $3$ base $(-1)$'s              & $1$--$2$ \\
$223$ & $\mathfrak{sp}_1\!\oplus\!\mathfrak{g}_2$         & one base $(-1)$                       & $1$ \\
\bottomrule
\end{tabular}
\caption{\textbf{NHC-cluster chain attachment.} A whole NHC chain ($23,232,223$) is
attached to $-1$ generators in the LST base (the $23$ can also be attached to $-2$ with \#(int.) 1), at \#(int.) $1$--$2$. The $223$ is restricted to a single
$(-1)$ per generator at \#(int.) $1$. (A $23$ also grows into a $232$ or $223$ when a later
$-2$/$\mathfrak{su}_2$ lands on its $-3$ or $-2$.)}
\label{tab:cluster_chain}
\end{table}

The allowed external attachments fall into three types: a single external generator attached to one or several base $-1$ generators (Table~\ref{tab:multi}), an external generator attached to one ``core'' base generator together with extra $-1$ legs (Table~\ref{tab:mixed}, we call this type mixed attachment), and a whole NHC cluster glued in as an external unit (Table~\ref{tab:cluster_chain}).
For each type of external generators, the code enumerates only the intersection multiplicities listed in the table. In particular, we impose a cap of two on single-target $\mathfrak{su}_2$ and $\mathfrak{su}_3$ external sectors attached only to a single $-1$ generator (see the last column in Table~\ref{tab:multi}). This is a deliberate chain-load bound: the C++ implementation generates supergravity candidate chains by increasing the number of external sectors, and this cap substantially reduces the computational cost. As a benchmark, at $T^H=10$, raising this single-target cap from $2$ to $5$ increases the number of distinct blocks by only $2.9$\%, from $8{,}191{,}978$ to $8{,}430{,}707$, while increasing the wall time by $95$\%, approximately a factor of $1.95$. Since this cost grows rapidly at larger $T^H$, we keep the single-target cap fixed at 2. For the same practical reason, we also exclude configurations in which an external $-2$ generator attached to a $23$ cluster is simultaneously attached to $-1$ generators in the LST base (see the last row in Table~\ref{tab:mixed}). Including such configurations nearly doubles the wall time in our benchmark runs. In addition, the code bounds the number of external attachments to any given generator by nine. This count refers to the number of external sectors, not to the number of tensors contained in them; for instance, a $223$ external sector counts as a single attachment. The only cases affected by this bound occur for LST bases with $T_H\leq 9$. We checked the $399$ corresponding bases explicitly, and none gives rise to a effective supergravity block.\footnote{The intuitive reason is as follows. The blocks with large number of external generators contain $\mathfrak{su}_2$ external sectors.
Then, $\mathfrak{su}_2$ external sectors attached only to $-1$ generators would have to be paired with an equal number of $\mathfrak{su}_2$ external sectors attached to $-3$ generators to form the required non-Higgsable cluster, whereas the resulting bases admit no such pairing.}
In all cases the code subsequently applies the E-string central charge bound on each attached $-1$ generator~\eqref{eq:E-string-bound}, the global external-current bounds~\eqref{eq:f-bound}, the Gram-matrix signature test~\eqref{eq:eigenvalue_constraint}, and the blowdown tests to the terminal bases~\eqref{eq:P2-Fn}. Moreover, when the block has too large value of $29T_{\rm min}+ H_{\rm ch} - V$, any gluing cannot lead to a consistent supergravity model due to gravitational anomaly bound~\eqref{eq:GS-conds-non-abelian}, and such blocks are thus discarded.

\begin{figure}[t]
\centering
\includegraphics[width=\textwidth]{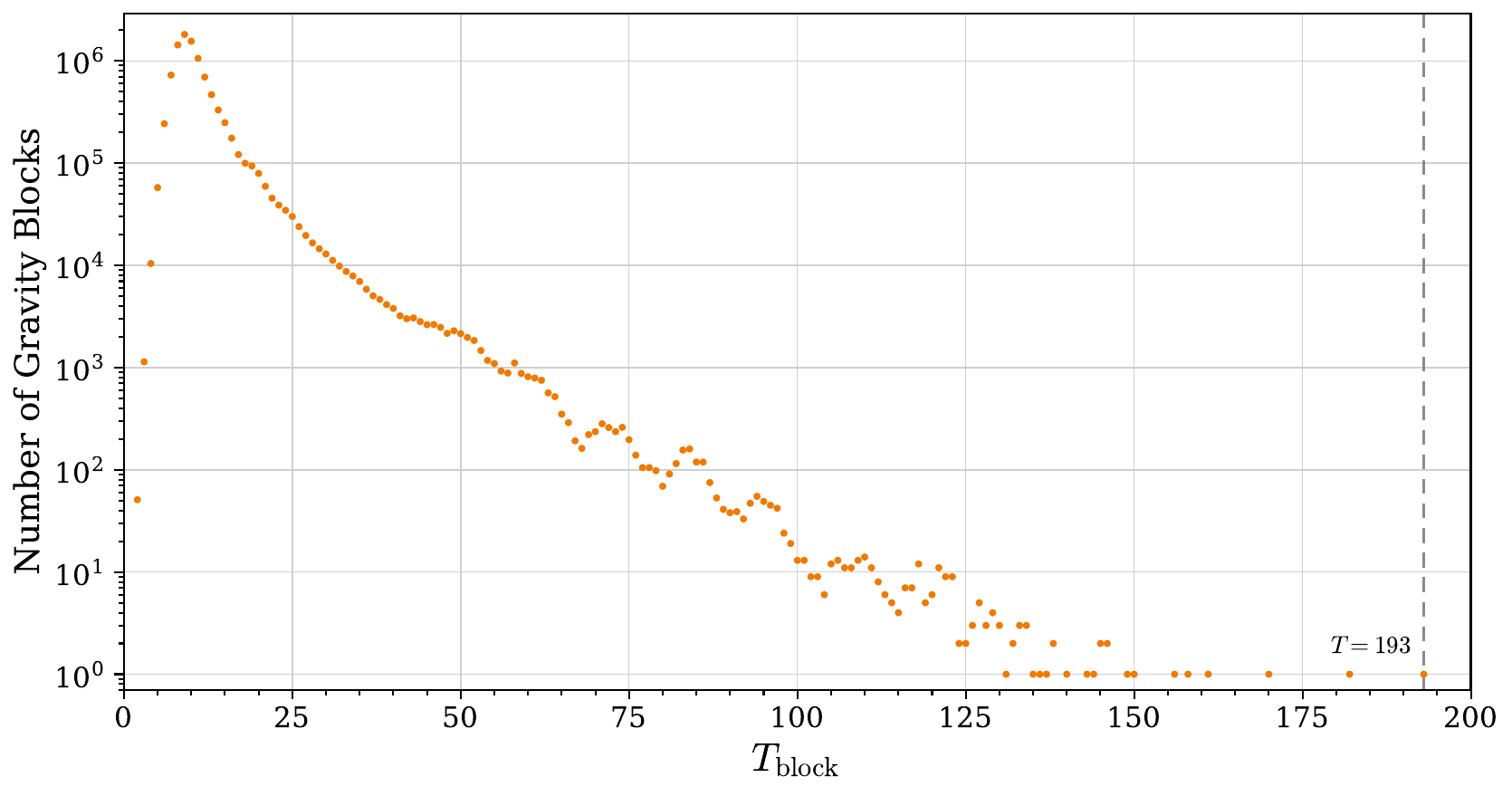}
\caption{The number of non-Higgsable gravity blocks as a function of $T_{\rm block}$, plotted on a logarithmic vertical scale.}
\label{fig:gravity_block_count_byT}
\end{figure}

\begin{figure}[t]
\centering
\includegraphics[width=\textwidth]{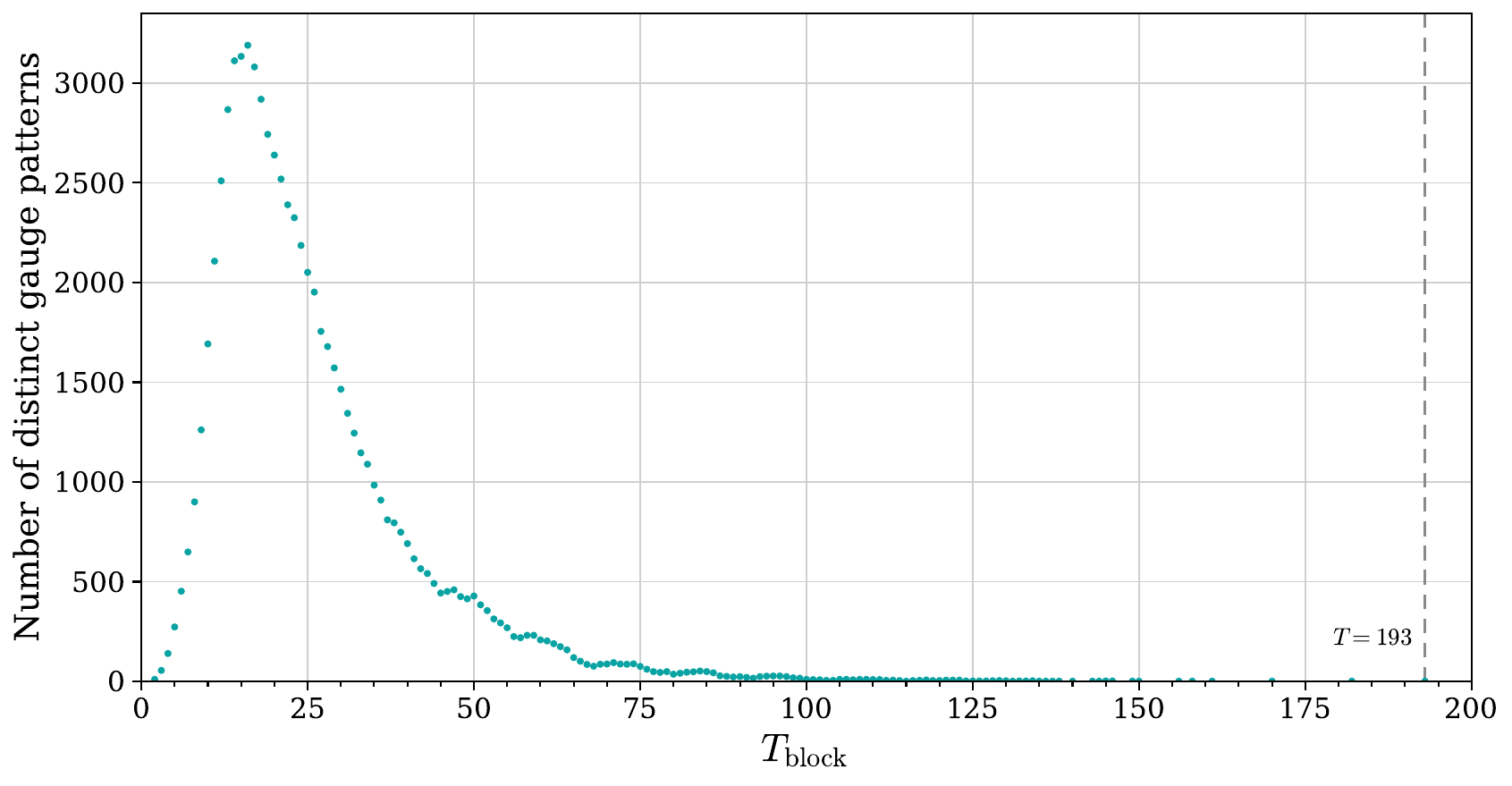}
\caption{Distinct gauge patterns of non-Higgsable gravity blocks. The total number of distinct gauge patterns is 34,747}
\label{fig:Number_distinct_gauge_patterns}
\end{figure}

\begin{table}[t]
\centering
\begin{tabular}{lrrr}
\hline
\textbf{gauge algebra} & \textbf{self-int.} & \textbf{\#blocks} & \textbf{\%} \\
\hline
$\mathfrak{su}_2$ & $-2$  & 9,180,808 & 95.33 \\
$\mathfrak{su}_3$ & $-3$  & 3,979,682 & 41.32 \\
$\mathfrak{so}_8$  & $-4$  &   1,072,370 &  11.14 \\
$\mathfrak{f}_4$ & $-5$  &   409,421 &  4.25 \\
$\mathfrak{g}_2$ & $-3$  & 316,603 & 3.29 \\
$\mathfrak{e}_6$ & $-6$  & 311,800 & 3.24 \\
$\mathfrak{e}_7$  & $-8$  &   180,064 &  1.87 \\
$\mathfrak{e}_7'$  & $-7$  &   144,475 &  1.50 \\
$\mathfrak{e}_8$  & $-12$ &    74,882 &  0.78 \\
$\mathfrak{so}_7$  & $-3$  &   1,821 &  0.02 \\
$\mathfrak{su}_8$  & $-1$ &    1,731 &  0.02 \\
$\mathfrak{su}_8$  & $-2$ &    9 &  0.00 \\
$\mathfrak{su}_{16}$  & $-1$ &    3 &  0.00 \\
$\mathfrak{so}_{16}$  & $-4$ &    3 &  0.00 \\
\hline
\end{tabular}
\caption{Simple Lie-algebra external generators consisting of a single tensor and appearing in the supergravity blocks.
The 3rd column counts the number of blocks containing the indicated external generator.}
\label{tab:simple-lie-algebra-externals}
\end{table}

We summarize several statistical features of the resulting non-Higgsable gravity blocks in Figs.~\ref{fig:gravity_block_count_byT} and \ref{fig:Number_distinct_gauge_patterns}. These figures show how the number of blocks (Fig.~\ref{fig:gravity_block_count_byT}), and the number of distinct external gauge-algebra patterns (Fig.~\ref{fig:Number_distinct_gauge_patterns}) vary with the number $T_{\mathrm{block}}$ of tensor multiplets. Here, $T_{\mathrm{block}}$ is the number of negative eigenvalues of the reduced Gram matrix obtained after excluding the gravitational entries $b_0\cdot C_i$. It should be distinguished from the total tensor-multiplet number $T$, which is determined only after gluing the supergravity blocks into the full intersection form. Together, these figures provide a statistical overview of the supergravity blocks obtained from the classification.

The accompanying tables give a more detailed breakdown of the external gauge-algebra data. Table~\ref{tab:simple-lie-algebra-externals} summarizes the simple Lie-algebra external generators consisting of a single tensor that occur in the catalog.  The percentages are normalized by $N_{\rm tot}=9{,}630{,}606$. The dominant external generator is the $-2$ tensor supporting $\mathfrak{su}_2$, which appears in $9{,}180{,}808$ blocks, or $95.33\%$ of the catalog.  The next most frequent simple external generator is the $-3$ tensor supporting $\mathfrak{su}_3$, which appears in $3{,}979{,}682$ blocks, or $41.32\%$ of the catalog.  Exceptional external generators are much rarer. For example, $\mathfrak{e}_8$ appears only in $74{,}882$ blocks, or $0.78\%$. Although the numerical scan finds three supergravity blocks with a $\hat{1}$ external generator supporting $\mathfrak{su}_{16}$, none of them is an effective supergravity block. Indeed, two are rejected as explained around \eqref{eq:hat1_external}, while the remaining case differs from them only by one additional external generator.

Table~\ref{tab:1-2-externals} shows the external generators supporting no gauge algebra, which are specifically the E-string and M-string, that we considered in our enumeration. For the E-string case, we only count supergravity blocks that contain a single external generator (the E-string itself). When all LSTs are blown down to their endpoints, keeping the external $-1$ tensor intact, the corresponding tensor base always reduces to that of $\mathbb{F}_1$. This fixes the intersection number to $C^{\rm ext}\cdot f=1$. We also considered the cases of a single $-2$ external tensor attached to the $23$ cluster, since the resulting $223$ configuration forms a non-Higgsable cluster and this $-2$ tensor cannot be removed by Weyl reflection.
Table~\ref{tab:nhc-external-units} records the cases in which an entire NHC is glued in as one external unit.  Among these, the $23$ cluster carrying $\mathfrak{su}_2\oplus\mathfrak{g}_2$ is the most frequent.
\begin{table}[t]
\centering
\begin{tabular}{ccrrr}
\hline
$C^{\rm ext}$ & \textbf{self-int.} & \textbf{target} & \textbf{\#blocks} & \textbf{\%} \\
\hline
E-string & $-1$ & $C^{\rm ext}\cdot f=1$ & 34,900 & 0.36 \\
M-string & $-2$ & $23$ cluster in the LST base  & 619,008 & 6.43 \\
\hline
\end{tabular}
\caption{External generators supporting no gauge algebra. The first row corresponds to the non-Higgsable  gravity blocks that contain only a single external $-1$ tensor.
In the 2nd row, the $-2$
external generator is attached to the $23$ cluster in the LST base, yielding the
$223$ enhancement.}
\label{tab:1-2-externals}
\end{table}
\begin{table}[t]
\centering
\begin{tabular}{llrrr}
\hline
\textbf{cluster} & \textbf{cluster enhancement} & \textbf{gauge algebra} & \textbf{\#blocks} & \textbf{\%} \\
\hline
\multirow{3}{*}{$23$}
   & none                  & $\mathfrak{su}_2\oplus\mathfrak{g}_2$                       & 1,592,605 & 16.54 \\
   & $2\to2\textcolor{red}{32}$     & $\mathfrak{su}_2\oplus\mathfrak{so}_7\oplus\mathfrak{su}_2$ &   152,349 &  1.58 \\
   & $2\to2\textcolor{red}{23}$     & $\mathfrak{sp}_1\oplus\mathfrak{g}_2$                       &    56,243 &  0.58 \\
\hline
$223$ & none & $\mathfrak{sp}_1\oplus\mathfrak{g}_2$                       & 454,552 & 4.72 \\
$232$ & none & $\mathfrak{su}_2\oplus\mathfrak{so}_7\oplus\mathfrak{su}_2$ & 433,453 & 4.50 \\
\hline
\end{tabular}
\caption{NHCs glued in as external units. The $23$ cluster can be attached to a standalone $-2$ generator in the LST base, yielding the $223$ enhancement, or the $232$ enhancement.}
\label{tab:nhc-external-units}
\end{table}

Table~\ref{tab:distribution-by-number-of-externals} summarizes the classification results in terms of the number of external generators.
Most blocks contain three, four, or five external generators, which account for $75.3\%$ of the total supergravity blocks. The largest number of external generators observed in this range is twelve, realized by $14$ bases. This result does not conflict with the practical cutoff of nine applied during our enumeration, since that cutoff is imposed on the number of attached external sectors, not on the total number of external generators contained in those sectors. For example, the $23$, $223$, and $232$ sectors are each counted as one attached external sector by the cutoff, although they contain two, three, and three external generators, respectively.
\begin{table}[t]
\centering
\begin{tabular}{rrr}
\hline
\textbf{\#externals} & \textbf{\#blocks} & \textbf{\%} \\
\hline
 1 &   233,252 &  2.4 \\
 2 & 1,055,928 & 11.0 \\
 3 & 2,513,867 & 26.1 \\
 4 & 2,878,780 & 29.9 \\
 5 & 1,854,076 & 19.3 \\
 6 &   774,594 &  8.0 \\
 7 &   241,135 &  2.5 \\
 8 &    61,684 &  0.6 \\
 9 &    14,473 &  0.2 \\
10 &     2,523 & $<0.1$ \\
11 &       280 & $<0.1$ \\
12 &        14 & $<0.1$ \\
\hline
\end{tabular}
\caption{Number of supergravity blocks with $\#$ of external generators.}
\label{tab:distribution-by-number-of-externals}
\end{table}

Finally, Table~\ref{tab:top-external-combinations} lists the most common external-generator combinations.  The leading pattern is $\mathfrak{su}_2 \oplus \mathfrak{su}_2 \oplus\mathfrak{su}_2$, which accounts for $849{,}343$ bases.  This dominance of small $\mathfrak{su}_2$ and $\mathfrak{su}_3$ external generators is expected, since the addition of external generators supporting large gauge algebra easily violates the central charge bound associated to the {\it H}-string.

\begin{table}[t]
\centering
\small
\begin{tabular}{rlrr}
\hline
\textbf{rank} & \textbf{external gauge algebra} & \textbf{\#blocks} & \textbf{\%} \\
\hline
 1 & $\mathfrak{su}_2 \oplus \mathfrak{su}_2 \oplus \mathfrak{su}_2$ & 849,343 & 8.82 \\
 2 & $\mathfrak{su}_2 \oplus \mathfrak{su}_2 \oplus \mathfrak{su}_2 \oplus \mathfrak{su}_2$ & 675,063 & 7.01 \\
 3 & $\mathfrak{su}_2 \oplus \mathfrak{su}_2 \oplus \mathfrak{su}_3$ & 610,263 & 6.34 \\
 4 & $\mathfrak{su}_2 \oplus \mathfrak{su}_2 \oplus \mathfrak{su}_2 \oplus \mathfrak{su}_3$ & 596,891 & 6.20 \\
 5 & $\mathfrak{su}_2 \oplus \mathfrak{su}_2$ & 520,943 & 5.41 \\
 6 & $\mathfrak{g}_2 \oplus \mathfrak{su}_2 \oplus \mathfrak{su}_2 \oplus \mathfrak{su}_2$ & 336,841 & 3.50 \\
 7 & $\mathfrak{su}_2 \oplus \mathfrak{su}_2 \oplus \mathfrak{su}_2 \oplus \mathfrak{su}_2 \oplus \mathfrak{su}_3$ & 266,083 & 2.76 \\
 8 & $\mathfrak{su}_2 \oplus \mathfrak{su}_2 \oplus \mathfrak{su}_2 \oplus \mathfrak{su}_2 \oplus \mathfrak{su}_2$ & 258,748 & 2.69 \\
 9 & $\mathfrak{g}_2 \oplus \mathfrak{su}_2 \oplus \mathfrak{su}_2$ & 244,944 & 2.54 \\
10 & $\mathfrak{su}_2 \oplus \mathfrak{su}_3$ & 242,726 & 2.52 \\
\hline
\end{tabular}
\caption{Ten most frequent external gauge-algebra patterns in NH gravity blocks. The dataset contains 1,293 distinct external gauge-algebra patterns in total.}
\label{tab:top-external-combinations}
\end{table}

The full list of non-Higgsable gravity blocks is available on Zenodo~\cite{Zenodo}. We also provide a python script to read the data and extract the supergravity bases with given properties, such as the number of tensor multiplets, the gauge algebra on the external generator, and so on.  The script is provided as the SUGRA toolkit~\cite{Github:SUGRAtoolkit}.

\subsection{Sketch of the classification up to \texorpdfstring{$T=9$}{T=9}}
For small values of $T\ge1$, the block construction can be implemented explicitly. In this subsection we sketch the resulting classification of supergravity theories built solely from non-Higgsable gravity blocks up to $T=9$. So all models in this section contain only NHCs. We will give the complete lists for $T\le 9$~\footnote{See the data file \texttt{Bases\_upto\_T9.wl} on Zenodo~\cite{Zenodo} for the data.} (up to remarks which we will clarify below) and present only the qualitatively new features for $T=10$, leaving their full enumeration to future work. The classification for larger values of $T$, together with possible gauge enhancements and additional matter content, is likewise left for future work. We expect that the tensor bases arising in these non-Higgsable constructions already provide the complete list of tensor bases for general 6d supergravity theories with $T \le 9$, even after accounting for gauge enhancements. As evidence for this expectation, we compare our results with the geometric base classification of \cite{Taylor:2015isa}, which is available up to $T\le 6$, and find complete agreement (for details, see Appendix~\ref{app:non-toric}).

In the following, we construct the supergravity theories based on the strategy outlined in Section~\ref{sec:construction}. First, we choose the external generators and then glue the non-Higgsable gravity blocks that share these external tensors. When the external generator is an M-string (without gauge algebra) intersecting only $-1$ tensors, we can remove it by Weyl reflections, and thus we do not consider such cases.
As remarked at the end of Section~\ref{sec:general_construction}, in the absence of external gauge algebras, a single external $-1$ tensor is necessary.
Such a supergravity base with single LST sector is collected in our block data.
It turns out that, as long as the local intersection rules are satisfied, all combinations of single external $-1$ blocks are consistent supergravity theories.
Since this class of constructions is numerous, but is easily understood from our data, we restrict our attention to the cases where the external is not a single $-1$ tensor.

When $T=1$, as explained in Section~\ref{sec:refined_structure_of_tensor_moduli_space}, the tensor bases are those of Hirzebruch surfaces $\mathbb{F}_n$, and the corresponding gravity models have a minimal gauge algebra on the external generator $\mathcal{C}$ satisfying $\mathcal{C}^2=-n$ and $\mathcal{C}\cdot b_0=2-n$. Thus we have $T=1$ gravity theories as
\begin{align}
    \textcolor{red}{\overset{\text{any}}{\mathcal{C}}}0,
\end{align}
where $0$ denotes the tensor multiplet for an {\it H}-string.  Throughout this subsection, external generators are marked in red and, unless otherwise stated, they satisfy $\mathcal{C}^2=-n$ and $\mathcal{C}\cdot b_0=2-n$. The superscript on an external generator specifies the gauge algebra it carries.
$\overset{\text{any}}{\mathcal{C}}$ allows any admissible external algebra, whereas $\overset{\leq\mathfrak{g}}{\mathcal{C}}$ means that $\mathcal{C}$ may support the gauge algebra $\mathfrak{g}$ or any smaller external algebra, i.e.\ any non-Higgsable algebra carried by an external curve whose self-intersection is no more negative than the one supporting $\mathfrak{g}$. These external algebras are ordered by the self-intersection $\mathcal{C}^2=-n$ as $\mathfrak{su}_2\,(n{=}2)<\mathfrak{su}_3<\mathfrak{so}_8<\mathfrak{f}_4<\mathfrak{e}_6<\mathfrak{e}_7<\mathfrak{e}_8\,(n{=}12)$, so that, for instance, $\overset{\leq\mathfrak{e}_6}{\mathcal{C}}$ allows $\mathfrak{e}_6$ or any of $\mathfrak{f}_4,\mathfrak{so}_8,\mathfrak{su}_3,\mathfrak{su}_2$. As we will see, every tensor base appearing at higher values of $T$ reduces, after a sequence of blowdowns of $-1$ tensors, to one of these $T=1$ bases.

For $T=2$, the supergravity theories are only
\begin{align}
    \textcolor{red}{\overset{\text{any}}{\mathcal{C}}}11 \ ,
    \label{eq:T2_bases}
\end{align}
where $11$ is an LST formed by two $-1$ tensors with mutual intersection number one. These are constructed by a single non-Higgsable block consisting of the simplest LST of $11$ and an external generator $\mathcal{C}$.

For $T=3$,  the allowed bases are
\begin{align}
    &11\textcolor{red}{\overset{\text{any}}{\mathcal{C}}}11,
    &&\textcolor{red}{\overset{\text{any}}{\mathcal{C}}}121,
    &&\textcolor{red}{\mathfrak{g}_2}212,
    &&1\overset{\textcolor{red}{\mathfrak{g}_2}}{2}1 \ ,
    \label{eq:T3_bases}
\end{align}
where, in the last two cases, the $-2$ tensor adjacent to the external generator supports the $\mathfrak{su}_2$ gauge algebra. 
The last three cases consist of single non-Higgsable blocks, whereas the first case is formed by gluing two non-Higgsable blocks of the same type along their common external generator $\textcolor{red}{\overset{\text{any}}{\mathcal{C}}}$ with self-intersection $-n$ as follows:
\begin{align}
    \left(\begin{array}{ccc}-n & 1 & 0\\1&-1&1\\0&1&-1\end{array}\right) \oplus \left(\begin{array}{ccc}-n & 1 & 0\\1&-1&1\\0&1&-1\end{array}\right) \quad \rightarrow \quad \Omega= \left(\begin{array}{ccccc}-n & 1 & 0 & 1 & 0\\1&-1&1&0&0\\0&1&-1&0&0 \\ 1 &0&0&-1&1 \\ 0&0&0&1&-1\end{array}\right) \ .
\end{align}

For $T=4$, the allowed supergravity bases constructed from the single LST are
\begin{align}
    &\textcolor{red}{\overset{\text{any}}{\mathcal{C}}}1221,
    &&\textcolor{red}{\overset{\leq\mathfrak{e}_{6}}{\mathcal{C}}}1\overset{1}{3}1,
    &&1\underset{\textcolor{red}{\mathfrak{su}_{2}}}{\overset{1}{3}}1,
    &&\textcolor{red}{\mathfrak{g}_2}\, \overset{1}{2}21,
    \nonumber\\
    &2131\textcolor{red}{\overset{\leq\mathfrak{e}_{6}}{\mathcal{C}}},
    &&\textcolor{red}{\mathfrak{g}_2}2131,
    &&21\overset{\textcolor{red}{\mathfrak{su}_{2}}}{3}1,
    &2213\textcolor{red}{\mathfrak{su}_{2}},
    &&\textcolor{red}{\mathfrak{g}_2}2213.
    \label{eq:T4_bases}
\end{align}
In addition, there are supergravity bases formed by gluing two or three non-Higgsable blocks along their common external generators. We denote these bases by
\begin{align}
    &(T=3)\textcolor{red}{\mathcal{C}}11,
    \label{eq:T4_bases_gluing}
\end{align}
where $(T=3)$ represents any one of the $T=3$ supergravity bases listed above~\eqref{eq:T3_bases}, and the external generator $\mathcal{C}$ is shared by the two/three blocks. 

For $T=5$, the allowed supergravity bases constructed from the single LST are
\begin{align}
    &\textcolor{red}{\overset{\text{any}}{\mathcal{C}}}12221,
    &&\textcolor{red}{\overset{\leq\mathfrak{so}_{8}}{\mathcal{C}}}{1 \overset{1}{\underset{1}{4}} 1},
    &&\textcolor{red}{\overset{\leq\mathfrak{e}_{7}}{\mathcal{C}}}{1 2 \overset{1}{3} 1},
    &&{1 2 \overset{1}{3} 1}\textcolor{red}{\overset{\leq\mathfrak{f}_{4}}{\mathcal{C}}},
    &&{1 2 \underset{\textcolor{red}{\mathfrak{su}_2}}{\overset{1}{3}} 1},
    &&{1 \overset{\textcolor{red}{2}}{2} \overset{1}{3} 1},
    \nonumber\\
    &2 1 \overset{1}{4} 1 \textcolor{red}{\overset{\leq\mathfrak{so}_{8}}{\mathcal{C}}},
    &&\textcolor{red}{\mathfrak{g}_{2}} {2 1 \overset{1}{4} 1},
    &&\textcolor{red}{\mathfrak{su}_{2}}{3 1 \overset{1}{3} 2},
    &&{3 1 \overset{1}{3} 2 \textcolor{red}{2}},
    &&\textcolor{red}{\overset{\leq\mathfrak{e}_{7}}{\mathcal{C}}}12312,
    &&12\overset{\textcolor{red}{\mathfrak{su}_{2}}}{3}12,
    &&1\overset{\textcolor{red}{2}}{2}312
    \nonumber\\
    &12312\textcolor{red}{\mathfrak{g}_2},
    &&\textcolor{red}{\overset{\leq\mathfrak{e}_{6}}{\mathcal{C}}}13131,
    &&1\overset{\textcolor{red}{\mathfrak{su}_{2}}}{3}131,
    &&\textcolor{red}{\mathfrak{g}_{2}}21412,
    &&22141\textcolor{red}{\overset{\leq\mathfrak{so}_{8}}{\mathcal{C}}},
    &&\textcolor{red}{\mathfrak{g}_2}22141,
    \nonumber\\
    &23123\textcolor{red}{\mathfrak{su}_2},
    &&\textcolor{red}{2}23123,
    &&31231\textcolor{red}{\overset{\leq\mathfrak{f}_{4}}{\mathcal{C}}},
    &&312\overset{\textcolor{red}{\mathfrak{su}_2}}{3}1,
    &&\textcolor{red}{\mathfrak{su}_2}31231,
    &&\textcolor{red}{\mathfrak{su}_2}31313,
    &&2\overset{1}{2}31\textcolor{red}{\overset{\leq\mathfrak{f}_{4}}{\mathcal{C}}}.
    \label{eq:T5_bases}
\end{align}
There are also supergravity bases formed by gluing multiple non-Higgsable blocks, which are given by
\begin{align}
    &(T=4)\textcolor{red}{\mathcal{C}}11,
    &&(T=3)\textcolor{red}{\mathcal{C}}(T=3).
\end{align}
Here, the first base is formed by gluing a $T=4$ block and $\textcolor{red}{\mathcal{C}}11$ block, and by $(T=4)$, we mean not only any bases from \eqref{eq:T4_bases} but also \eqref{eq:T4_bases_gluing}.
Note that the second base is an allowed supergravity base since, for all the $T=3$ blocks, an external intersects with the {\it H}-string at one point, and the resulting Gram matrix has the correct signature.

Similarly, for $T=6$, the allowed supergravity bases constructed from a single supergravity block are
\begin{align}
    &\textcolor{red}{\overset{\text{any}}{\mathcal{C}}}122221,
    &&\textcolor{red}{\overset{\leq\mathfrak{su}_3}{\mathcal{C}}}1^{\otimes 5}5
    &&\textcolor{red}{\overset{\leq\mathfrak{su}_3}{\mathcal{C}}}{1 \overset{1}{\underset{1}{5}} 1 2},
    &&{1 \overset{1}{\underset{1}{5}} 1 2}\textcolor{red}{\mathfrak{g}_2},
    &&\textcolor{red}{\overset{\text{any}}{\mathcal{C}}}{1 2 2 \overset{1}{3} 1},
    &&{1 2 2 \overset{1}{3} 1}\textcolor{red}{\overset{\leq\mathfrak{f}_{4}}{\mathcal{C}}},
    &&\textcolor{red}{\overset{\leq\mathfrak{e}_{7}}{\mathcal{C}}}{1 2 \overset{1}{3} 2 1},
    \nonumber\\
    &{1 2 \overset{\overset{\textcolor{red}{\overset{\leq\mathfrak{so}_{8}}{\mathcal{C}}}}{1}}{3} 2 1},
    &&\textcolor{red}{\overset{\leq\mathfrak{e}_{6}}{\mathcal{C}}}{1 3 1 \overset{1}{4} 1},
    &&{1 3 1 \overset{1}{4} 1}\textcolor{red}{\overset{\leq\mathfrak{so}_{8}}{\mathcal{C}}},
    &&{1 3 1 \overset{\overset{\textcolor{red}{\overset{\leq\mathfrak{so}_{8}}{\mathcal{C}}}}{1}}{4} 1},
    &&{1 \overset{\textcolor{red}{\mathfrak{su}_2}}{3} 1 \overset{1}{4} 1},
    &&\textcolor{red}{\mathfrak{g}_2}{2 1 \overset{1}{5} 1 2},
    &&{2 1 \overset{\overset{\textcolor{red}{\overset{\leq\mathfrak{su}_3}{\mathcal{C}}}}{1}}{5} 1 2},
    \nonumber\\
    &{2 \overset{1}{3} 1 4 1}\textcolor{red}{\overset{\leq\mathfrak{so}_{8}}{\mathcal{C}}},
    &&\textcolor{red}{2}{2 \overset{1}{3} 1 4 1},
    &&\textcolor{red}{\mathfrak{su}_{2}}{3 1 \overset{1}{4} 1 3},
    &&\textcolor{red}{\mathfrak{su}_{2}}{3 1 \overset{2}{3} 2 1},
    &&\textcolor{red}{\overset{\text{any}}{\mathcal{C}}}{1 2 2 3 1 2},
    &&{1 2 2 3 1 2}\textcolor{red}{\mathfrak{g}_2},
    &&\textcolor{red}{\overset{\leq\mathfrak{e}_{6}}{\mathcal{C}}}{1 3 1 3 2 1},
    \nonumber\\
    &{1 3 1 3 2 1} \textcolor{red}{\overset{\leq\mathfrak{e}_{7}}{\mathcal{C}}},
    &&{1 \overset{\textcolor{red}{\mathfrak{su}_2}}{3} 1 3 2 1},
    &&{1 3 1 \overset{\textcolor{red}{\mathfrak{su}_2}}{3} 2 1},
    &&{1 3 1 3 \overset{\textcolor{red}{2}}{2} 1},
    &&\textcolor{red}{\overset{\leq\mathfrak{e}_{6}}{\mathcal{C}}}{1 3 1 4 1 2},
    &&{1 3 1 4 1 2}\textcolor{red}{\mathfrak{g}_2},
    &&{1 \overset{\textcolor{red}{\mathfrak{su}_2}}{3} 1 4 1 2},
    \nonumber\\
    &\textcolor{red}{\overset{\leq\mathfrak{f}_{4}}{\mathcal{C}}}{1 3 2 1 4 1},
    &&{1 3 2 1 4 1}\textcolor{red}{\overset{\leq\mathfrak{so}_{8}}{\mathcal{C}}},
    &&{1 \overset{\textcolor{red}{\mathfrak{su}_2}}{3} 2 1 4 1},
    &&\textcolor{red}{\overset{\leq\mathfrak{f}_{4}}{\mathcal{C}}}{1 3 2 2 1 4},
    &&\textcolor{red}{\mathfrak{g}_2}{2 2 1 5 1 2},
    &&{2 2 1 5 1 2}\textcolor{red}{\mathfrak{g}_2},
    &&{2 2 2 1 5 1}\textcolor{red}{\overset{\leq\mathfrak{su}_3}{\mathcal{C}}},
    \nonumber\\
    &\textcolor{red}{\mathfrak{su}_2}{3 1 2 3 2 1},
    &&{3 1 2 3 2 1}\textcolor{red}{\overset{\leq\mathfrak{e}_{7}}{\mathcal{C}}},
    &&\textcolor{red}{\mathfrak{su}_2}{3 1 3 1 4 1},
    &&{3 1 3 1 4 1}\textcolor{red}{\overset{\leq\mathfrak{so}_{8}}{\mathcal{C}}},
    &&\textcolor{red}{\mathfrak{su}_2}{3 1 3 2 1 4},
    &&\textcolor{red}{\mathfrak{su}_2}{3 2 1 4 1 3},
    &&{3 2 1 4 1 3}\textcolor{red}{\mathfrak{su}_2},
    \nonumber\\
    &\textcolor{red}{2}231314.
\end{align}
The supergravity bases formed by gluing multiple non-Higgsable blocks are given by
\begin{align}
    &(T=5)\textcolor{red}{\mathcal{C}}11,
    &&(T=4)\textcolor{red}{\mathcal{C}}(T=3).
\end{align}
In the second base, we remark that not all choices of the $T=3$~\eqref{eq:T3_bases} and $T=4$~\eqref{eq:T4_bases} bases are allowed, but only those whose local intersection pattern is an allowed one. For instance, the following choices are not allowed:
\begin{align}
    \text{Not allowed:}\quad 3122\textcolor{red}{\mathfrak{g}_2}212,
\end{align}
since the apperance of the pattern $2232$ is not locally allowed. On the other hand, as long as the local intersection pattern is allowed, the resulting supergravity base is automatically allowed. This is again because, for all the bases in \eqref{eq:T3_bases} and \eqref{eq:T4_bases}, an external generator intersects with the {\it H}-string at one point.

For $T=7$, the allowed supergravity bases constructed from a single non-Higgsable block are
\begingroup\allowdisplaybreaks
\begin{align}
    &\textcolor{red}{\overset{\text{any}}{\mathcal{C}}}1222221,
    &&{\textcolor{red}{\overset{\leq\mathfrak{so}_8}{\mathcal{C}}} 1 \overset{1}{4} 1 \overset{1}{4} 1}, 
    &&{\textcolor{red}{\overset{\leq\mathfrak{su}_3}{\mathcal{C}}} 1 \overset{1}{5} 1 \overset{1}{3} 2},
    &&{ 1 \overset{1}{5} 1 \overset{1}{3} 2}\textcolor{red}{2},
    &&{\textcolor{red}{\overset{\leq\mathfrak{su}_3}{\mathcal{C}}} 1 \overset{1}{\underset{1}{5}} 1 3 1},
    &&{1 \overset{1}{\underset{1}{5}} 1 3 1 \textcolor{red}{\overset{\leq\mathfrak{e}_6}{\mathcal{C}}}}, 
    \nonumber\\
    &{1 \overset{\textcolor{red}{\overset{\leq\mathfrak{su}_3}{\mathcal{C}}}}{\overset{1}{\underset{1}{5}}} 1 3 1},
    &&{1 \overset{1}{\underset{1}{5}} 1 \overset{\textcolor{red}{\mathfrak{su}_2}}{3} 1},
    &&{\textcolor{red}{\mathfrak{g}_2} 2 1 \underset{1}{\overset{1}{6}} 1 2},
    &&{2 1 \underset{1}{\overset{\textcolor{red}{\overset{\leq\mathfrak{su}_3}{\mathcal{C}}}}{\overset{1}{6}}} 1 2}, 
    &&{\textcolor{red}{\mathfrak{g}_2} 2 1 \overset{2}{\overset{1}{6}} 1 2},
    &&{\textcolor{red}{\mathfrak{g}_2} 2 2 1 \underset{1}{\overset{1}{6}} 1},
    \nonumber\\
    &{2 2 1 \underset{1}{\overset{1}{6}} 1 \textcolor{red}{\overset{\leq\mathfrak{su}_3}{\mathcal{C}}}}, 
    &&{\textcolor{red}{\mathfrak{su}_2} 3 1 \underset{1}{\overset{1}{5}} 1 3},
    &&{\textcolor{red}{\mathfrak{su}_2} 3 1 \overset{2}{\overset{1}{5}} 1 3},  
    &&{\textcolor{red}{\overset{\leq\mathfrak{e}_7}{\mathcal{C}}} 1 2 3 1 \overset{1}{4} 1}, 
    &&{1 2 3 1 \overset{1}{4} 1 \textcolor{red}{\overset{\leq\mathfrak{so}_8}{\mathcal{C}}}}, 
    &&{1 2 \overset{\textcolor{red}{\mathfrak{su}_2}}{3} 1 \overset{1}{4} 1},
    \nonumber\\
    &{1 \overset{\textcolor{red}{2}}{2} 3 1 \overset{1}{4} 1},
    &&{1 2 \overset{2}{3} 1 4 1 \textcolor{red}{\overset{\leq\mathfrak{so}_8}{\mathcal{C}}}}, 
    &&{\textcolor{red}{\overset{\leq\mathfrak{su}_3}{\mathcal{C}}} 1 5 1 2 \overset{1}{3} 2},
    &&{\textcolor{red}{\overset{\leq\mathfrak{su}_3}{\mathcal{C}}} 1 5 1 \overset{1}{3} 2 2},
    &&{\textcolor{red}{\overset{\leq\mathfrak{so}_8}{\mathcal{C}}} 1 \overset{1}{4} 1 4 1 2}, 
    &&{1 \overset{1}{4} 1 4 1 2 \textcolor{red}{\mathfrak{g}_2}}, 
    \nonumber\\
    &{\textcolor{red}{\overset{\leq\mathfrak{su}_3}{\mathcal{C}}} 1 \overset{1}{5} 1 2 3 1},
    &&{1 \overset{1}{5} 1 2 3 1 \textcolor{red}{\overset{\leq\mathfrak{f}_4}{\mathcal{C}}}}, 
    &&{1 \overset{1}{5} 1 2 \overset{\textcolor{red}{\mathfrak{su}_2}}{3} 1}, 
    &&{\textcolor{red}{\overset{\leq\mathfrak{su}_3}{\mathcal{C}}} 1 \overset{1}{5} 1 3 1 3},
    &&{1 \overset{1}{5} 1 3 1 3 \textcolor{red}{\mathfrak{su}_2}}, 
    &&{\textcolor{red}{\mathfrak{g}_2} 2 1 5 1 \overset{1}{3} 2},
    \nonumber\\
    &{2 1 5 1 \overset{1}{3} 2 \textcolor{red}{2}},
    &&{\textcolor{red}{\mathfrak{g}_2} 2 1 \overset{1}{5} 1 3 1},
    &&{2 1 \overset{1}{5} 1 3 1 \textcolor{red}{\overset{\leq\mathfrak{e}_6}{\mathcal{C}}}}, 
    &&{2 1 \overset{\textcolor{red}{\overset{\leq\mathfrak{su}_3}{\mathcal{C}}}}{\overset{1}{5}} 1 3 1}, 
    &&{2 1 \overset{1}{5} 1 \overset{\textcolor{red}{\mathfrak{su}_2}}{3} 1}, 
    &&{\textcolor{red}{\mathfrak{g}_2} 2 2 1 \overset{1}{6} 1 2},
    \nonumber\\
    &{2 2 1 \overset{1}{6} 1 2 \textcolor{red}{\mathfrak{g}_2}}, 
    &&{2 2 1 \overset{\textcolor{red}{\overset{\leq\mathfrak{su}_3}{\mathcal{C}}}}{\overset{1}{6}} 1 2}, 
    &&{2 2 2 1 \overset{1}{6} 1 \textcolor{red}{\overset{\leq\mathfrak{su}_3}{\mathcal{C}}}}, 
    &&{\textcolor{red}{\mathfrak{su}_2} 3 1 \overset{1}{4} 1 4 1},
    &&{3 1 \overset{1}{4} 1 4 1 \textcolor{red}{\overset{\leq\mathfrak{so}_8}{\mathcal{C}}}}, 
    &&{\textcolor{red}{\mathfrak{su}_2} 3 1 \overset{1}{5} 1 2 3},
    \nonumber\\
    &{3 1 \overset{1}{5} 1 2 3 \textcolor{red}{\mathfrak{su}_2}}, 
    &&{\textcolor{red}{\overset{\leq\mathfrak{e}_7}{\mathcal{C}}} 1 2 3 1 3 2 1}, 
    &&{1 2 \overset{\textcolor{red}{\mathfrak{su}_2}}{3} 1 3 2 1},
    &&{1 \overset{\textcolor{red}{2}}{2} 3 1 3 2 1}, 
    &&{\textcolor{red}{\overset{\leq\mathfrak{e}_7}{\mathcal{C}}} 1 2 3 1 4 1 2}, 
    &&{1 2 3 1 4 1 2 \textcolor{red}{\mathfrak{g}_2}}, 
    \nonumber\\
    &{1 2 \overset{\textcolor{red}{\mathfrak{su}_2}}{3} 1 4 1 2}, 
    &&{\textcolor{red}{\overset{\leq\mathfrak{e}_7}{\mathcal{C}}} 1 2 3 2 1 4 1}, 
    &&{1 2 3 2 1 4 1 \textcolor{red}{\overset{\leq\mathfrak{so}_8}{\mathcal{C}}}}, 
    &&{\textcolor{red}{\overset{\leq\mathfrak{e}_6}{\mathcal{C}}} 1 3 1 3 2 2 1}, 
    &&{1 3 1 3 2 2 1 \textcolor{red}{\overset{\leq\mathfrak{e}_8}{\mathcal{C}}}}, 
    &&{1 \overset{\textcolor{red}{\mathfrak{su}_2}}{3} 1 3 2 2 1}, 
    \nonumber\\
    &{\textcolor{red}{\overset{\leq\mathfrak{e}_6}{\mathcal{C}}} 1 3 1 4 1 3 1}, 
    &&{1 \overset{\textcolor{red}{\mathfrak{su}_2}}{3} 1 4 1 3 1}, 
    &&{\textcolor{red}{\overset{\leq\mathfrak{so}_8}{\mathcal{C}}} 1 4 1 3 1 4 1}, 
    &&{\textcolor{red}{\overset{\leq\mathfrak{so}_8}{\mathcal{C}}} 1 4 1 4 1 2 3}, 
    &&{1 4 1 4 1 2 3 \textcolor{red}{\mathfrak{su}_2}}, 
    &&{\textcolor{red}{\overset{\leq\mathfrak{su}_3}{\mathcal{C}}} 1 5 1 2 2 3 1},
    \nonumber\\
    &{1 5 1 2 2 3 1 \textcolor{red}{\overset{\leq\mathfrak{f}_4}{\mathcal{C}}}}, 
    &&{\textcolor{red}{\overset{\leq\mathfrak{su}_3}{\mathcal{C}}} 1 5 1 2 3 1 3},
    &&{1 5 1 2 3 1 3 \textcolor{red}{\mathfrak{su}_2}}, 
    &&{\textcolor{red}{\mathfrak{g}_2} 2 1 4 1 4 1 2}, 
    &&{\textcolor{red}{\mathfrak{g}_2} 2 1 5 1 2 3 1},
    &&{2 1 5 1 2 3 1 \textcolor{red}{\overset{\leq\mathfrak{f}_4}{\mathcal{C}}}}, 
    \nonumber\\
    &{2 1 5 1 2 \overset{\textcolor{red}{\mathfrak{su}_2}}{3} 1}, 
    &&{\textcolor{red}{\mathfrak{g}_2} 2 1 5 1 3 1 3},
    &&{2 1 5 1 3 1 3 \textcolor{red}{\mathfrak{su}_2}}, 
    &&{\textcolor{red}{\mathfrak{g}_2} 2 2 1 5 1 3 1},
    &&{2 2 1 5 1 3 1 \textcolor{red}{\overset{\leq\mathfrak{e}_6}{\mathcal{C}}}}, 
    &&{2 2 1 5 1 \overset{\textcolor{red}{\mathfrak{su}_2}}{3} 1}, 
    \nonumber\\
    &{\textcolor{red}{\mathfrak{g}_2} 2 2 1 6 1 2 2}, 
    &&{2 2 2 1 6 1 2 \textcolor{red}{\mathfrak{g}_2}}, 
    &&{2 2 2 2 1 6 1 \textcolor{red}{\overset{\leq\mathfrak{su}_3}{\mathcal{C}}}}, 
    &&{2 3 1 3 1 5 1 \textcolor{red}{\overset{\leq\mathfrak{su}_3}{\mathcal{C}}}}, 
    &&{\textcolor{red}{\mathfrak{su}_2} 3 2 1 5 1 2 3}, 
    &&{3 2 2 1 5 1 3 \textcolor{red}{\mathfrak{su}_2}}, 
    \nonumber\\
    &{4 1 2 3 1 4 1 \textcolor{red}{\overset{\leq\mathfrak{so}_8}{\mathcal{C}}}}, 
    &&{4 1 3 1 4 1 3 \textcolor{red}{\mathfrak{su}_2}}, 
    &&{5 1 2 2 3 1 3 \textcolor{red}{\mathfrak{su}_2}}, 
    &&{5 1 2 3 1 3 2 \textcolor{red}{2}},
    &&{1 5 1 3 1 3 2 \textcolor{red}{2}},
    &&{1 \overset{\textcolor{red}{2}}{2} 3 1 4 1 2},
    \nonumber\\
    &{\textcolor{red}{2} 2 3 1 \overset{1}{4} 1 4},
\end{align}
The supergravity bases consisting of multiple non-Higgsable blocks are given by
\begin{align}
    &(T=6)\textcolor{red}{\mathcal{C}}11,
    &&(T=5)\textcolor{red}{\mathcal{C}}(T=3),
    &&(T=4)\textcolor{red}{\mathcal{C}}(T=4).
    \label{eq:T7_blocks_gluing}
\end{align}
All the above bases are consistent supergravity bases as long as the local intersection pattern is an allowed one.
Starting from $T=7$, we have consistent supergravity bases with two external generators, which are given by
\endgroup
\begin{align}
    &{\textcolor{red}{\mathfrak{su}_3} 1 2 2 2 2 2 1 \textcolor{red}{\mathfrak{su}_3}},
    &&\text{L}: \textcolor{red}{\mathfrak{su}_{3}}1\underbrace{2\cdots 2}_{n_1}1\textcolor{red}{\mathfrak{su}_{3}}1\underbrace{2\cdots 2}_{n_2}1\quad(n_{1,2}\ge0, n_1+n_2=3).
\end{align}
Here L denotes a loop structure, meaning that the leftmost and rightmost curves intersect at one point.
The bases with more than two LST blocks (up to 6 blocks) sharing the two external generators are also allowed. 

We can carry out an analogous enumeration for higher values of $T=8,9,10$.
For brevity, we refer the reader to Zenodo for the list of $T=8,9,10$ supergravity bases constructed from a single LST and a single external curve.\footnote{For supergravity bases with a single LST and a single external curve, it is easy to extract from the data of supergravity blocks.}
Here, we present the qualitatively new features for $T\geq8$.
One of the new features starting at $T=8$ is that a single external generator can intersect the {\it H}-string at two points. The following are all the allowed supergravity bases in this class for $T=8,9,10$:
\begin{align}
    &T=8: {2 \overset{2}{2} 2 2 2 2 1 \textcolor{red}{\mathfrak{su}_3}},
    &&T=9: {2 \overset{2}{2} 2 2 2 2 2 1 \textcolor{red}{\mathfrak{so}_8}},
    \nonumber\\
    &T=10: {2 \overset{2}{2} 2 2 2 2 2 1 \textcolor{red}{\mathfrak{f}_4}},
    &&{\textcolor{red}{\mathfrak{su}_3} 1 2 2 3 1 4 1 \overset{2}{3} 2}.
\end{align}
Furthermore, for $T=10$, we have a supergravity base where the single external tensor intersects with the {\it H}-string at four points, which is given by 
\begin{align}
    &T=10: {2 \overset{2}{2} 2 2 2 2 2 1 = \textcolor{red}{\mathfrak{so}_8}}.
\end{align}
Note that this is consistent with the restriction we have observed in \eqref{eq:external_intersection_patterns}.

As in \eqref{eq:T7_blocks_gluing}, we can construct the $T=8,9,10$ supergravity bases by gluing multiple supergravity bases that share a single external generator.
For $T_0=8,9,10$, we obtain
\begin{align}
    &(T=T_0-1)\textcolor{red}{\mathcal{C}}11,
    &&(T=T_0-2)\textcolor{red}{\mathcal{C}}(T=3),
    &&\cdots,
    &&(T=\lceil T_0/2 \rceil)\textcolor{red}{\mathcal{C}}(T=\lfloor T_0/2 \rfloor).
    \label{eq:T8_9_10_bases_gluing}
\end{align}
However, for $T\geq8$, this construction does not exhaust all consistent supergravity bases in this class. For $T=8$, adding the following bases to the above list gives all the consistent supergravity bases with a single external curve shared by multiple LSTs:
\begin{align}
    &\text{Gluing through common }\textcolor{red}{\mathfrak{su}_{3}}
    \quad
    &\left(\text{L:}1\underbrace{2\cdots2}_{n_1}1\textcolor{red}{\mathfrak{su}_{3}},\,
    2\underbrace{\overset{2}{2}\cdots2}_{n_2}1\textcolor{red}{\mathfrak{su}_{3}}\right), 
    \label{eq:T8_bases_gluing_extra}
\end{align}
where the integers $n_i$ are nonnegative.
The consistent supergravity bases are obtained by taking multiple copies of the blocks~\eqref{eq:T8_bases_gluing_extra}, with multiplicities chosen in such a way that $T=8$ is realized, following the formula~\eqref{eq:number-of-tensors}.
This construction is not restricted to two LST sectors; it can produce supergravity bases with up to seven LST sectors.
We emphasize that this is not the form of \eqref{eq:T8_9_10_bases_gluing}.
For instance, in the case of two LST sectors, choosing $n_1=0$ and $n_2=4$, neither L:$11\textcolor{red}{\mathfrak{su}_{3}}$ nor $2\overset{2}{2}2221\textcolor{red}{\mathfrak{su}_{3}}$ is a consistent supergravity base for $T=2$ and $T=7$, respectively. Nevertheless, the gluing produces a consistent supergravity base for $T=8$. This is another feature that first appears at $T=8$.  

The $T=8$ supergravity bases with two external curves and multiple LSTs are given by
\begin{align}
    &\text{Gluing through common }(\textcolor{red}{\mathfrak{su}_{3}},\textcolor{red}{\mathfrak{so}_{8}})
    \quad
    &\textcolor{red}{\mathfrak{su}_{3}}12\cdots21\textcolor{red}{\mathfrak{so}_{8}}, 
    \nonumber\\
    &\text{Gluing through common }(\textcolor{red}{\mathfrak{su}_{2}},\textcolor{red}{\mathfrak{g}_{2}})
    \quad
    &\text{L:}\,\textcolor{red}{\mathfrak{su}_{2}}12\cdots21\textcolor{red}{\mathfrak{g}_{2}}. 
\end{align}
In both cases, consistent supergravity bases are obtained by taking multiple copies of the displayed blocks, with multiplicities chosen so that $T=8$ is realized.
One can similarly glue through a common $(\textcolor{red}{\mathfrak{su}_2},\textcolor{red}{\mathfrak{su}_2})$ pair of external curves:
\begin{align}
    &221\overset{\textcolor{red}{\mathfrak{su}_2}}{3}\textcolor{red}{2},
    &&
    21\underset{\textcolor{red}{\mathfrak{su}_2}}{\overset{\textcolor{red}{\mathfrak{su}_2}}{3}}1,
    &&
    1^{\otimes 3}
      \overset{\textcolor{red}{\mathfrak{su}_2}}{3}\textcolor{red}{\mathfrak{su}_2},
    &&\textcolor{red}{\mathfrak{su}_2}312\overset{\textcolor{red}{\mathfrak{su}_2}}{3}1,
    &&
    1\overset{\textcolor{red}{\mathfrak{su}_2}}{3}1
      \overset{\textcolor{red}{\mathfrak{su}_2}}{3}1,
    &&
    \textcolor{red}{\mathfrak{su}_2}31313\textcolor{red}{\mathfrak{su}_2},
    \label{eq:T8_bases_gluing_2}
\end{align}
where $T=8$ is realized by adding A-type dummy LSTs of the form
\begin{align}
    &\textcolor{red}{\mathfrak{su}_2}12\cdots21\textcolor{red}{\mathfrak{su}_2} \ ,
    \label{eq:T8_bases_gluing_2_extra}
\end{align}
to the above bases, following the formula~\eqref{eq:number-of-tensors}. Here two external generators are attached to the leftmost and rightmost $-1$ generators of the dummy LSTs.

Likewise, one can glue through a common $(\textcolor{red}{\mathfrak{su}_2},\textcolor{red}{3})$ pair of external curves among the following bases:
\begin{align}
    &21\overset{\textcolor{red}{\mathfrak{su}_2}}{3}1\textcolor{red}{3},
    &&
    \textcolor{red}{3}21\overset{\textcolor{red}{\mathfrak{su}_2}}{3}1,
    &&
    \textcolor{red}{3}2213\textcolor{red}{\mathfrak{su}_2},
    &&
    1\underset{1}{\overset{\textcolor{red}{\mathfrak{su}_2}}{3}}1\textcolor{red}{3},
    &&21\overset{\textcolor{red}{\mathfrak{su}_2}}{3}21\textcolor{red}{3},
    &&
    1\underset{\textcolor{red}{\mathfrak{su}_2}}{\overset{1}{3}}21\textcolor{red}{3},
    \notag\\[1ex]
    &\textcolor{red}{3}13213\textcolor{red}{\mathfrak{su}_2},
    &&
    1\overset{\textcolor{red}{\mathfrak{su}_2}}{3}131\textcolor{red}{3},
    &&\textcolor{red}{\mathfrak{su}_2}312321\textcolor{red}{3},
    &&
    1\overset{\textcolor{red}{\mathfrak{su}_2}}{3}1321\textcolor{red}{3},
    &&1\overset{\textcolor{red}{\mathfrak{su}_2}}{3}13221\textcolor{red}{3}.
    \label{eq:two_three_gluing}
\end{align}
Again, one block is chosen from \eqref{eq:two_three_gluing} and any number of A-type dummy LSTs are added (external generators attach to the leftmost and rightmost $-1$ generators) to realize $T=8$.

We also have supergravity bases with two external generators including $-2$ external generators attached to a $23$ cluster.
The list of supergravity bases in this class is given by
\begin{align}
    &\textcolor{red}{\mathfrak{su}_3}{1 2 2 3 1 3 \overset{\textcolor{red}{2}}{2} 1},
    &&\textcolor{red}{\mathfrak{su}_3}{1 2 3 1 3 \overset{\textcolor{red}{2}}{2} 1},
    &&{1 \overset{\textcolor{red}{2}}{2} 3 1 3 \overset{\textcolor{red}{2}}{2} 1},
    &&{1 \overset{\textcolor{red}{2}}{2} 3 1 3 1}\textcolor{red}{\mathfrak{su}_3},
    &&{1 \overset{\textcolor{red}{2}}{2} 3 1 \overset{\textcolor{red}{\mathfrak{su}_2}}{3} 1},
    &&{1 \underset{\textcolor{red}{2}}{2} \overset{1}{3} 1}\textcolor{red}{\mathfrak{su}_3},
    \nonumber\\
    &{1 \overset{\textcolor{red}{2}}{2} 3 1 2}\textcolor{red}{\mathfrak{su}_3},
    &&\textcolor{red}{2}{2 \overset{1}{3} 1 3}\textcolor{red}{\mathfrak{su}_2},
    &&\textcolor{red}{2}{2 3 1 2 3}\textcolor{red}{\mathfrak{su}_2},
    \label{eq:two_three_gluing_2}
\end{align}
where $T=8$ is realized by adding A-type dummy LSTs, with the external generators attached to the leftmost and rightmost $-1$ generators.

Finally, supergravity bases can be constructed by gluing through a common $(\textcolor{red}{3},\textcolor{red}{3})$ pair of external curves among the following bases:
\begin{align}
&  {1 \overset{\textcolor{red}{3}}{1}} \textcolor{red}{3},
&& {1 \overset{\textcolor{red}{3}}{2}} 1 \textcolor{red}{3},
&& \textcolor{red}{3} {2 1 2} \textcolor{red}{3},
&& \textcolor{red}{3} {1 2 \overset{\textcolor{red}{3}}{2} 1}, \nonumber\\
& \textcolor{red}{3} {2 1 3 1} \textcolor{red}{3},
&& \textcolor{red}{3} {1 \overset{1}{3} 1} \textcolor{red}{3},
&& \textcolor{red}{3} {1 3 1 3 1} \textcolor{red}{3},
&& \textcolor{red}{3} {1 2 3 1 2} \textcolor{red}{3}, \nonumber\\
& \textcolor{red}{3} {1 2 \overset{1}{3} 1} \textcolor{red}{3}
&& \textcolor{red}{3} {1 3 1 3 2 1} \textcolor{red}{3}
&& \textcolor{red}{3} {1 2 2 3 1 2} \textcolor{red}{3}
&& \textcolor{red}{3} {1 2 \overset{1}{3} 2 1} \textcolor{red}{3} \nonumber\\
& \textcolor{red}{3} {1 2 2 \overset{1}{3} 1} \textcolor{red}{3}
&& \textcolor{red}{3} {1 3 1 3 2 2 1} \textcolor{red}{3},
&& \textcolor{red}{3} {1 2 3 1 3 2 1} \textcolor{red}{3},
&& \textcolor{red}{3} {1 2 3 1 3 2 2 1} \textcolor{red}{3}.
\label{eq:su3_su3_gluing_2}
\end{align}
\begin{align}
& \textcolor{red}{3} {1 2\cdots2 1} \textcolor{red}{3}
\label{eq:su3_su3_gluing}
\end{align}
In this case, one block is chosen from \eqref{eq:su3_su3_gluing_2} and any number of blocks from \eqref{eq:su3_su3_gluing}. There are no $T=8$ bases with more than two external curves.

For $T=9$, many of the qualitative features are similar to $T=8$, and we refer the reader to the data file \texttt{Bases\_upto\_T9.wl} on Zenodo~\cite{Zenodo} for the complete list of the supergravity bases including multiple LSTs and/or multiple external curves.

For $T=10$, there are two new features.
One is the appearance of supergravity bases with three external generators. An example of such a base is
\begin{align}
    &\left(\textcolor{red}{\mathfrak{su}_3}{1 \overset{\textcolor{red}{\mathfrak{su}_3}}{\overset{1}{3}} 1}\,\textcolor{red}{\mathfrak{su}_3}\right)^3,
\end{align}
where there are three copies of the block sharing the same three external tensors $\textcolor{red}{(\mathfrak{su}_3)^3}$. The second genuinely new feature at $T=10$ is the appearance of supergravity bases with frozen singularity, which we discuss in the next subsection. 

Let us finally clarify which bases are \emph{not} included in the lists above (nor in the accompanying data file \texttt{Bases\_upto\_T9.wl}~\cite{Zenodo}):
\begin{itemize}
    \item Bases with a single $-1$ external tensor intersecting the {\it H}-string at one point. As noted above, such configurations always yield consistent supergravity theories and are easily identified from our block data, but they are too numerous to enumerate explicitly and are therefore omitted.
    \item Bases in which the leftmost $-2$ generator of a cluster $2\,\overset{\mathfrak{sp}_1}{2}\,\overset{\mathfrak{g}_2}{3}$ is taken as an external generator. Configurations whose intersections are already realized by the blocks obtained in Section~\ref{sec:non_higgsable_gravity_blocks_numerics} are included in our lists. However, since this $-2$ generator does not host a gauge algebra, it may also admit intersections that are not realized by those blocks. A complete treatment of such additional configurations as well as the classification for higher $T$ is left for future work.
\end{itemize}
Apart from these two classes, our classification is complete up to $T=9$.

\subsection{Supergravity with frozen singularity}\label{sec:frozen_singularity}

Our construction framework can also be used to generate candidate 6d supergravity theories that are difficult to realize using conventional string constructions. A typical class of such examples consists of theories involving frozen singularities. While our framework does not by itself establish the existence of a UV completion, it naturally produces tensor bases containing frozen NHCs, which do not arise in ordinary F-theory compactifications or standard heterotic string constructions. The relevant frozen clusters are
\begin{align}\label{eq:frozen-clusters}
    \overset{\mathfrak{su}_8}{\hat{1}} \,, \qquad
    \overset{\mathfrak{su}_8}{2}\ \overset{\mathfrak{su}_{16}}{\hat{1}} \,, \qquad
    \overset{\mathfrak{su}_8}{2}\!\!=\!\!\! \overset{\mathfrak{so}_{16}}{4} \ ,
\end{align}
which can be realized locally by allowing frozen singularities \cite{Bhardwaj:2019hhd}. Among these NHCs, the second one cannot be embedded in any supergravity theory, as argued around \eqref{eq:hat1_external}. As recorded in Table~\ref{tab:simple-lie-algebra-externals} of Section~\ref{sec:non_higgsable_gravity_blocks_numerics}, the first cluster in \eqref{eq:frozen-clusters} is contained in $1{,}731$ blocks, whereas the last cluster appears in twelve blocks. For three of these configurations, the $-4$ generator is taken as the external generator, while the remaining nine feature the external $-2$ generator supporting the $\mathfrak{su}_8$ algebra. The three blocks with the external $-4$ generator are explicitly
\begin{align}\label{eq:so16-three-blocks}
    \textcolor{red}{\overset{\mathfrak{so}_{16}}{4}}\!=\!\overset{\mathfrak{su}_8}{2}\,1\,2 \,, \qquad
    \textcolor{red}{\overset{\mathfrak{so}_{16}}{4}}\!=\!\overunderset{\mathfrak{su}_8}{1}{2}\,1 \,, \qquad
    \textcolor{red}{\overset{\mathfrak{so}_{16}}{4}}\!=\!\overunderset{\mathfrak{su}_8}{1}{2}\,1\,\textcolor{red}{\overset{\mathfrak{su}_2}{2}} \ ,
\end{align}
while the nine blocks with the external $-2$ generator are
\begin{align}\label{eq:su8-nine-blocks}
    & 1\,\overunderset{\textcolor{red}{\overset{\mathfrak{su}_8}{||}}}{1\,1}{4}\,1 \,, \qquad
    2\,1\,\overunderset{\textcolor{red}{\overset{\mathfrak{su}_8}{||}}}{1}{4}\,1 \,, \qquad
    1\,\overset{\textcolor{red}{\overset{\mathfrak{su}_8}{||}}}{4}\,1\,2\,2 \,, \qquad
    2\,1\,\overset{\textcolor{red}{\overset{\mathfrak{su}_8}{||}}}{4}\,1\,2 \,,
    \nonumber \\
    & \textcolor{red}{\overset{\mathfrak{su}_8}{2}}\!=\!\overset{\mathfrak{so}_{16}}{4}\,1\,2\,2\,2 \,, \qquad
    \textcolor{red}{\overset{\mathfrak{su}_3}{3}}\,2\,2\,1\,\overset{\textcolor{red}{\overset{\mathfrak{su}_8}{||}}}{4}\,1 \,, \qquad
    \textcolor{red}{\overset{\mathfrak{su}_8}{2}}\!=\!\overset{\mathfrak{so}_{16}}{4}\,1\,2\,2\,3\,1 \,,
    \nonumber \\
    & \textcolor{red}{\overset{\mathfrak{su}_8}{2}}\!=\!\overset{\mathfrak{so}_{16}}{4}\,1\,2\,2\,3\,1\,\textcolor{red}{\overset{\mathfrak{su}_2}{2}} \,, \qquad
    \textcolor{red}{\overset{\mathfrak{su}_8}{2}}\!=\!\overset{\mathfrak{so}_{16}}{4}\,1\,2\,2\,3\,1\,\textcolor{red}{\overset{\mathfrak{su}_3}{3}} \ .
\end{align}
Here the first six blocks have $T^H=4$, and the last three have $T^H=5$.
We now present two examples generated by our framework that involve the frozen NHCs.

The first example has $T=10$ and contains an external $\hat{1}$ generator supporting an $\mathfrak{su}_8$ gauge algebra:
\begin{align}\label{eq:frozen-example-hat-one}
    \textcolor{red}{\overset{\mathfrak{su}_8}{\hat{1}}}\,
    1\,2\,2\,2\,2\,2\,2\,\overset{2}{2}\,2 \ .
\end{align}
This tensor base can be viewed as a configuration obtained from ten blowups of $\mathbb{P}^2$, and hence has the intersection form of a K\"ahler surface that can occur as an F-theory base. The canonical class $b_0$ of this surface satisfies $b_0^2=-1$ and is represented by a genus-one curve. It can therefore serve as an external $\hat{1}$ generator supporting the $\mathfrak{su}_8$ gauge algebra with a symmetric hypermultiplet. This external generator intersects the little string charge $f$ of the LST, formed by one $-1$ curve and nine $-2$ curves, at two points. Nevertheless, because the gauge algebra is supported on the canonical class $b_0$, this theory does not admit a realization within ordinary F-theory geometry. This theory is formed by a single supergravity block containing an external $\hat{1}$ generator as well as the LST of the $-1$ and $-2$ generators.

The second example occurs at $T=11$ and involves the frozen cluster $\overset{\mathfrak{su}_8}{2}\!\!=\!\!\overset{\mathfrak{so}_{16}}{4}$:
\begin{align}\label{eq:frozen-example-two-four}
    2\,1\,\overset{\mathfrak{su}_8}{2}\textcolor{red}{=\!\!\overset{\mathfrak{so}_{16}}{4}}\,
    1\,2\,2\,2\,2\,2\,\overset{2}{2}\,2 \ .
\end{align}
Here, the $-4$ curve is an external generator that intersects the fiber class $f$ at two points. The charge $f$ here is shared by two LST sectors: one with tensor base $212$, and another formed by one $-1$ tensor and eight $-2$ tensors. So the theory is assembled from two supergravity blocks, which contain the first and second LST sectors, respectively. Notably, the intersection $2=4$ in this theory is also not realized by an ordinary F-theory geometry~\cite{Bhardwaj:2018jgp,Morrison:2023hqx}.

These examples demonstrate how tensor bases containing frozen NHCs naturally emerge within our framework. Since such configurations do not correspond to ordinary F-theory bases, the existence of global UV completions remains an open question. Nevertheless, they exhibit several properties expected of consistent quantum gravity theories. First, every local SCFT or LST sector appearing in these constructions is itself a consistent local quantum theory. This includes not only the external frozen NHCs and the {\it P}-type LSTs associated with the charge $f$, but also composite local configurations such as $\hat{1}\!-\!1$ and $1\!-\!2\!=\!4$, which appear in the known classifications of 6d local theories \cite{Bhardwaj:2015oru,Bhardwaj:2018jgp,Bhardwaj:2019hhd}. Second, all gauge algebras are supported exclusively on NHC tensors, so the associated instanton strings admit known consistent local worldsheet descriptions. In particular, these examples do not place any gauge algebra, including Abelian factors, on tensors with positive self-intersection.  This avoids the primary source of uncertainty associated with instanton strings for such gauge sectors, whose UV completions are generally unknown unless they arise from an explicit F-theory geometry or a conventional string compactification. Taken together, these observations suggest that the frozen configurations described above are plausible candidates for UV-completable 6d supergravity theories. Establishing a global string compactification that realizes them remains an interesting open problem.

\section{Conclusion}\label{sec:conclusion}

In this paper we have proposed a systematic framework for constructing 6d ${\cal N}=(1,0)$ supergravity theories in terms of their tensor multiplet configurations. The guiding input is the refined structure of tensor moduli space developed in \cite{Kim:2024eoa}, where boundaries of the tensor moduli space are controlled by tensionless BPS strings.  A central role is played by the distinguished {\it H}-string charge shared by suitable LST sectors, which naturally leads to a decomposition of a tensor base into {\it supergravity blocks}. Each block consists of an LST sector associated with the {\it H}-string together with external generators that intersect the {\it H}-string positively. These blocks contain intrinsically gravitational tensor multiplets whose BPS string tensions remain strictly positive everywhere in the moduli space. Therefore, in this picture, supergravity blocks serve as the elementary building blocks from which general 6d supergravity theories can be assembled.

As a first concrete application of this framework, we focused on non-Higgsable gravity blocks, namely blocks built entirely from NHCs appearing in the classifications of local SCFTs and LSTs. We began by classifying the allowed {\it P}-type LST sectors that can share a common {\it H}-string charge and found only 12,762 such sectors that can arise in 6d supergravity theories. By combining these sectors with all admissible external tensor attachments, we classified all possible non-Higgsable gravity blocks.  We also enumerated all tensor bases completely up to $T=9$, identified the qualitatively new features arising at $T=10$, and verified that, for $T\le 6$, our results agree with the known geometric classification of F-theory bases.

There are several natural directions for future work. The most immediate extension is to move beyond the non-Higgsable subclass by allowing gauge enhancements and additional charged matter on the tensor bases constructed here. To achieve this, our classification of non-Higgsable gravity blocks must be promoted to a full classification of massless spectra, subject to the cancellation of gauge, mixed, and gravitational anomalies. This extension also requires a classification of gauge algebras, including Abelian factors, on tensor multiplets with positive self-intersection. The anomaly constraints for such Abelian sectors have been analyzed systematically in \cite{Park:2011wv}, and incorporating these constraints into the block construction should be important for sharpening its completeness for general supergravity theories.

Another interesting direction concerns tensor bases involving frozen NHCs. While our framework naturally yields candidate bases of this type, these examples do not correspond to standard F-theory bases, and our current analysis does not establish their UV completions. Nevertheless, we can construct many such examples that pass important local consistency tests. Notably, all local SCFT and LST components are known consistent sectors. Furthermore, all gauge instanton strings have well-defined local descriptions with known UV completions for their moduli spaces, mainly because no gauge algebra is assigned to a tensor with positive self-intersection. These frozen examples therefore raise the broader question of whether intrinsically non-geometric supergravities satisfying all known local and non-perturbative consistency conditions can always be completed in quantum gravity. We hope that the block framework developed here provides a useful starting point for addressing this question.

Finally, in the present classification we imposed practical upper bounds, for computational tractability, on the intersection numbers between generators that either carry no gauge algebra or support only small $\mathfrak{su}_2$ or $\mathfrak{su}_3$ gauge algebras. It would be highly desirable to replace these practical cutoffs with sharp, physically motivated bounds. We expect that the equivalence relations and dualities discussed in Section~\ref{sec:equivalence_relations} may provide a more systematic approach to identifying redundant tensor intersections and thereby determine the true range of allowed intersection numbers.

\bigskip

\acknowledgments
We would like to thank Cumrun Vafa and Kai Xu for helpful discussions.
Y.H.\ was supported in part by JSPS KAKENHI Grant Nos.\ JP24H00976, JP24K07035, JP24KF0167, and by JST BOOST Program Japan Grant No.\ JPMJBY25E1. S.J. and H.K. are supported by the National
Research Foundation of Korea (NRF) grant funded by the Korean government (MSIT)
(2023R1A2C1006542).
We acknowledge the KEK Computing Research Center for providing the KEK Central Computing System.
Y.H. and H.K. further thank the Harvard Swampland Initiative and its Visitors
Program for hospitality during part of this
work.

\newpage

\appendix

\section{Structure of 6d SCFTs and LSTs}\label{app:SCFTsLSTs}

We briefly review the structure of 6d $(1,0)$ SCFTs and LSTs used in the construction of 6d supergravity theories. Our discussion is based on the classifications of SCFTs and LSTs established in the literature \cite{DelZotto:2014hpa,Heckman:2015bfa,Bhardwaj:2015xxa,Bhardwaj:2015oru,Bhardwaj:2019hhd}, and focuses on the aspects relevant for our purposes.

In particular, we concentrate on LSTs that can share an {\it H}-string charge in 6d supergravity. As explained in Section~\ref{sec:LST}, these LSTs have {\it P}-type endpoints and admit F-theory realizations without frozen singularities. These theories are constructed from basic building blocks, known as ``atoms,’’ which can be consistently glued together to form complete LSTs. There are two types of (non-Higgsable) atoms:
\begin{align} 
\text{DE type}: & \quad \stackrel{\mathfrak{so}_8}{4},\, \stackrel{\mathfrak{e}_6}{6},\, \stackrel{\mathfrak{e}_7'}{7},\, \stackrel{\mathfrak{e}_7}{8},\, \stackrel{\mathfrak{e}'''_8}{9},\, \stackrel{\mathfrak{e}_8''}{10},\, \stackrel{\mathfrak{e}_8'}{11},\, \stackrel{\mathfrak{e}_8}{12} \\ \text{non-DE type}: & \quad \stackrel{\mathfrak{su}_3}{3},\ \stackrel{\mathfrak{su}_2}{2}\stackrel{\mathfrak{g}_2}{3},\ \stackrel{\mathfrak{su}_2}{2}\stackrel{\mathfrak{so}_7}{3}\stackrel{\mathfrak{su}_2}{2},\ \stackrel{}{2}\stackrel{\mathfrak{sp}_1}{2}\stackrel{\mathfrak{g}_2}{3},\ \stackrel{\mathfrak{f}_4}{5},\ {\rm ADE\ graphs} \ , 
\end{align}
where $\overset{\mathfrak{g}}{n}$ denotes a tensor multiplet with self-intersection $-n$ supporting a gauge algebra $\mathfrak{g}$. These building blocks are NHCs, meaning that their gauge algebras cannot be Higgsed to smaller ones. The gauge algebra $\mathfrak{e}_7'$ is coupled to a fundamental half-hypermultiplet, while $\mathfrak{e}_8'$, $\mathfrak{e}_8''$, and $\mathfrak{e}_8'''$ denote one, two, and three small $\mathfrak{e}_8$ instantons, respectively. The non-Higgsable LST sectors used in this paper are obtained by gluing these atoms together through tensor multiplets with self-intersection $-1$, corresponding to E-string theory. The gluing rules are essentially fixed by the requirement that the gauge algebras supported on tensor multiplets adjacent to the $-1$ tensor must admit an embedding into $\mathfrak{e}_8$ flavor symmetry of the
E-string.

The DE-type atoms are also referred to as ``nodes'' and are denoted by $g_i$ in the atomic classification of \cite{Heckman:2015bfa}. On the other hand,  the non–DE-type atoms are glued together via $-1$ tensor multiplets to form chains of tensor multiplets known as ``links'', denoted by $L_{i,i+1}$ or $S_i$. An interior link $L_{i,i+1}$ connects the nodes $g_i$ and $g_{i+1}$, while a side link $S_i$ attaches to a single node $g_i$. Both nodes and links are systematically classified in \cite{Heckman:2015bfa}. With these ingredients, all 6d SCFTs realized in F-theory with bases containing one or more nodes can be described by the following structural forms \cite{Heckman:2015bfa}:
\begin{align}\label{eq:CFT-bases}
     &\text{\#(nodes)}=1,
     &&S_{0,1}\underset{\mathbf{I}^{\oplus u}}{\overset{S_{1}}{g_{1}}}S_{1,2},
     \\ 
     &\text{\#(nodes)}=2,
     &&S_{0,1}\overset{S_{1}}{g_{1}}L_{1,2}\overset{\mathbf{I}^{\oplus u}}{g_{2}}S_{2,3},
     \\
     &\text{\#(nodes)}=3,
     &&S_{0,1}\overset{S_{1}}{g_{1}}L_{1,2}\overset{\mathbf{I}^{\oplus s}}{g_{2}}L_{2,3}\overset{\mathbf{I}^{\oplus u}}{g_{3}}S_{3,4},
     \\
     &\text{\#(nodes)}=4,
     &&S_{0,1}\overset{S_{1}}{g_{1}}L_{1,2}\overset{\mathbf{I}^{\oplus s}}{g_{2}}L_{2,3}\overset{\mathbf{I}^{\oplus t}}{g_{3}}L_{3,4}\overset{\mathbf{I}^{\oplus u}}{g_{4}}S_{4,5},
     \\
     &\text{\#(nodes)}=5,
     &&S_{0,1}\overset{S_{1}}{g_{1}}L_{1,2}\overset{\mathbf{I}^{\oplus s}}{g_{2}}L_{2,3}\overset{\mathbf{I}^{\oplus r}}{g_{3}}L_{3,4}\overset{\mathbf{I}^{\oplus t}}{g_{4}}L_{4,5}\overset{\mathbf{I}^{\oplus u}}{g_{5}}S_{5,6},
     \\
     &\text{\#(nodes)}\geq6,
     &&S_{0,1}\overset{S_{1}}{g_{1}}L_{1,2}\overset{\mathbf{I}^{\oplus s}}{g_{2}%
     }L_{2,3}\cdots\overset{\mathbf{I}^{\oplus t}}{g_{k-1}%
     }L_{k-1,k}\overset{\mathbf{I}^{\oplus u}}{g_{k}}S_{k,k+1}.
\end{align}
where $\cdots$ stands for a chain of nodes and interior links, and $\mathbf{I}^{\oplus s}$ denotes small instanton nodes of the form $\underbrace{122\cdots2}_{s}$ with the $1$ attached to the adjacent node $g_i$. These structures follow from the negative definiteness of the intersection form, which is necessary for the simultaneous shrinking of tensor multiplets, together with the gluing rules for pairs of atoms.

Internal links correspond to the conformal matters introduced in \cite{DelZotto:2014hpa}, which we list as 
\begin{align}
     \mathfrak{so}_{2n} \overset{1,1}{\otimes} \mathfrak{so}_{2n} &: \ \mathfrak{so}_{2n}1\mathfrak{so}_{2n}\ ,  && \mathfrak{e}_6 \overset{3,3}{\bigcirc} \mathfrak{e}_6 & : \ \mathfrak{e}_61\overset{\mathfrak{su}_3}{3}1\overset{\mathfrak{f}_4}{5}1\overset{\mathfrak{su}_3}{3}1\mathfrak{e}_6 \ , \nonumber\\
     \mathfrak{e}_7 \overset{3,2}{\otimes} \mathfrak{so}_{2n} &: \ \mathfrak{e}_71\overset{\mathfrak{su}_2}{2}\overset{\mathfrak{g}_2}{3}1\mathfrak{so}_{2n} \ , && \mathfrak{e}_7 \overset{4,3}{\otimes} \mathfrak{e}_6 & : \ \mathfrak{e}_7 1\overset{\mathfrak{su}_2}{2}\overset{\mathfrak{g}_2}{3}1\overset{\mathfrak{f}_4}{5}1\overset{\mathfrak{su}_3}{3}1 \mathfrak{e}_6\ ,\nonumber\\
     \mathfrak{e}_8 \overset{4,2}{\otimes} \mathfrak{so}_{2n} &: \ \mathfrak{e}_81\overset{}{2}\overset{\mathfrak{sp}_1}{2}\overset{\mathfrak{g}_2}{3}1\mathfrak{so}_{2n} \ , && \mathfrak{e}_7 \overset{4,4}{\otimes} \mathfrak{e}_7 & : \ \mathfrak{e}_7 1\overset{\mathfrak{su}_2}{2}\overset{\mathfrak{g}_2}{3}1\overset{\mathfrak{f}_4}{5}1\overset{\mathfrak{g}_2}{3}\overset{\mathfrak{su}_2}{2}1 \mathfrak{e}_7\ ,\nonumber\\
     \mathfrak{e}_6 \overset{2,2}{\otimes} \mathfrak{e}_6 &: \ \mathfrak{e}_61\overset{\mathfrak{su}_3}{3}1\mathfrak{e}_6 \ , && \mathfrak{e}_8 \overset{5,3}{\otimes} \mathfrak{e}_6 & : \ \mathfrak{e}_8 1\overset{}{2}\overset{\mathfrak{sp}_1}{2}\overset{\mathfrak{g}_2}{3}1\overset{\mathfrak{f}_4}{5}1\overset{\mathfrak{su}_3}{3}1 \mathfrak{e}_6\ , \nonumber\\
     \mathfrak{e}_7 \overset{3,3}{\otimes} \mathfrak{e}_7 &: \ \mathfrak{e}_71\overset{\mathfrak{su}_2}{2}\overset{\mathfrak{so}_7}{3}\overset{\mathfrak{su}_2}{2}1\mathfrak{e}_7 \ , && \mathfrak{e}_8 \overset{5,4}{\otimes} \mathfrak{e}_7 & : \ \mathfrak{e}_8 1\overset{}{2}\overset{\mathfrak{sp}_1}{2}\overset{\mathfrak{g}_2}{3}1\overset{\mathfrak{f}_4}{5}1\overset{\mathfrak{g}_2}{3}\overset{\mathfrak{su}_2}{2}1 \mathfrak{e}_7\ , \nonumber\\
     \mathfrak{e}_8 \overset{5,5}{\otimes} \mathfrak{e}_8 & : \ \mathfrak{e}_8 1\overset{}{2}\overset{\mathfrak{sp}_1}{2}\overset{\mathfrak{g}_2}{3}1\overset{\mathfrak{f}_4}{5}1\overset{\mathfrak{g}_2}{3}\overset{\mathfrak{sp}_1}{2}\overset{}{2}1 \mathfrak{e}_8 \ . \label{eq:conformal-matters}
 \end{align}
Here, the superscripts $p, q$ above $\otimes$ and $\bigcirc$ means the shifts in the intersection numbers of two adjacent nodes induced by blowing down the conformal matter, such that $(-n,-m)\rightarrow (-n+p,-m+q)$. If both nodes attached to an interior link carry the gauge algebras specified above, then such a link is called a minimal interior link. Otherwise, it is referred to as a non-minimal interior link. A non-minimal interior link can appear only at the ends of the base \cite{Heckman:2015bfa}.

We construct LST bases with the {\it P}-type endpoints following Ref.~\cite{Bhardwaj:2015oru}. We start from SCFT bases, and then add one generator to realize the {\it P}-type endpoints. 
The tensor decoupling criterion, removing any one of the generators gives a SCFT base, is imposed.
For the construction of SCFT bases, we use Table for side links $S$ and noble molecules $N$ (SCFT without nodes) in Appendix D of \cite{Heckman:2015bfa} with the following modifications:
\begin{itemize}
    \item The side link $3\overset{2}{2}1$ induces the $-7$ blowdown instead of $-4$.
    \item We add the side link $15131$ which can be attached to $4$ and $6$ nodes.
    \item We add the following noble molecules:
    \begin{align}
        &5 1,
        &&1 5 1,
        &&2 1 5 1,
        &&3 1 5 1,
        &&3 2 1 5 1,
        &&2 3 1 5 1,
        \nonumber \\
        &2 2 1 5 1,
        &&3 2 2 1 5 1,
        &&2 3 2 1 5 1,
        &&2 2 3 1 5 1,
        &&3 1 3 1 5 1,
        &&5 1 3 1 5 1,
        \nonumber \\
        &5 1 3 2 1 5 1,
        &&1 5 1 3 1 5 1,
        &&5 1 2 3 1 5 1,
        &&1 5 1 2 3 1 5 1,
        &&5 1 2 3 2 1 5 1,
        &&2 1 5 1 3 1 5 1,
        \nonumber \\
        &3 1 5 1 3 1 5 1,
        &&5 1 2 2 3 1 5 1,
        &&3 1 5 1 3 2 1 5 1,
        &&2 3 1 5 1 3 1 5 1,
        &&5 1 3 1 5 1 3 1 5 1,
        &&2 3 1 5 1 3 2 1 5 1,
        \nonumber \\
        &2 2 3 1 5 1 3 1 5 1,
        &&2 2 3 1 5 1 3 2 1 5 1,
        &&1 \overset{1}{5} 1 3 1 5 1,
        &&1 \overset{1}{5} 1 3 1 5,
        &&1 5 1 3 2 1 5 1,
        &&2 1 5 1 3 1 5 1,
        \nonumber \\
        &1 5 1 3 1 5 1 3 1 5,
        &&5 1 2 3 2 1 5 1,
    \end{align}
\end{itemize}

\section{Long bases} \label{app:long_base}
Here we list the long bases of 6d LSTs with {\it P}-type endpoints, which are relevant for the construction of non-Higgsable gravity blocks.\footnote{For sufficiently large $n$ (equivalently, large $T^H$), these long bases violate the gravitational-anomaly bound given by the first equation in~\eqref{eq:GS-conds-non-abelian}, and therefore do not appear as part of any consistent supergravity base.}
In this appendix, $\mathfrak{so}$ denotes $\mathfrak{so}_8$.
First, there are the $A$- and $D$-types long bases:
\begin{align}
    &1\underbrace{2\cdots2}_{n}1,
    &&2\overset{2}{2}\underbrace{2\cdots2}_n1,
    \label{eq:LST_AD_type}
\end{align}
where $n=0,1,\cdots$ (in this appendix, $n$ is always a nonnegative integer).
Next, the long bases without decoration or with only left side links are given by
\begingroup\allowdisplaybreaks
{\footnotesize
\begin{align}
&\left.
\begin{array}{l}
  \mathfrak{so} \\
  1 \mathfrak{so} \overset{2,3}{\otimes} \mathfrak{e}_7 \\
  \overset{,2}{\otimes} \mathfrak{e}_6 \\
  \overset{,3}{\otimes} \mathfrak{e}_7^{\prime} \\
  \overset{,4}{\otimes} \mathfrak{e}_7
\end{array}
\right\}
\left(\overset{3,3}{\otimes} \mathfrak{e}_7 \right)^{n}\overset{3,3}{\otimes} \mathfrak{so},
&&\left.
\begin{array}{l}
  \overset{,6}{\otimes} \\
  \mathfrak{so} \overset{3,5}{\otimes} \mathfrak{e}_8 \\
  1 \mathfrak{so} 1 \mathfrak{so} \overset{2,4}{\otimes} \mathfrak{e}_8 \\
  31 \mathfrak{so} \overset{2,4}{\otimes} \mathfrak{e}_8 \\
  \overset{,2}{\otimes} \mathfrak{e}_6 \overset{3,5}{\otimes} \mathfrak{e}_8 \\
  1 \mathfrak{e}_6 \overset{4,5}{\otimes} \mathfrak{e}_8 \\
  \mathfrak{e}_6 \\
  \overset{,2}{\otimes} \mathfrak{e}_7^{\prime} \overset{4,5}{\otimes} \mathfrak{e}_8 \\
  1 \mathfrak{e}_7^{\prime} \\
  \overset{,3}{\otimes} \mathfrak{e}_7 \overset{4,5}{\otimes} \mathfrak{e}_8 \\
  \overset{,2}{\otimes} \mathfrak{e}_7
\end{array}
\right\}
\left(\overset{5,5}{\otimes} \mathfrak{e}_8 \right)^{n}
\begin{cases}
\overset{5,3}{\otimes} \mathfrak{so} \\
\overset{5,5}{\otimes} \mathfrak{e}_6
\end{cases},
\end{align}
}
The long bases with both left and right decorations are as follows:
{\footnotesize
\begin{align}
    &{\overset{,2}{\otimes} \mathfrak{so} \left(1 \mathfrak{so} \right)^{n}\overset{2,}{\otimes}},
    \qquad
    {\overset{,2}{\otimes} \mathfrak{so} \left(1 \mathfrak{so} \right)^{n}1 \overset{2}{3} 2},
    \qquad
    {\overset{,3}{\otimes} \mathfrak{e}_6 \left(\overset{2,2}{\otimes} \mathfrak{e}_6 \right)^{n}\overset{3,}{\otimes}},
    \qquad
    {1 \mathfrak{so} \left(\overset{2,2}{\otimes} \mathfrak{e}_6 \right)^{n}\overset{2,2}{\otimes} \mathfrak{so} 1}, \\
    &
    {1 \mathfrak{so} \left(\overset{2,2}{\otimes} \mathfrak{e}_6 \right)^{n}\overset{3,}{\otimes}},
    \quad
    {31 \overset{3}{\overset{1}{\mathfrak{e}_6}} \left(\overset{2,2}{\otimes} \mathfrak{e}_6 \right)^{n}\overset{3,}{\otimes}},
    \quad
    {31 \overset{3}{\overset{1}{\mathfrak{e}_6}} \left(\overset{2,2}{\otimes} \mathfrak{e}_6 \right)^{n}\overset{2,2}{\otimes} \mathfrak{so} 1}, 
    \quad
    L_{\mathfrak{e}_7} \left(\overset{3,3}{\otimes} \mathfrak{e}_7 \right)^{n} R_{\mathfrak{e}_7},
    \quad
    L_{\mathfrak{e}_8} \left(\overset{5,5}{\otimes} \mathfrak{e}_8 \right)^{n} R_{\mathfrak{e}_8},
\end{align}
}
where
{\footnotesize
\begin{align}
    &L_{\mathfrak{e}_7} = \overset{,4}{\otimes}\mathfrak{e}_7, \, 1 \mathfrak{so} \overset{2,3}{\otimes} \mathfrak{e}_7, \, \overset{,2}{\otimes}\mathfrak{e}_6, \, \overset{,3}{\otimes}\mathfrak{e}_7^{\prime},
    \qquad
    R_{\mathfrak{e}_7} = \overset{4,}{\otimes}, \, \overset{3,2}{\otimes} \mathfrak{so} 1, \, \overset{3,3}{\otimes} \mathfrak{e}_6 \overset{2,}{\otimes}, \, \overset{3,3}{\otimes}\mathfrak{e}_7^{\prime} \overset{3,}{\otimes},
    \\
    &L_{\mathfrak{e}_8} = \overset{,6}{\otimes}\mathfrak{e}_8, \,
    31 \mathfrak{so} \overset{2,4}{\otimes} \mathfrak{e}_8, \,
    1 \mathfrak{so} 1 \mathfrak{so} \overset{2,4}{\otimes} \mathfrak{e}_8, \,
    \overset{,2}{\otimes} \mathfrak{e}_6 \overset{3,5}{\otimes} \mathfrak{e}_8, \,
    1 \mathfrak{e}_6 \overset{4,5}{\otimes} \mathfrak{e}_8, \,
    \overset{,2}{\otimes} \mathfrak{e}_7^{\prime} \overset{4,5}{\otimes} \mathfrak{e}_8, \,
    1 \mathfrak{e}_7^{\prime}, \overset{,3}{\otimes} \mathfrak{e}_7 \overset{4,5}{\otimes} \mathfrak{e}_8, \,
    \overset{,2}{\otimes} \mathfrak{e}_7,
    \\
    &R_{\mathfrak{e}_8} = \overset{6,}{\otimes}, \,
    \overset{4,2}{\otimes} \mathfrak{so} 13, \,
    \overset{4,2}{\otimes} \mathfrak{so} 1 \mathfrak{so} 1, \,
    \overset{5,3}{\otimes} \mathfrak{e}_6 \overset{2,}{\otimes}, \,
    \overset{5,4}{\otimes} \mathfrak{e}_6 1, \,
    \overset{5,4}{\otimes} \mathfrak{e}_7^\prime \overset{2,}{\otimes}, \,
    \overset{5,5}{\otimes} \mathfrak{e}_7^\prime 1, \,
    \overset{5,4}{\otimes} \mathfrak{e}_7 \overset{3,}{\otimes}, \,
    \overset{5,5}{\otimes} \mathfrak{e}_7 \overset{2,}{\otimes}. 
\end{align}
}
The long bases without decoration are also found in Appendix~C of \cite{Bhardwaj:2015oru}, where the following seems to be missing:
{\footnotesize
\begin{align}
&{\mathfrak{so} \overset{3,5}{\otimes} \mathfrak{e}_8 \left(\overset{5,5}{\otimes} \mathfrak{e}_8 \right)^{n}\overset{5,5}{\otimes} \mathfrak{e}_6}.
\end{align}
}

The long bases with the instanton at the middle are given by
{\footnotesize
\begin{align}
    &{1 \mathfrak{so} \overset{2,4}{\otimes} \overset{1}{\mathfrak{e}_8} \left(\overset{5,5}{\otimes} \mathfrak{e}_8 \right)^{n}\overset{5,5}{\otimes} \overset{1}{\mathfrak{e}_8} \overset{4,2}{\otimes} \mathfrak{so} 1},
    &&{1 \mathfrak{so} \overset{2,4}{\otimes} \overset{1}{\mathfrak{e}_8} \left(\overset{5,5}{\otimes} \mathfrak{e}_8 \right)^{n}}
    \begin{cases}
    \overset{6,}{\otimes} \\
    \overset{5,3}{\otimes} \mathfrak{so} \\
    \overset{4,2}{\otimes} \mathfrak{so} 13 \\
    \overset{4,2}{\otimes} \mathfrak{so} 1 \mathfrak{so} 1 \\
    \overset{5,3}{\otimes} \mathfrak{e}_6 \overset{2,}{\otimes} \\
    \overset{5,4}{\otimes} \mathfrak{e}_6 1 \\
    \overset{5,5}{\otimes} \mathfrak{e}_6 \\
    \overset{5,4}{\otimes} \mathfrak{e}_7^{\prime} \overset{2,}{\otimes} \\
    \overset{5,5}{\otimes} \mathfrak{e}_7^{\prime} 1 \\
    \overset{5,4}{\otimes} \mathfrak{e}_7 \overset{3,}{\otimes} \\
    \overset{5,5}{\otimes} \mathfrak{e}_7 \overset{2,}{\otimes}
    \end{cases}.
\end{align}
}
In the above, we use the following shorthand definitions:
{\footnotesize
\begin{align}
    \overset{,2}{\otimes} \mathfrak{so} =
    &1 \overset{1}{\mathfrak{so}}, \quad
    \mathbf{I}^{\oplus 2} \mathfrak{so}, \quad
    \overset{2,2}{\otimes} \mathfrak{so}, \quad
    \overset{3,2}{\otimes} \mathfrak{so}, \quad
    \overset{4,2}{\otimes} \mathfrak{so}, 
    \\
    \overset{,3}{\otimes} \mathfrak{so} = 
    &\mathbf{I}^{\oplus 2} \overset{1}{\mathfrak{so}}, \quad
    \overset{2,2}{\otimes} \overset{1}{\mathfrak{so}}, \quad
    \mathbf{I}^{\oplus 3} \mathfrak{so}, \quad
    \overset{2,3}{\otimes} \mathfrak{so}, \quad
    \overset{3,3}{\otimes} \mathfrak{so}, \quad
    \overset{3,3}{\bigcirc} \mathfrak{so}, \quad
    \overset{4,3}{\otimes} \mathfrak{so}, \quad
    \overset{5,3}{\otimes} \mathfrak{so}, \quad
    2\overset{1}{3}1 \mathfrak{so}, \quad
    12\overset{2}{3}1 \mathfrak{so},
    \nonumber\\
    &
    3131 \mathfrak{so}, \quad
    1\overset{1}{5}131 \mathfrak{so}, \quad
    151231 \mathfrak{so}, \quad
    512231 \mathfrak{so}, \quad
    215131 \mathfrak{so},
    \\
    \overset{,2}{\otimes} \mathfrak{e}_6 =
    &1 \overset{1}{\mathfrak{e}}_6, \quad
    \mathbf{I}^{\oplus 2} \mathfrak{e}_6, \quad
    \overset{2,2}{\otimes} \mathfrak{e}_6, 
    \\
    \overset{,3}{\otimes} \mathfrak{e}_6 = 
    &\overset{,2}{\otimes} \overset{1}{\mathfrak{e}}_6, \quad
    \mathbf{I}^{\oplus 3} \mathfrak{e}_6, \quad
    \overset{2,3}{\otimes} \mathfrak{e}_6, \quad
    \overset{3,3}{\otimes} \mathfrak{e}_6, \quad
    \overset{3,3}{\bigcirc} \mathfrak{e}_6 , \quad
    \overset{4,3}{\otimes} \mathfrak{e}_6, \quad
    \overset{5,3}{\otimes} \mathfrak{e}_6, \quad
    3131 \mathfrak{e}_6, \quad
    1\overset{1}{5}131 \mathfrak{e}_6, \quad
    215131 \mathfrak{e}_6,
    \\
    \overset{,4}{\otimes} \mathfrak{e}_6 =
    &\overset{,2}{\otimes} \overset{\overset{,2}{\otimes}}{\mathfrak{e}_6}, \quad
    \overset{,3}{\otimes} \overset{1}{\mathfrak{e}}_6, \quad
    \mathbf{I}^{\oplus 4} \mathfrak{e}_6, \quad
    \overset{2,4}{\otimes} \mathfrak{e}_6, \quad
    \overset{3,4}{\otimes} \mathfrak{e}_6, \quad
    \overset{4,4}{\otimes} \mathfrak{e}_6, \quad
    \overset{5,4}{\otimes} \mathfrak{e}_6, \quad
    2\overset{1}{3}21 \mathfrak{e}_6, \quad
    31321 \mathfrak{e}_6, \quad
    23131 \mathfrak{e}_6, \quad    
    1\overset{1}{5}1321 \mathfrak{e}_6,
    \nonumber\\
    &
    31\overset{1}{5}131 \mathfrak{e}_6, \quad
    1512321 \mathfrak{e}_6, \quad
    3215131 \mathfrak{e}_6, \quad
    2151321 \mathfrak{e}_6, \quad
    151315131 \mathfrak{e}_6, \quad
    512315131 \mathfrak{e}_6,
    \\
    \overset{,2}{\otimes} \mathfrak{e}_7 =
    &1 \overset{1}{\mathfrak{e}}_7, \quad
    \mathbf{I}^{\oplus 2} \mathfrak{e}_7,
    \\
    \overset{,3}{\otimes} \mathfrak{e}_7 =
    &
    \overset{,2}{\otimes} \overset{1}{\mathfrak{e}}_7, \quad
    \mathbf{I}^{\oplus 3} \mathfrak{e}_7, \quad
    \overset{2,3}{\otimes} \mathfrak{e}_7, \quad
    \overset{3,3}{\otimes} \mathfrak{e}_7,
    \\
    \overset{,4}{\otimes} \mathfrak{e}_7 =
    &\overset{,2}{\otimes} \overset{\overset{,2}{\otimes}}{\mathfrak{e}_7}, \quad
    \overset{,3}{\otimes} \overset{1}{\mathfrak{e}}_7, \quad
    \mathbf{I}^{\oplus 4} \mathfrak{e}_7, \quad
    \overset{2,4}{\otimes} \mathfrak{e}_7, \quad
    \overset{3,4}{\otimes} \mathfrak{e}_7, \quad
    \overset{4,4}{\otimes} \mathfrak{e}_7, \quad
    \overset{5,4}{\otimes} \mathfrak{e}_7, \quad
    2\overset{1}{3}21 \mathfrak{e}_7, \quad
    31321 \mathfrak{e}_7,
    \nonumber\\
    &    
    1\overset{1}{5}1321 \mathfrak{e}_7, \quad
    1512321 \mathfrak{e}_7, \quad
    2151321 \mathfrak{e}_7,
    \\
    \overset{,6}{\otimes} \mathfrak{e}_8 =
    &
    \mathbf{I}^{\oplus 3} \overset{\mathbf{I}^{\oplus 3}}{\mathfrak{e}_8}, \quad
    \mathbf{I}^{\oplus 4} \overset{\mathbf{I}^{\oplus 2}}{\mathfrak{e}_8}, \quad
    \overset{2,4}{\otimes} \overset{\mathbf{I}^{\oplus 2}}{\mathfrak{e}_8}, \quad
    \mathbf{I}^{\oplus 5} \overset{1}{\mathfrak{e}}_8, \quad
    \overset{3,5}{\otimes} \overset{1}{\mathfrak{e}}_8, \quad
    \overset{4,5}{\otimes} \overset{1}{\mathfrak{e}}_8, \quad
    \overset{5,5}{\otimes} \overset{1}{\mathfrak{e}}_8, \quad
    313221 \overset{1}{\mathfrak{e}}_8, \quad
    \nonumber\\
    &
    1\overset{1}{5}13221 \overset{1}{\mathfrak{e}}_8, \quad
    21513221 \overset{1}{\mathfrak{e}}_8, \quad
    \mathbf{I}^{\oplus 6} \mathfrak{e}_8, \quad
    2313221 \mathfrak{e}_8, \quad
    31\overset{1}{5}13221 \mathfrak{e}_8, \quad
    321513221 \mathfrak{e}_8.
\end{align}
}
and similarly for $\overset{\bullet,}{\otimes}$. $\overset{,\bullet}{\otimes} \mathfrak{e}_7^{\prime}$ is defined in the same way as $\overset{,\bullet}{\otimes} \mathfrak{e}_7$.

\endgroup

\section{Short LST bases with {\it P}-type endpoints up to 10 curves} \label{app:LST_base}
This appendix records the short LST bases with {\it P}-type endpoints that enter the non-Higgsable block construction.  We list the bases with $T^H=7,8,9$ that are not contained in the long families of Appendix~\ref{app:long_base}. The number of short bases for each value of $T^H$ is summarized in Table~\ref{tab:short-P-summary}. The bases with $T^H\leq 6$ are displayed in Section~\ref{sec:classification}.  The enumeration uses the SCFT/LST building blocks and gluing rules summarized in Appendix~\ref{app:SCFTsLSTs}, together with the tensor-decoupling criterion for {\it P}-type LST endpoints: deleting any one tensor curve from a listed base gives an admissible SCFT base.  This is the criterion used in the LST classification of \cite{Heckman:2015bfa,Bhardwaj:2015oru}. In this appendix, $\mathfrak{so}$ denotes $\mathfrak{so}_8$.

\begin{table}[t]
\centering
\begin{tabular}{c c c}
\toprule
$T^H$ & number of short bases & displayed in \\
\midrule
$7$  & $41$ & \eqref{eq:short-P-8} \\
$8$  & $62$ & \eqref{eq:short-P-9} \\
$9$ & $74$ & \eqref{eq:short-P-10} \\
\bottomrule
\end{tabular}
\caption{Short {\it P}-type LST bases with $7\leq T^H\leq 9$, after removing the long families listed in Appendix~\ref{app:long_base}.}
\label{tab:short-P-summary}
\end{table}

\begingroup\allowdisplaybreaks
\subsection*{$T^H=7$}
There are $41$ short {\it P}-type LST bases with $T^H=7$.  They are
\begin{align}
&{2 \overset{2}{3} 1 3 2 2 1},
{1 2 3 1 3 2 2 1},
{1 \overset{1}{\underset{1}{5}} 1 \overset{2}{3} 2},
{1 \overset{1}{\underset{1}{5}} \overset{2,3}{\otimes}},
{3 1 \overset{3}{\overset{1}{5}} \overset{2,2}{\otimes}},
{\overset{2,2}{\otimes} \overset{1}{5} \overset{2,2}{\otimes}},
{1 \overset{1}{5} \overset{3,3}{\otimes}},
{2 1 \overset{1}{5} \overset{2,3}{\otimes}},
{2 \overset{1}{3} 1 5 1 3 1},
{1 3 2 1 5 1 3 1},
\nonumber\\
&{2 1 5 \overset{3,3}{\otimes}},
{2 2 1 5 1 3 2 1},
{3 1 3 1 5 1 3 1},
{1 \overset{1}{5} 1 3 2 1 \mathfrak{so}},
{3 1 \overset{1}{5} \overset{2,2}{\otimes} \mathfrak{so}},
{1 5 \overset{3,3}{\otimes} \mathfrak{so}},
{2 1 5 1 3 2 1 \mathfrak{so}},
{2 3 1 \mathfrak{so} \overset{2,2}{\otimes} 5},
{3 1 \mathfrak{so} 1 3 1 5 1},
{3 1 \mathfrak{so} 1 3 2 1 5},
\nonumber\\
&{3 1 5 1 2 3 1 \mathfrak{so}},
{3 2 1 5 1 3 1 \mathfrak{so}},
{\overset{2,3}{\otimes} \overset{1}{\underset{1}{\mathfrak{e}_6}} 1},
{3 \overset{2,2}{\otimes} \overset{1}{\underset{1}{\mathfrak{e}_6}} 1},
{3 2 1 \overset{1}{\underset{1}{\mathfrak{e}_6}} 1 3},
{3 2 1 \overset{3}{\overset{1}{\mathfrak{e}_6}} \mathbf{I}^{\oplus 2}},
{3 2 1 \overset{1}{\mathfrak{e}_6} 1 2 3},
{3 2 2 1 \overset{1}{\mathfrak{e}_6} 1 3},
{2 2 3 1 3 1 \mathfrak{e}_6 1},
{2 2 3 1 3 2 1 \mathfrak{e}_6},
\nonumber\\
&{2 3 1 3 2 1 \mathfrak{e}_6 1},
{2 3 1 3 2 2 1 \mathfrak{e}_6},
{3 1 3 2 2 1 \mathfrak{e}_6 1},
{3 2 2 1 \mathfrak{e}_6 1 2 3},
{\mathbf{I}^{\oplus 5} \overset{1}{\mathfrak{e}_7^{\prime}} 1},
{\mathbf{I}^{\oplus 5} \mathfrak{e}_7^{\prime} \mathbf{I}^{\oplus 2}},
{\mathbf{I}^{\oplus 7} \mathfrak{e}_7^{\prime}},
{1 \overset{1}{\mathfrak{so}} 1 \mathfrak{so} \overset{2,2}{\otimes}},
{1 \mathfrak{so} 1 \overset{1}{\mathfrak{so}} 1 \mathfrak{so} 1},
{3 1 \mathfrak{so} 1 \overset{1}{\mathfrak{so}} 1 \mathfrak{so}},
\nonumber\\
&{1 \mathfrak{so} 1 \mathfrak{so} \overset{2,2}{\otimes} \mathfrak{so}}.
\label{eq:short-P-8}
\end{align}

\subsection*{$T^H=8$}
There are $62$ short {\it P}-type LST bases with $T^H=8$.  They are
\begin{align}
&{1 2 2 3 1 3 2 2 1},
{1 \overset{1}{\underset{1}{5}} \overset{2,4}{\otimes}},
{3 1 \overset{3}{\overset{1}{5}} \overset{2,3}{\otimes}},
{\overset{3,2}{\otimes} \overset{1}{5} \overset{2,2}{\otimes}},
{2 1 \overset{1}{5} \overset{2,4}{\otimes}},
{2 \overset{1}{3} 1 5 1 3 2 1},
{1 2 3 1 5 1 3 1 3},
{1 3 1 5 \overset{3,3}{\otimes}},
\nonumber\\
&{1 3 2 1 5 1 3 2 1},
{2 2 1 5 1 3 2 2 1},
{1 5 1 2 3 1 \overset{1}{5} 1},
{1 \overset{1}{5} 1 3 2 2 1 5},
{2 1 5 1 3 1 \overset{1}{5} 1},
{2 3 1 \overset{1}{5} 1 3 1 5},
{3 1 \overset{1}{5} 1 3 1 5 1},
{3 1 \overset{1}{5} 1 3 2 1 5},
\nonumber\\
&{1 5 \overset{3,3}{\otimes} 5 1},
{2 1 5 1 3 1 5 1 2},
{2 1 5 1 3 2 1 5 1},
{2 1 5 1 3 2 2 1 5},
{2 3 1 5 1 2 3 1 5},
{2 3 2 1 5 1 3 1 5},
{3 1 5 1 2 3 1 5 1},
{3 1 5 \overset{3,3}{\otimes} 5},
\nonumber\\
&{3 2 1 5 1 3 1 5 1},
{3 2 1 5 1 3 2 1 5},
{\overset{3,4}{\otimes} \mathfrak{so}},
{\overset{2,2}{\otimes} \overset{3}{\overset{1}{\mathfrak{e}_6}} 1 2 3},
{3 2 1 \overset{1}{\underset{1}{\mathfrak{e}_7^{\prime}}} 1 2 3},
{3 2 1 \overset{\mathbf{I}^{\oplus 2}}{\mathfrak{e}_7^{\prime}} 1 2 3},
{3 2 2 1 \overset{1}{\mathfrak{e}_7^{\prime}} 1 2 3},
{2 2 3 1 3 2 2 1 \mathfrak{e}_7^{\prime}},
\nonumber\\
&{2 3 1 3 2 1 \mathfrak{e}_7^{\prime} \mathbf{I}^{\oplus 2}},
{3 1 3 2 2 1 \mathfrak{e}_7^{\prime} \mathbf{I}^{\oplus 2}},
{3 2 2 1 \mathfrak{e}_7^{\prime} 1 2 2 3},
{\mathbf{I}^{\oplus 5} \mathfrak{e}_7 \mathbf{I}^{\oplus 3}},
{\mathbf{I}^{\oplus \mathfrak{e}_7^{\prime}} \mathfrak{e}_7 1},
{\mathbf{I}^{\oplus \mathfrak{e}_7} \mathfrak{e}_7},
{1 \mathfrak{so} 1 \mathfrak{so} 1 3 1 5 1},
{1 \mathfrak{so} 1 \mathfrak{so} 1 3 2 1 5},
\nonumber\\
&{3 1 \mathfrak{so} 1 \mathfrak{so} \overset{2,2}{\otimes} 5},
{3 1 5 1 3 1 \mathfrak{so} 1 \mathfrak{so}},
{\mathfrak{so} \overset{3,3}{\bigcirc} \mathfrak{so}},
{3 1 \overset{\mathbf{I}^{\oplus 2}}{\mathfrak{e}_6} \overset{2,2}{\otimes} \mathfrak{so}},
{1 \overset{1}{\mathfrak{e}_6} \overset{3,2}{\otimes} \mathfrak{so} 1},
{1 \overset{1}{\mathfrak{e}_6} \overset{2,2}{\otimes} \mathfrak{so} 1 3},
{\mathbf{I}^{\oplus 2} \overset{1}{\mathfrak{e}_6} \overset{2,2}{\otimes} \mathfrak{so} 1},
{3 1 \overset{1}{\mathfrak{e}_6} \overset{3,2}{\otimes} \mathfrak{so}},
\nonumber\\
&{3 2 1 \overset{1}{\mathfrak{e}_6} \overset{2,2}{\otimes} \mathfrak{so}},
{1 \mathfrak{so} \overset{2,2}{\otimes} \mathfrak{e}_6 \mathbf{I}^{\oplus 3}},
{1 \mathfrak{so} \overset{2,3}{\otimes} \mathfrak{e}_6 \mathbf{I}^{\oplus 2}},
{1 \mathfrak{so} \overset{2,4}{\otimes} \mathfrak{e}_6 1},
{1 \mathfrak{e}_6 \overset{3,2}{\otimes} \mathfrak{so} 1 3},
{1 \mathfrak{e}_6 \overset{2,2}{\otimes} \mathfrak{so} 1 3 2},
{3 1 \mathfrak{so} \overset{2,2}{\otimes} \mathfrak{e}_6 \mathbf{I}^{\oplus 2}},
{\mathfrak{so} \overset{2,2}{\otimes} \mathfrak{e}_6 1 2 2 3},
\nonumber\\
&{\mathfrak{so} \overset{2,3}{\otimes} \mathfrak{e}_6 1 2 3},
{\mathfrak{so} \overset{2,4}{\otimes} \mathfrak{e}_6 1 3},
{\mathfrak{e}_6 \overset{4,2}{\otimes} \mathfrak{so} 1 3},
{\mathfrak{e}_6 \overset{3,2}{\otimes} \mathfrak{so} 1 3 2},
{\mathfrak{e}_6 \overset{2,2}{\otimes} \mathfrak{so} 1 3 2 2},
{1 \mathfrak{so} 1 \mathfrak{so} 1 \overset{1}{\mathfrak{so}} 1 \mathfrak{so}}.
\label{eq:short-P-9}
\end{align}
\subsection*{$T^H=9$}
There are $74$ short {\it P}-type LST bases with $T^H=9$.  They are
\begin{align}
&{3 1 \overset{3}{\overset{1}{5}} \overset{2,4}{\otimes}},
{\overset{4,2}{\otimes} \overset{1}{5} \overset{2,2}{\otimes}},
{\overset{3,2}{\otimes} \overset{1}{5} \overset{2,3}{\otimes}},
{2 \overset{1}{3} 1 5 1 3 2 2 1},
{\overset{5,3}{\otimes} 3},
{\overset{3,3}{\otimes} 5 1 3 2 1},
{1 3 2 1 5 1 3 2 2 1},
\nonumber\\
&{1 \overset{1}{5} \overset{3,3}{\bigcirc}},
{1 5 \overset{4,3}{\otimes}},
{2 1 5 \overset{3,3}{\bigcirc}},
{5 \overset{5,3}{\otimes}},
{1 2 3 1 5 1 3 2 1 \mathfrak{so}},
{1 5 \overset{3,3}{\bigcirc} \mathfrak{so}},
{5 \overset{4,3}{\otimes} \mathfrak{so}},
\nonumber\\
&{3 1 \overset{1}{\underset{1}{\mathfrak{e}_6}} 1 3 1 5 1},
{3 1 \overset{\mathbf{I}^{\oplus 2}}{\mathfrak{e}_6} 1 3 1 5 1},
{1 \overset{1}{5} \overset{2,4}{\otimes} \mathfrak{e}_6 1},
{2 2 3 1 \overset{1}{5} \overset{2,2}{\otimes} \mathfrak{e}_6},
{2 3 1 \overset{1}{5} \overset{2,2}{\otimes} \mathfrak{e}_6 1},
{2 3 1 \overset{1}{5} 1 3 2 1 \mathfrak{e}_6},
{2 3 2 1 \overset{1}{\mathfrak{e}_6} \overset{2,2}{\otimes} 5},
\nonumber\\
&{3 1 \overset{1}{5} 1 3 2 1 \mathfrak{e}_6 1},
{3 1 \overset{1}{5} \overset{2,4}{\otimes} \mathfrak{e}_6},
{3 1 \overset{1}{\mathfrak{e}_6} 1 2 3 1 5 1},
{3 1 \overset{1}{\mathfrak{e}_6} \overset{3,3}{\otimes} 5},
{3 2 1 \overset{1}{\mathfrak{e}_6} 1 3 1 5 1},
{2 1 5 1 3 2 2 1 \mathfrak{e}_6 1},
{2 3 1 5 \overset{3,3}{\otimes} \mathfrak{e}_6},
\nonumber\\
&{2 3 2 1 5 1 3 1 \mathfrak{e}_6 1},
{2 3 2 1 5 1 3 2 1 \mathfrak{e}_6},
{2 3 2 1 \mathfrak{e}_6 \overset{3,2}{\otimes} 5},
{3 1 5 \overset{3,3}{\otimes} \mathfrak{e}_6 1},
{3 1 \mathfrak{e}_6 1 2 2 3 1 5 1},
{3 2 1 5 1 3 2 1 \mathfrak{e}_6 1},
{3 2 1 5 1 3 2 2 1 \mathfrak{e}_6},
\nonumber\\
&{3 2 1 \mathfrak{e}_6 1 2 3 1 5 1},
{3 2 1 \mathfrak{e}_6 \overset{3,3}{\otimes} 5},
{3 2 2 1 \mathfrak{e}_6 1 3 1 5 1},
{3 2 1 \overset{\mathbf{I}^{\oplus 2}}{\underset{1}{\mathfrak{e}_7}} 1 2 3},
{3 2 1 \overset{3}{\overset{2}{\overset{1}{\mathfrak{e}_7}}} \mathbf{I}^{\oplus 3}},
{3 2 2 1 \overset{1}{\underset{1}{\mathfrak{e}_7}} 1 2 3},
{3 2 2 1 \overset{\mathbf{I}^{\oplus 2}}{\mathfrak{e}_7} 1 2 3},
\nonumber\\
&{3 2 2 1 \overset{1}{\mathfrak{e}_7} 1 2 2 3},
{\overset{2,3}{\otimes} \mathfrak{e}_7 \mathbf{I}^{\oplus 5}},
{2 2 3 1 3 2 2 1 \mathfrak{e}_7 1},
{2 3 1 3 2 1 \mathfrak{e}_7 \mathbf{I}^{\oplus 3}},
{3 1 3 2 2 1 \mathfrak{e}_7 \mathbf{I}^{\oplus 3}},
{\overset{2,2}{\otimes} \overset{3}{\overset{1}{\mathfrak{e}_6}} \overset{2,2}{\otimes} \mathfrak{so}},
{1 \mathfrak{so} \overset{2,3}{\otimes} \mathfrak{e}_6 \overset{2,2}{\otimes}},
\nonumber\\
&{3 1 \mathfrak{so} \overset{2,2}{\otimes} \mathfrak{e}_6 \overset{2,2}{\otimes}},
{1 \overset{1}{\mathfrak{e}_7^{\prime}} \overset{4,2}{\otimes} \mathfrak{so} 1},
{1 \overset{1}{\mathfrak{e}_7^{\prime}} \overset{3,2}{\otimes} \mathfrak{so} 1 3},
{3 2 1 \overset{1}{\mathfrak{e}_7^{\prime}} \overset{3,2}{\otimes} \mathfrak{so}},
{1 \mathfrak{so} \overset{2,4}{\otimes} \mathfrak{e}_7^{\prime} \mathbf{I}^{\oplus 2}},
{1 \mathfrak{e}_7^{\prime} \overset{4,2}{\otimes} \mathfrak{so} 1 3},
{1 \mathfrak{e}_7^{\prime} \overset{3,2}{\otimes} \mathfrak{so} 1 3 2},
\nonumber\\
&{3 1 \mathfrak{so} \overset{2,3}{\otimes} \mathfrak{e}_7^{\prime} \mathbf{I}^{\oplus 2}},
{\mathfrak{so} \overset{2,3}{\otimes} \mathfrak{e}_7^{\prime} 1 2 2 3},
{\mathfrak{so} \overset{2,4}{\otimes} \mathfrak{e}_7^{\prime} 1 2 3},
{\mathfrak{e}_7^{\prime} \overset{4,2}{\otimes} \mathfrak{so} 1 3 2},
{\mathfrak{e}_7^{\prime} \overset{3,2}{\otimes} \mathfrak{so} 1 3 2 2},
{1 \mathfrak{so} 1 \mathfrak{so} 1 \mathfrak{so} \overset{2,2}{\otimes} 5},
{1 \overset{1}{\mathfrak{e}_6} \overset{2,2}{\otimes} \mathfrak{so} 1 \mathfrak{so} 1},
\nonumber\\
&{3 1 \overset{1}{\mathfrak{e}_6} \overset{2,2}{\otimes} \mathfrak{so} 1 \mathfrak{so}},
{\mathfrak{so} \overset{2,2}{\otimes} \overset{1}{\mathfrak{e}_6} \overset{2,2}{\otimes} \mathfrak{so}},
{1 \mathfrak{so} 1 \mathfrak{so} \overset{2,2}{\otimes} \mathfrak{e}_6 \mathbf{I}^{\oplus 2}},
{1 \mathfrak{so} 1 \mathfrak{so} \overset{2,3}{\otimes} \mathfrak{e}_6 1},
{1 \mathfrak{so} 1 \mathfrak{so} \overset{2,4}{\otimes} \mathfrak{e}_6},
{1 \mathfrak{e}_6 \overset{2,2}{\otimes} \mathfrak{so} 1 \mathfrak{so} 1 3},
{2 3 1 \mathfrak{so} 1 \mathfrak{so} \overset{2,2}{\otimes} \mathfrak{e}_6},
\nonumber\\
&{3 1 \mathfrak{so} 1 \mathfrak{so} \overset{2,3}{\otimes} \mathfrak{e}_6},
{3 1 \mathfrak{e}_6 \overset{3,2}{\otimes} \mathfrak{so} 1 \mathfrak{so}},
{3 2 1 \mathfrak{e}_6 \overset{2,2}{\otimes} \mathfrak{so} 1 \mathfrak{so}},
{\mathfrak{so} \overset{2,3}{\otimes} \mathfrak{e}_6 \overset{2,2}{\otimes} \mathfrak{so}}.
\label{eq:short-P-10}
\end{align}
\endgroup

\section{Comparison with geometric F-theory bases} \label{app:non-toric}
In this Appendix, we show how the F-theory bases for elliptic CY 3-folds classified in \cite{Taylor:2015isa} are fitted in our classification, which is available on the webpage~\cite{Taylor:webdata}.
We start from $T=2$ supergravity bases. The bases in \cite{Taylor:2015isa} are
\begin{align}
    \textcolor{red}{\overset{1,\cdots,12}{\mathcal{C}}}11,
\end{align}
which is the same as the one in \eqref{eq:T2_bases}. Note that, in \cite{Taylor:2015isa}, the external and LST curves are not distinguished, but here we have distinguished them by coloring the external curves in red to make the comparison with our results easier. We will follow this convention later as well.

For $T=3$ supergravity bases, the bases in \cite{Taylor:2015isa} are
\begin{align}
    &1\overset{\textcolor{red}{\overset{1,2,3}{\mathcal{C}}}}{2}1,
    &&121\textcolor{red}{\overset{2,\cdots,12}{\mathcal{C}}},
    &&212\textcolor{red}{\overset{2,3}{\mathcal{C}}},
    &&11\textcolor{red}{\overset{2,\cdots,12}{\mathcal{C}}}11,
    &&\text{L}:111111.
\end{align}
The first four bases are the same as the ones in \eqref{eq:T3_bases}. 
Furthermore, as we have discussed in the main text, we can remove $-2$ external when it does not host any gauge algebra. For instance, $1\overset{\textcolor{red}{2}}{2}1$ is Weyl equivalent to $11\textcolor{red}{1}11$.
The last one is the topology of ${\rm dP}_3$, the del Pezzo surface obtained by blowing up $\mathbb{P}^2$ at three points, which can also be understood as two $11$ LSTs sharing a common external $-1$ curve, such as $11\textcolor{red}{1}11$, where the remaining $-1$ curve is not an independent generator but is fixed uniquely by the others. In the following, we omit such redundant curves to avoid complications.

For $T=4$ supergravity bases, omitting redundant curves for simplicity, the bases in \cite{Taylor:2015isa} are
\begin{align}
    &\textcolor{red}{\overset{2,3,4,5,6}{\mathcal{C}}}1\overset{1}{3}1,
    &&1\underset{\textcolor{red}{\underset{1,2}{\mathcal{C}}}}{\overset{1}{3}}1,
    &&21\overset{\textcolor{red}{2}}{3}1,
    &&\textcolor{red}{\overset{2,3}{\mathcal{C}}}\overset{1}{2}21,
    &&\textcolor{red}{2}2\overset{1}{2}2,
    \nonumber\\
    &1\overset{1}{2}\textcolor{red}{\overset{2,3}{\mathcal{C}}}11,
    &&11\textcolor{red}{\overset{2,3}{\mathcal{C}}}212,
    &&11\textcolor{red}{\overset{3,\cdots,12}{\mathcal{C}}}121,
    &&11\textcolor{red}{1}121,
    &&11\overset{1}{\overset{1}{\textcolor{red}{\underset{\text{any}}{\mathcal{C}}}}}11, \nonumber\\
    &\textcolor{red}{\overset{3,\cdots,12}{\mathcal{C}}}1221,
    &&2131\textcolor{red}{\overset{2,3,4,5,6}{\mathcal{C}}},
    &&\textcolor{red}{3}2131,
    &&2213\textcolor{red}{\overset{1,2}{\mathcal{C}}},
    &&\textcolor{red}{\overset{2,3}{\mathcal{C}}}2213.    
\end{align}
Here we again see several bases where the external is a $-2$ curve without gauge algebra.
Let us spell out one representative example of this equivalence. 
Consider the last configuration in the first line of the bases above, $\textcolor{red}{2}2\overset{1}{2}2$.
We denote the four $-2$ curves from left to right by
$\alpha_1,\alpha_2,\alpha_3,\alpha_4$, with $\alpha_1$ the external
curve, and denote by $E$ the $-1$ curve attached to $\alpha_3$. Thus
\begin{align}
    \alpha_i^2=-2,\qquad E^2=-1,\qquad
    \alpha_i\cdot\alpha_{i+1}=1,\qquad E\cdot\alpha_3=1 .
\end{align}
Since all the $-2$ curves in this configuration are M-strings, we can
remove them by Weyl reflections. The resulting extended BPS cone contains
the following $-1$ generators:
\begin{align}
    &P_0 = E, 
    &&P_1 = E+\alpha_3, 
    &&P_2 = E+\alpha_3+\alpha_4,\nonumber\\
    &P_3 = E+\alpha_2+\alpha_3, 
    &&P_4 = E+\alpha_2+\alpha_3+\alpha_4, 
    &&P_5 = E+\alpha_2+2\alpha_3+\alpha_4,\nonumber\\
    &P_9 = E+\alpha_1+2\alpha_2+2\alpha_3+\alpha_4 .
\end{align}
They satisfy $P_i^2=-1$, and the three pairs $(P_0,P_5),\, (P_1,P_4),\, (P_2,P_3)$ form three $11$ LST sectors with the same little-string charge $f=P_0+P_5=P_1+P_4=P_2+P_3$.
The remaining generator $P_9$ is an external $-1$ generator, since
\begin{align}
    P_9\cdot P_0=P_9\cdot P_1=P_9\cdot P_2=1,\qquad
    P_9\cdot P_5=P_9\cdot P_4=P_9\cdot P_3=0 .
\end{align}
Consequently, we view the base after Weyl reflection as three $11$ LSTs sharing a common external $-1$ generator $P_9$.
In this way, all the bases from \cite{Taylor:2015isa} are included in our construction~\eqref{eq:T4_bases}.

For $T=5$ supergravity bases, the bases in \cite{Taylor:2015isa} are
\begingroup\allowdisplaybreaks
\begin{align}
    &12\overset{\textcolor{red}{2}}{2}21,
    &&\textcolor{red}{2}2\overset{2}{2}21,
    &&12\overset{\textcolor{red}{\mathfrak{su}_{2}}}{3}12,
    &&\textcolor{red}{2}2\overset{1}{3}13,
    &&12\overset{1}{3}1\textcolor{red}{\overset{3,4,5}{\mathcal{C}}},
    \nonumber\\
    &21\overset{\textcolor{red}{1}}{4}12,
    &&2 1 \overset{1}{4} 1 \textcolor{red}{\overset{3,4}{\mathcal{C}}},
    &&121\textcolor{red}{\overset{2,3}{\mathcal{C}}}121,
    &&11\textcolor{red}{2}2\overset{2}{2}1,
    &&1\overset{1}{2}\textcolor{red}{\overset{2,3}{\mathcal{C}}}212,
    \nonumber\\
    &1\overset{1}{2}\textcolor{red}{3}121,
    &&1\overset{1}{3}1\textcolor{red}{\overset{3,4,5,6}{\mathcal{C}}}11,
    &&212\textcolor{red}{\overset{2,3}{\mathcal{C}}}212,
    &&11\textcolor{red}{\overset{2,3}{\mathcal{C}}}2213,
    &&2213\textcolor{red}{2}11,
    \nonumber\\
    &212\textcolor{red}{3}121,
    &&2131\textcolor{red}{\overset{3,4,5,6}{\mathcal{C}}}11,
    &&11\textcolor{red}{3}2131,
    &&11\textcolor{red}{\overset{4,\cdots,12}{\mathcal{C}}}1221,
    &&121\textcolor{red}{\overset{4,\cdots,12}{\mathcal{C}}}121,
    \nonumber\\
    &11\textcolor{red}{1}1221,
    &&1221\textcolor{red}{2}11,
    &&121\textcolor{red}{2}121,
    &&11\textcolor{red}{3}1221,
    &&11\textcolor{red}{1}2131,
    \nonumber\\
    &212\overset{1}{\overset{1}{\textcolor{red}{\underset{1,2,3}{\mathcal{C}}}}}11,
    &&121\overset{1}{\overset{1}{\textcolor{red}{\underset{\text{any}}{\mathcal{C}}}}}11,
    &&11\underset{1}{\underset{1}{\overset{1}{\overset{1}{\textcolor{red}{\overset{\text{any}}{\mathcal{C}}}}}}}11,
    &&1\underset{\textcolor{red}{2}}{2}\overset{1}{3}1,
    &&{\textcolor{red}{2} 1 3 \overset{1}{2} 2},
    \nonumber\\
    &1\overset{\textcolor{red}{\mathfrak{su}_{2}}}{3}131,
    &&{\textcolor{red}{\overset{3,4,5,6,7,8}{\mathcal{C}}} 1 2 \overset{1}{3} 1},
    &&{\textcolor{red}{\overset{3,4,5}{\mathcal{C}}} 1 3 \overset{2}{2} 1},
    &&\textcolor{red}{\mathfrak{su}_{2}}{3 1 \overset{1}{3} 2},
    &&312\overset{\textcolor{red}{\mathfrak{su}_2}}{3}1,
    \nonumber\\
    &{\textcolor{red}{\overset{2,3}{\mathcal{C}}} 2 1 \overset{1}{4} 1},
    &&11\textcolor{red}{\mathfrak{su}_2}\underset{1}{\overset{1}{3}}1,
    &&1\underset{1}{2}\overset{1}{\overset{1}{\textcolor{red}{3}}}11,
    &&12\underset{\textcolor{red}{\mathfrak{su}_2}}{\overset{1}{3}}1,
    &&\textcolor{red}{\overset{2,3,4}{\mathcal{C}}}{1 \overset{1}{\underset{1}{4}} 1},
    \nonumber\\
    &{\textcolor{red}{\overset{2,3,4,5}{\mathcal{C}}} 1 3 2 1 3},
    &&12312\textcolor{red}{3},
    &&\textcolor{red}{\mathfrak{su}_2}31231,
    &&{\textcolor{red}{2} 2 3 1 2 3},
    &&23123\textcolor{red}{\mathfrak{su}_2},
    \nonumber\\
    &{\textcolor{red}{\overset{3,4,5,6}{\mathcal{C}}} 1 3 1 3 1},
    &&\textcolor{red}{2}31313,
    &&{1 2 2 2 1 \textcolor{red}{\overset{4,\cdots,12}{\mathcal{C}}}},
    &&{1 4 1 2 2 \textcolor{red}{2}},
    &&{4 1 2 2 2 \textcolor{red}{2}},
    \nonumber\\
    &{2 2 1 4 1 \textcolor{red}{\overset{2,3,4}{\mathcal{C}}}},
    &&\textcolor{red}{3}22141,
    &&\textcolor{red}{3}21412,
    &&{2 1 3 2 1 \textcolor{red}{\overset{4,5,6,7,8}{\mathcal{C}}}},
    &&1^{\otimes 4}4\textcolor{red}{1}, 
\end{align}
\endgroup
where $1^{\otimes 4}4\textcolor{red}{1}$ represents the configuration of four $-1$ curves attached to a $-4$ curve, and the remaining $-1$ curve is an external curve.
We have also confirmed that all the bases above are included in our classification after applying Weyl reflections appropriately. 

Finally, the $T=6$ supergravity bases in \cite{Taylor:2015isa} are included in our classification as well, though we do not write all patterns for brevity.

\bibliographystyle{JHEP}
\bibliography{refs}
\end{document}